\documentclass[11pt,a4paper]{article}
\usepackage[left=3cm, right=3cm, top=3.5cm]{geometry}
\usepackage{amssymb, amsmath, amsthm, amsfonts, rotating}
\usepackage{mathrsfs, color,bbold,url}
\usepackage{stmaryrd}
\usepackage{tikz}
\usepackage[all]{xypic}
\usepackage{enumerate}
\usepackage{siunitx} 
\usepackage{multirow} 
\usepackage{lscape}
\usepackage{multirow}
\usepackage[]{mdframed}

\newtheorem{theorem}{Theorem}[section]
\newtheorem{lemma}[theorem]{Lemma}
\newtheorem{corollary}[theorem]{Corollary}
\newtheorem{proposition}[theorem]{Proposition}

\theoremstyle{definition}
\newtheorem{definition}[theorem]{Definition}

\newtheorem{example}[theorem]{Example}

\newcommand{\fmodels}{=\!\!\!|}

\usepackage{adjustbox}
\usepackage[most]{tcolorbox}

\usepackage{tikz-qtree}
\usepackage{tikz-qtree-compat}

\tikzset{every tree node/.style={align=center,anchor=north}}

\makeatother

\begin{document}

\title{  Normal Nested Answer Set Programs: \\ 
Syntactics, Semantics and  Logical Calculi }
\author{Gonzalo E. Imaz\\ [2mm] %
{\small  Artificial Intelligence Research Institute (IIIA - CSIC)}\\ {\small Campus UAB,
 Bellaterra 08193,  Barcelona, Spain}\\
{\small \tt email: {gonzalo}@iiia.csic.es}\\[1mm]
}

\date{}
\maketitle

\begin{abstract}
 
Nested answer set   programming  
(NASP; Lifschitz et {\em al.},  1999)  
generalizes answer set   programming (ASP) by admitting
nested expressions in rule bodies and heads,
and thus, NASP allows to  exploit  
program succinctness. 
Yet, although NASP expressiveness
is undoubtedly superior to  ASP one, 
     the former's  reasoning capabilities 
   remain unexplored.  
This reality seems   subsequent  to 
the next existing wide-ranging  gap:
normal nested   programs (NNPs) are not known, or in other words,
   the  nested   normal-disjunctive   boundary  
is unidentified thus far. Such an unfavorable situation is 
yet antagonistic to that of ASP as  its
   normal     programs  (NPs)   have been vital  
    for propelling  ASP.
We  will fill  such a gap by  defining the NNPs,   their semantics 
  and their associated nested logical calculi.  Besides,
  {while the unique known way to  compute 
 nested  programs is 
  unfold them back,  
   we  propose to do so in their  original   form.}
 
       By faithfully emulating   ASP  in which  positive-Horn 
  clauses  underpin  NPs, we
   propose positive-Horn nested-expressions to underpin NNPs and   furnish 
     heterogeneous arguments  in support
 of our defined NNPs being the  rightful generalization 
 of  NPs. We  synthesize the syntactical,
     semantical  and  computational 
      relationships between NNP and NP programs    
  in the following four    features: {\em   (a)
  NP  heads are  single  literals, while  NNP  heads  can have 
 infinitely many literals; 
      (b)  any NNP  program is strongly equivalent 
 to an NP program  but the former can be
    exponentially smaller; (c)  NNPs  comply with most
     semantical properties of NPs,
        particularly,  a {\em not}-free    NNP program 
   has a least model;  and 
   (d)  NNP programs  behave complexitywise  like   NP ones.}
  
 \underline{Firstly}, we give the    syntax of NNPs.
 For that, we   initially   define the positive-Horn nested-expressions and 
 then  an NNP rule as one whose 
 head (resp. body) is  a  positive-Horn 
 (resp. general) nested-expression.  
 \underline{Secondly}, we set up the 
 semantics of  NNPs  by  
 lifting  to the nesting level,  classical NP notions     
           including:  answer set,  minimal and least model, 
 closedness,  supportedness, 
     immediate consequence
   operator  and program consistency. 
   We besides show that NNP restricted to ASP coincides with NP.     
 {\underline{Thirdly},} we introduce nested logical  calculi, concretely, {\em nested unit-resolution}  
 and {\em nested hyper unit-resolution,} 
   proving that   they  
  recover unit-resolution and hyper unit-resolution in the   ASP setting.  
    We also show how both nested logical calculi allow to process the least model of   {\em not}-free  
 NNP programs. To end, we   demonstrate that computing answer sets of  
   (resp. {\em not}-free) NNP programs    is   
  (resp. $\mathcal{P}$-complete)    $\mathcal{NP}$-complete.

 
\vspace{.25cm} 
\end{abstract}



\newenvironment{niceproof}{\trivlist\item[\hskip
\labelsep{\it Proof.\/}]\ignorespaces}{\hfill$\blacksquare$\endtrivlist}

\newenvironment{proofsketch}{\trivlist\item[\hskip
\labelsep{\it Proof Sketch.\/}]\ignorespaces}{\hfill$\blacksquare$\endtrivlist}

\newenvironment{niceproofsketch}{\trivlist\item[\hskip
       \labelsep{\bf Proof Sketch.\/}]\ignorespaces}{\hfill$\blacksquare$\endtrivlist}



\section{Introduction} \label{sec:INTRODUCTION}

\subsection{Research Framework} \label{subsec:RESEARCH-FRAME}

The invention of the  logic programming  (LP)  paradigm   \cite{Colmerauer73,Kowalski74,EmdenK76,Colmerauer83}  was   pioneer 
in artificial intelligence 
(AI),  having supported the latter along its history \cite{Brewka2005,lallouet2020logic} as shown
 in the  comprehensive recently   published survey  \cite{KornerLBCDHMWDA22} over its fifty years of  
 developments.  Nowadays, 
LP lies at the core of a family of state-of-the-art  AI
fields such as 
  answer set programming (ASP) \cite{CBaral03,Lifschitz19},
constraint logic programing \cite{JaffarL87,Rossi06}, constraint handling rules \cite{Fruhwirth94,Fruhwirth18}, 
constrained Horn clauses \cite{de2021analysis,Gurfinkel22,KSG22} 
and abductive logic programming \cite{KakasKT92}, 
  all of which share, as a common   endeavor, 
  to create modeling and  reasoning techniques
  allowing for  effectively solving both commonsense and combinatorial problems.

\vspace{.05cm}
In addition, the  theoretical foundations and solving mechanisms of LP and those of 
its aforesaid related fields 
  were initially constructed      on   first-order logic, but 
  later on, they  were also   exported to description logic \cite{EiterILST08}, possibilistic logic \cite{NievesOC13},
multivalued logic \cite{DohertySzalas18}, fuzzy logic \cite{KrajciLV04},  
modal logic \cite{Nguyen09}, 
temporal logic \cite{GaintzarainL13}, rough sets \cite{DohertyS21} and more.

 \vspace{.05cm}  
 This  paper provides contributions to  ASP, 
  but given that ASP is the non-monotonic prolongation of  LP,
  the latter paradigm  is indirectly at the center of this  work as well.  
 And this also means that our proposal is connected  
 with the above mentioned family of AI fields and related  group of 
logics, and  thereby,  its scope can be, to some extent,
of benefit to a variety of current AI scenarios.  
Next we  circumscribe  our ASP research context.

\vspace{.05cm} 
   The literature shows that LP  under  
 the answer set semantics  paradigm  \cite{GelfondL88,GelfondL91,Lifschitz99,MarekT99,Niemela99} is widely
 acknowledged as a versatile formalism for 
 knowledge representation and deduction  \cite{Lifschitz02,GelfondL02a,CBaral03,Lifschitz08,EiterIK09,Gebser2012,Lifschitz19} 
 and, in fact,  interest in its theoretical
  and practical  advances
 has  grown  considerably over the last twenty years. 
Among the practical realms in which ASP has been used to solve complex
problems, one can mention: Artificial Intelligence \cite{GagglMRWW15}, Bioinformatics \cite{CampeottoDP15}, Databases \cite{MannaRT15}, 
Game Theory \cite{AmendolaGLV16}, Diagnosis \cite{WotawaK22}, etc. We stress  that 
ASP has also been  applied to solve industrial applications e.g. \cite{DodaroLNR15,DodaroGLMRS16}.

\vspace{.05cm} 
 Concerning the  enhancements that
  research    in declarative semantics has performed, 
   the  one  experienced  in  augmentation of expressiveness 
    is noteworthy. 
   Within this concern,  we will concentrate on  a general  formalization that 
    was authored by Lifschitz et {\em al.} \cite{LifschitzTT99} and    
  called nested answer set programming (NASP), wherein
 rule  heads and bodies   are allowed to be {\em nested   expressions}, i.e.
  non-clausal formulas without the connectives 
 for implication and logical equivalence, 
 i.e.  constituted by an unlimited nesting of the connectives of
  negation, conjunction and   disjunction. 

 \vspace{.05cm} 
We accept   both  classical   and default negations  
but limit   their {\bf scope   to only literals},
that is,  we compute programs in the so-called  negation
normal form (NNF). This is not a proper limiting restriction
as a non-NNF  program    can be  translated  into a strongly-equivalent NNF one
by using the  transformations already developed  \cite{AguadoCF0PV19}.
Likewise,   the nesting of default literals  can also be eliminated
with the  transformation rules proposed in reference \cite{LifschitzTT99}.
On the other hand,  we permit   default literals to occur    in rule heads only.

\vspace{.05cm}  
We    consider nested programs as propositional objects; as usual, rules with variables
are viewed as schemata   representing their ground instances and
will appear only in some   
 illustrating  (borrowed from the literature) examples. 
This is because the difference between a rule with variables
and the set of its ground instances is not essential semantically.

 \vspace{.05cm}
The only option contemplated in the literature to compute   a   nested program 
  is converting it into  an  unnested  one, and then, 
  make use of the available techniques developed for solving classical programs.
  To unfold the nested program,   two      translation types exist.

  \vspace{.05cm}
  The first one   applies distributivity
 to the input nested program until   an 
   equivalent unnested program is obtained, e.g. \cite{LifschitzTT99,Janhunen01,CabalarPV09}. 
   This translation is  obviously  impractical because   
   it unavoidably leads to an exponential blow-up of  the final unnested-program size.
   
   \vspace{.05cm}
The second    type of  translation 
 is   based on  recursively replacing sub-expressions by  fresh literals,  
as was originally proposed  in  \cite{Tseitin83}. Currently, it
turns out to be the unique real  means  considered thus far  in the state of the art
to process nested programs and, in fact,
a  large number of its variants e.g. \cite{ FerrarisL05,PearceSSTW02,Cabalar02,LinkeTW04,CabalarPV09,NievesL15,AguadoCF0PV19,Fandinno21}
 can be found.  

  \vspace{.05cm}
  This translation, however,  loses    the   original structure 
 of the input nested program and increases  its size and  
 number of variables, deficiencies that have been
 experimentally assessed in diverse fields (mentioned below) as   harmful to global  efficiency.
This translation   provokes  another  drawback:  
     clausal   programs
  hard to compute and solve can  be   obtained,   
but unfortunately,  no means exists to optimally guide  
 the translation   towards   
``easy programs" as this depends on  the  
   input nested program
and      employed ASP solver.

   \vspace{.05cm}
  Due to employing translators, a last operation   is still required
  consisting in   recovering the answers sets of the
  original  nested program by extracting them from those obtained 
  for the  transformed program. 
  This operation, even if linearly, increases the final overhead.

   \vspace{.05cm}
  
    The aforementioned
drawbacks of this second translation have been repeatedly reported
within many  AI domains, and just as an example, we mention the recent    
references \cite{EglySW06,books/daglib/0020523,EglySW09} whose authors
   have done so  within the    quantified boolean formulas domain.
 
  \vspace{.05cm} 
In view of the exposed shortcomings, we abandon 
the assumption of employing  translators
  and will  process   programs  directly   
in their original nested structure.

\subsection{Motivation: A  Gap }

Concerning the pragmatical side, 
two   easily verifiable observations are outlined below.
The first one is that  NASP arises as a serious  
remedy for significantly mitigating ASP redundancies as it 
opens the possibility to take advantage of high succinctness       
of programs.  Specifically, NASP, vs.    ASP,  allows 
  for diminishing   up to an exponential factor  
  the number of literals and  connectives to be evaluated.   
 We illustrate this issue 
with an  example: 
   
\begin{example} \label{ex:INTRO-compact}  
The twelve ASP rules  in the first three lines  
 are equivalent to the   nested  
 $r$. 
  An overlined   literal  amounts to a negated one,
 its accurate meaning is explained later.
\begin{gather*}
  b \leftarrow a, \quad  b \leftarrow \mbox{\em not} \, e,
   \quad  b \leftarrow m, 
 \quad  \neg c \leftarrow a \wedge b,   \quad  
 \neg c \leftarrow  \mbox{\em not} \, e \wedge b,
   \quad  \neg c \leftarrow m \wedge b,  \\
     g \leftarrow a \wedge d, \ \ 
     g \leftarrow  \mbox{\em not} \, e \wedge d,
      \ \      g \leftarrow m \wedge d,\\
   g \leftarrow a \wedge \mbox{\em not} \, f,
 \ \ g \leftarrow  
 \mbox{\em not} \, e \wedge \mbox{\em not} \, f,
  \ \ g \leftarrow m \wedge \mbox{\em not} \, f. 
\end{gather*} 

\vspace{-.3cm}
$$ r= \bigwedge \ [ \ \vee (\overline{b}   \ \ \neg c )  \quad \
  b \ \quad  \bigvee \, ( g   \ \ \ \wedge  
  [ \overline{d} \ \  \overline{\mbox{\em not} \, f} \, ] \ )  \ \ ]   \longleftarrow   \  \vee ( a \ \, \mbox{\em not} \,  e  \ \,  m ).$$ 

\noindent One    notes the substantial savings   both  in  literal occurrences, from 33   in
  ASP     to  9    in the nested   $ r $, and  
 connectives   ({\em not}, $ \wedge $, $\vee $ and $\leftarrow$),  
 from    28     to 8. Furthermore,  in this particular  case,   
we can observe that $r$  has no redundancies,  viz. $r$
has not any repeated literal.
 \qed 
\end{example} 
  
  
Giving permission to users to   unlimitedly nest connectives imposes,
when designing    NASP solving mechanisms,
contemplating    syntactically involved nested programs,        
 but at the same time, one can benefit from     the  advantage   of 
 having   programs with great succinctness. Thus, the first
 practical observation makes it evident that
      efficiently computing   involved   programs is not
    theoretically straightforward but 
     is practically  necessary. 

 \vspace{.05cm}
The second and pending to be mentioned  pragmatical observation  
  is that the real-world problem-solving  experience accumulated  over decades and   
   reflected in many  articles  published 
   across a variety of AI areas,   e.g. \cite{EglySW06,books/daglib/0020523,EglySW09},
    reveals that   encoding 
   in some practical environments 
   does not lead to  a clausal  
 but to a nested  formula. 
 
 \vspace{.05cm}
 Both previous  pragmatical observations allude to
 two     entities,  i.e. nested programs  and  formulas,    
 which were proven  to be logically synonyms in  \cite{CabalarF07}.  
 That being so,    towards  tackling   both ASP-redundancies 
    and real-world problem-solving   
   via NASP, studying NASP in depth    and
   from   practical and theoretical  perspectives turns out to be an promising research line.  
   Our investigation has led  us to  a main  outcome 
   briefly described below. 

\vspace{.05cm}
Regarding the state of the art on NASP, it  witnesses  a high research 
activity since   numerous approaches addressing 
 diverse  objectives  have been published:
discovering  relationships between  NASP programs and weighted constraints \cite{Turner03},  
proposing     normal forms  \cite{ BriaFL08,  BriaFL09,Bria09c,Fandinno21},
 checking strong equivalence
of programs \cite{LifschitzPV01},
 lifting nested programs to possibilistic logic (within   real-world 
 environments)
 \cite{NievesL12,NievesL15}, 
showing the  equivalence between  arbitrary
propositional theories and nested programs \cite{CabalarF07},
embedding functions in NASP programs \cite{WangYZ13},
 envisaging nested disjunctions  with   preferences  
\cite{ConfalonieriN11},  
 extending  NASP with belief operators \cite{WangZ05}, 
and many other results surveyed in 
  related work. 
  
    \vspace{.05cm}
  We also stress after reviewing the literature, e.g. \cite{LloydT84,ErdemL01,Turner03,YouYM03}, that 
 rules  with
  a single literal as   head and a nested expression as body
  have   been  considered as normal nested.  Yet, 
  heads with a  literal is a proper    
 ASP limitation, underestimating the expressiveness of NASP   
  which also authorises allocating   nested expressions  to heads. 
 We will attempt to shed light on this matter and  show  that 
   NNPs are incomparably richer than those  
with a single literal in  heads as have exclusively  been  
 envisaged up to now. 

 \vspace{.05cm}
In spite of the shown interest in theoretical issues on NASP and
the  discussed pragmatical potential  of
its   high   succinctness, 
a   primary question is still 
open in the quest for obtaining competitive and scalable NASP efficiency:   
whether  normal  nested programs ({\em NNPs}) exist.
Namely, currently it is unknown whether a  neat frontier 
 separating normal from disjunctive nested      
   programs ({\em DNPs}) is drawable. 
 
 \vspace{.05cm}
   On the contrary, 
   looking at its ASP counterpart,  
 one   witnesses an opposite landscape since 
    its  normal    programs   (NPs) 
      have been fundamental for  enhancing  in many facets
       modeling and deducing with ASP. 
Therefore, it is reasonable to suppose that characterizing NNPs  
as well as their logical and computational
features would push NASP  forward  
just as the corresponding knowledge about NPs has done within ASP.

  \vspace{.05cm}
However, we should mention here that  opposite to the  visual distinction in ASP  
  between normal and  disjunctive  programs by heads being single literals and    
    literal disjunctions, respect.,  such a separation    is far from being
   visual and immediate  in NASP and this is so to the extent  
 of having remained   unsolved   to date.  Actually, 
 when elucidating such a distinction, 
   the faced  difficulty 
   has been   that    heads and bodies  are  nested expressions
   and yet   no  subclass of them
   has been discovered in the literature, which, consequently, 
     has made it infeasible 
        to draw borderlines and discern subclasses of nested programs.   

\subsection{Main Contribution}

The recently presented  Horn nested  expressions   
   \cite{Imaz2021horn,Imaz2023a,Imaz2023b} (named  
    Horn non-clausal formulas in  \cite{Imaz2021horn,Imaz2023a,Imaz2023b})
   as far as we know
  compose the first typified  meaningful  subclass of nested expressions.  
 Its sub-class  of positive-Horn expressions, 
    presented here for the first time,
  allows for unblocking 
  the above described bottleneck. 
 
 \vspace{.05cm}
  So,  first of all, we depict the Venn diagram in Fig. 1 
  to visualizing the  new  class  
     $\mathcal{H}^+$ of positive-Horn nested expressions 
    and its set relationships with other related classes
    (the classes are  syntactically defined, so the Venn diagram is meant to be of syntactical kind):
  
  \vspace{.2cm}  
   $\bullet$   {\bf Unnested}:  Horn  ($h$), positive-Horn ($h^+$)
      and    clausal formulas   ($\mathcal{C}$).

 \vspace{.05cm}       
    $\bullet$    {\bf Nested}: Horn ($\mathcal{H}$), positive-Horn  ($\mathcal{H}^+$)
 and  nested expressions ($\mathcal{NE}$).

\begin{center}
 \begin{tikzpicture}
  \begin{scope}[blend group=soft light]
    \fill[black!40!white]  (280:.9) ellipse  (6.cm and 2.4cm);
    \fill[green!100!white] (230:1.2) circle (1.5);
    \fill[red!100!white]    (310:1.2) circle (1.8);
    \fill[blue!100!white]   (310:1.2) circle (1.);
\end{scope}
  \node at (10:-3.5)       {\bf \ $\ \mathcal{NE}$};
    \node at (-3.:3.5)       {\bf \ $\ \mathcal{NE}$};
  \node at (200:1.6)      {$\mathcal{C}$};
    \node at (355:1.9)      {\bf $\mathcal{H}$};
     \node at (195:.3)    {\bf $h$};
  \node at (335:1.2)    {\bf $\mathcal{H}^+$};
 \node at (280:1.2)         {\bf $h^+$};
\end{tikzpicture}

\vspace{.15cm}
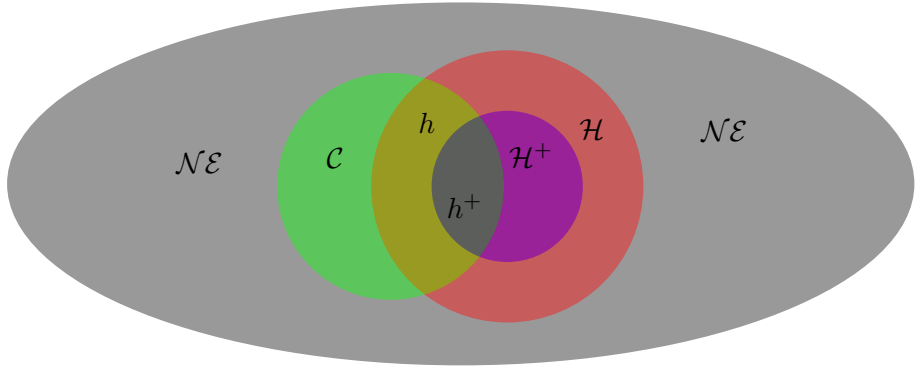
\small{{\bf Fig. 1.} The  class $\mathcal{H^+}$  and 
its related nested and unnested classes.}
\end{center}


With such a new result at hand,
 we   propose to project onto   NASP, the existing  criterion in ASP
  distinguishing
   normal from disjunctive   rules. Thus,
  by projecting positive-Horn clauses  corresponding to NP   rules
  onto NASP leads us to put forward the 
 {\em positive-Horn} nested  expressions
  as the  basis of NNP rules.
And this basic  Horn-like feature, 
common to  NPs and NNPs, interestingly 
allows us to demonstrate that NNPs  share  with NPs   many fundamental  logical and
computational features, namely,   
 that many well-known properties
of  NPs can be  extrapolated to the more general  
nesting level as they are retained by NNPs too, and thereby, that 
the proposed NNPs appear 
to be the legitimate  extension of    NPs.

\vspace{.05cm}
In particular, we
  unambiguously characterize syntactically  
    the NNP    (resp. DNP) 
    rules as those   whose   head is a positive-Horn (resp. positive)  nested 
  expression and 
 whose body is a general nested expression. 
 To put it another way, we neatly  draw  the 
 boundary separating    Horn-like or NNP 
 from non-Horn-like or DNP programs.

 \vspace{.05cm}
 It is worth mentioning here that, 
 just as  happens within the ASP framework, 
 exclusively the  head of a given nested rule makes 
 the difference to whether   the rule is either NNP or DNP. 
 After having characterized  NNPs, we establish their   associated semantical concepts 
and prove that, if   particularized to  ASP,   
they coincide with the ASP concepts.

\vspace{.05cm}
The    characterization  of  NNP programs and  
the  proof of  their main  properties 
 lead us to  summarize the  connections between NNP and NP as follows:  
  1) {\em syntactically}, 
  a  major  disparity exists between   NP and NNP heads as
the former are just  single literals  whereas the latter 
  can have    infinitely many  literals;
  2)  {\em semantically}, (a) every NNP program  
   is  strongly equivalent to an NP program but the former is up to
    exponentially     more succinct,  and (b)  
   NNPs exhibit essential semantical properties of NPs, and especially, 
   {\em not}-free   NNP  programs  just as   {\em not}-free  NP ones have
 a  least model (their potential unique answer set);    
 and 3)  {\em computationally}, if  NNP is restricted to NP, its polynomial and exponential complexities 
are those of   NP. 

\vspace{.05cm}
Among the future   possibilities  
gained thanks to    the  definition  of  the NNPs
and just  as an  example, we    mention  
 the  possibility of considering  to split  DNP programs into  NNPs,
 just as it is done with  disjunctive programs  and  NPs. 
  Afterwards, and  also mimicking ASP,
one can devise efficient mechanisms for computing NNPs, 
being simpler to conceive and implement than  for DNPs.  
We will furnish some illustrating examples of such an splitting, 
although   the  methodology will be formally described  
           in a forthcoming work.

\subsection{Presented Contributions}

Conveniently, our presentation will  introduce 
the   language   {\em flat-NNP}, 
placed  in between  NP and NNP, due to
  its provided support  
  to   more progressively describe   the somewhat sophisticated 
   NNP programs.   Altogether,  
  we define successively the   following three 
 {\em normal}   program classes:
  (i) NPs, based on positive-Horn clauses,  
(ii) flat-NNPs, 
based on positive-Horn clausal formulas, 
and (iii) NNPs, 
based on positive-Horn nested expressions. 
Below, we   describe  more technically 
the  contributions contained in the paper.

\vspace{.05cm}
\underline{Firstly},  
we introduce  the syntax of  NNPs.  For that,  
we   present, as the first contribution,  
the   positive-Horn nested expressions.  
 Then,   we stipulate that the head (resp. body) 
 of an NNP rule  is a positive-Horn  (resp. general) 
 nested expression, and
 that  as usual, a set of   NNP   rules 
   is an   NNP   program.  
 Subsequently, we will define  extended NNP programs,  
 which  generalize   
  extended NP programs  \cite{GelfondL91}  
      and  whose   nested head-consistency 
   will also be analyzed. 
   A rule whose head is a positive (non-Horn) nested expression   is DNP.          
  The following    example shows an overall combination
  of the preceding new syntactical items:
  
\begin{example} \label{ex:INTRO-rule}  
We give below   the body $\mbox{\it B}$ and the head $\mbox{\it H}$     
    of a  nested rule $ r $. 
   Classical and default literals
   are positive and they are negative when  are overlined.
  $ \phi_1 $ to $ \phi_3 $ are 
nested expressions. 
 We will show that if  $ \phi_2 $ is 
 fully-negative (contains solely negative literals)
  and    $ \phi_1 $ and $ \phi_3 $ are positive-Horn 
  nested,   then $H$ is positive-Horn  nested, which in turn implies 
 $$\mbox{\it B}=\bigwedge \, [\mbox{\em not} \, c  \ \  \ 
 \bigvee \, (d   \ \ \bigwedge \,  
[ \, \vee  (\neg a \ \    b  \ \ c \,) \ 
\ \bigvee \, (\neg d \ \ \wedge   [ \mbox{\em not} \, a  \ \  e \,] \, )
 \ \ \  b ] \ ) \ \ \ \  \neg b \ ]  $$
$$\mbox{\it H}=\bigwedge  \, [ f   \ \ \  \bigvee \, ( \overline{ g}  
 \ \ \ \bigwedge \ [ \
\vee \, ( \overline{f}  \ \ \overline{ not g}  \ \ c \,) \ 
\ \bigvee \,  (\phi_1 \ \ \wedge \,  
[\phi  \ \  \overline{not  f }\,] \, )
 \ \ \ g \,] \ \ ) \ \  \ {\phi_3} \, ]  $$

  
that  $ r $ is (extended)  NNP. Contrarily, if $\phi_1$ to $\phi_3$ 
 do not fulfill such conditions, then $H$ is non-Horn and
  $r$ would be (extended)  DNP.
 Returning to {\bf Example \ref{ex:INTRO-compact}}, we will verify that
$r$ is   an (extended) NNP rule since its head is  a positive-Horn nested expression.
\qed
\end{example}

 
\underline{Secondly}, we supply the semantics of  
 NNPs. To this aim, we    upgrade well-known semantical notions 
 of classical  NP 
     to the nesting level, which encompass: 
 answer set,  minimal and least model,  model
closed under and supported by a program,
immediate consequence operator and
 program consistency \cite{Turner94}.  Conversely, it is proven that the newly introduced  NNP  notions, if   
particularized to     ASP,  coincide
with  those of NP.
For instance,   
 the    nested immediate consequence operator, 
 denoted $ NT_P  $,  
 and  the  traditional operator, denoted  $ T_P  $,  
 fulfill the equality:  
  $ NT_P(I)= T_P(I) $  for classical NP programs.

\vspace{.05cm} 
\underline{Thirdly},  
we present  nested logical calculi to compute
 NNPs. Specifically, we establish
  {\em  nested   unit-resolution}  and 
{\it nested hyper unit-resolution } (given for      
possibilistic logic in \cite{Imaz2023a}  and   regular many-valued logic
 in \cite{Imaz2023b}) 
 showing  that, within  ASP,
they   recover  unit-resolution and hyper unit-resolution.
Subsequently, we prove that both logical calculi
 allows us to deduce  the least model
of a {\em not}-free NNP program 
and to do so
in polynomial time.   Since {\em not}-free NNP programs cover 
{\em not}-free  NP ones and NP logic programming 
  is $\mathcal{P} $-complete  \cite{DantsinEGV01}, so is NNP logic programming. 
This complexity and the fact  
that  an NNP  program  can be exponentially smaller than 
  its  strongly equivalent NP programs  enables us to state  that NNP efficiency can outperform
  NP efficiency up to an exponential factor. Also, such a polynomiality 
  allows us to  trivially demonstrate
  that finding   answer sets  of NNPs
  is $\mathcal{NP}$-complete. On the whole, one can check that
  the same polynomial and exponential complexities are met when tackling
  NNP as well as   NP programs \cite{MarekT91,DantsinEGV01}.

\vspace{.05cm}
 A further innovative aspect of  this work is 
that it paves the way for computing
 programs in their original nested  form,  
  {\em not requiring  translators.} 
In fact, NASP programs $P$ are computed  
  analogously as  ASP ones are:  
(a) for a given  interpretation $I$, obtaining the $I$-reduct  $P^I$ of  $P$;  (b) 
with  {\em  nested   unit-resolution},  determining the least model $LM(P^I)$  and then
 checking whether $I=LM(P^I)$ holds.

\vspace{.1cm}
 The   principal properties  exhibited by
   NNPs and   also applicable to head-consistent 
   extended NNPs, just as 
   one can do with NPs and extended NPs,   are summed up next: 
\begin{itemize} 
   
   \vspace{-.05cm}
   \item {\em not}-free NNP programs  have a least model.
 
     \vspace{-.25cm}
 \item Non-extended  NNP programs are head-consistent.
 

   \vspace{-.25cm}
 \item  Finding  the least model of a {\em not}-free   NNP is polynomial.
  
 \vspace{-.25cm}
 \item NNP logic programing is $\mathcal{P}$-complete.
 
 \vspace{-.25cm}
 \item  Finding   answer sets  of an NNP is $\mathcal{NP}$-complete.
 
   \vspace{-.25cm}
 \item Models of an NNP program  $ P $  are closed under $ P $.
 
  \vspace{-.25cm}
 \item Answer sets of an NNP  program $ P $  are supported by $ P $.
 
  \vspace{-.25cm}
  \item  Every NNP program is strongly equivalent to some  NP program.
 
  \vspace{-.25cm}
 \item  An NNP may compact exponentially their 
 strongly equivalent NPs.
 

 \vspace{-.25cm} 
 \item Nested  and classical immediate consequence operators   coincide  in ASP.

   \vspace{-.25cm}
 \item NNP and NP logical calculi coincide within  ASP.

 \vspace{-.25cm}
 \item NNP programs    embed
           NP ones as particular cases.
    

   \vspace{-.25cm}
 \item DNP programs are decomposable into   NNP ones.
  
\end{itemize}

  Such results hopefully signify a  starting point
 for   research aimed
 at achieving  NASP standards, 
 relative to   
  theoretical comprehension  and  practical efficiency,
  comparable to  ASP ones, since, as we understand it,
  the current  gap between them is barely explainable.

\subsection{Structure of the Paper}

  Section \ref{sec:NCbasis}  
 provides notation and   background.
  Section \ref{sec:General-Presentation} presents   the principle  
 of  our proposal.
  Section \ref{sec:NDs}    expresses  NP programs    
   from a slightly different   syntactical viewpoint.   
 Section  \ref{sec:FLAT-programs} describes 
 flat-NASP, an   in-between ASP and NASP language.
        Section \ref{sec:Positive-Horn-Nested-Form} presents, 
      first  informally and then giving   their (inductive)  
  formal specifications, the classes of Horn, 
       positive-Horn and positive-non-Horn
         nested expressions. We highlight that Subsection  \ref{sec:PROTERTIES-HORN-NEXPRESS}
   provides a  property of positive-Horn  nested expressions, which will be crucial
to  upgrade  semantic notions from NP to NNP.
    Sections \ref{sec:DefiniteNLP-Syntax} and \ref{sec:DefiniteNLP-Semantics}   
    present, respect.,  the     syntax and semantics of NNP programs.     
   Section  \ref{sec:PUR-HPUR}  introduces 
   nested logical calculi,       
         shows how to compute NNP programs in their nested original form
         and provides      complexity results.
     Section \ref{sec:related-work}  and \ref{sec:future-work}
     focus on  related and future work, respectively. The last section    
        summarizes our   results.
        Formal proofs requiring some development are relegated 
        to an appendix.

\section{Background and Notations} \label{sec:NCbasis}

We   supply   background on: $(i)$   normal ASP 
including classical concepts, mostly  from \cite{Lifschitz96},
 to   be lifted to the nesting level later on; and  
 $(ii)$   nested ASP;
 the reader is referred
to e.g.   \cite{BenAri2012} for a  
complete background on  nested expressions (non-clausal formulas).

\subsection{Normal ASP} \label{subsec:Background-LP}

\begin{definition} \label{def:Basic-Basic} Our language has
 the   sets of symbols:  
   individual constants   and variables;
relations $\mathcal{R}$; truth constants $ \{\bf \bot, \top\} $;
     connectives  $\{\wedge,  \vee, \neg, \mbox{\em not}, \leftarrow\}$ 
 and auxiliary symbols 
``(",  ``)",    ``["   and   ``]". Connective
  $  \neg $ (resp. $\mbox{\em not}$) is called
   classical (resp. default) negation.  
  $r ( \overline{\tau})$ (resp. $\neg r ( \overline{\tau})$), where
  $r \in \mathcal{R}$ and $\overline{\tau}$ consists of individual constants and/or variables,
  is a   positive (resp. negative) classical literal. Positive literals are also 
  called  atoms.     
    A fully-instantiated literal  is ground. 
    $\mathcal{A}=\{a,b, \ldots\} $ (resp.  
 $\mathcal{L}  =\mathcal{A} \cup 
 \{\neg a, \neg b   , \ldots\}$) denotes
        the set of ground atoms (resp. 
          literals).        
$\mathcal{A}_{not}=\{\mbox{\em not} \, a : a \in \mathcal{A}\}$        
       (resp. $\mathcal{L}_{not}=\{\mbox{\em not} \, \ell : \ell \in \mathcal{L}\}$)
        denotes the set of default ground atoms (resp. literals). 
We will refer  to   literals   and truth constants 
 $ \mathcal{E}_{\mathcal{A}}^1=\mathcal{A} \cup \mathcal{A}_{not} \cup \{\top,\bot\}$ 
       and $\mathcal{E}_{\mathcal{L}}^1 = \mathcal{L} \cup \mathcal{L}_{not} \cup \{\top,\bot\}$ 
    as  elementary expressions. 
\end{definition}

   We  generally deal   
  with  fully-instantiated or ground literals, rules and 
  programs (non-ground ones will only be employed  in
a few examples), and so,  
will omit the term {\em ground}.

\begin{definition} \label{def:LP-rule-BASIC}
A formula
$ h_1,   \ldots, h_n  \leftarrow
   b_1,     \ldots,    b_k,  
    \mbox{\em not} \, a_{1},    \ldots   \mbox{\em not} \, a_{m}$, 
     with 
   $h_i \in  \mathcal{A} \cup \{\bot\}$, $b_j,a_l \in   \mathcal{A}$, is an ASP rule $r$.  
   $H=h_1,    \ldots, h_n $     
and 
$B=b_1,    \ldots, \mbox{\em not} \, a_{m}  $ are 
  its  head  and body, respect., and we will denote
   $B^+=\{b_1,   \ldots, b_k\} $ and 
   $B^-=\{a_1,   \ldots, a_m\} $. Rule $r$ 
     is   normal ASP or NP$_\mathcal{A}$
  if $n=1$, i.e. $H=h$,  and    disjunctive ASP or DP$_\mathcal{A}$ otherwise.   An   NP$_\mathcal{A}$ (resp. DP$_\mathcal{A}$) program  
is a  set of $ \geq 1$ NP$_\mathcal{A}$ 
(resp. $  \geq 0$ NP$_\mathcal{A}$  and $ \geq 1$ DP$_\mathcal{A}$) rules. 
 An {\em NP$_\mathcal{A}$ rule}:

\vspace{.2cm}
            
\ \  -- $r$ is a  {\em   fact},  \ if      $ B=\emptyset $ 
 and $ h  \in  \mathcal{A} $. 
 
 \ \  -- $r$ is  \mbox{\em not}-{\em free},   
\  if    $ B^-=\emptyset $.

  \ \    -- $r$ is a  {\em  constraint}, \ if   $ h  = \bot $.   
\end{definition}

\begin{definition} Extended normal (NP$_\mathcal{L}$)  and 
  disjunctive (DP$_\mathcal{L}$)   ASP  programs 
 are defined as NP$_\mathcal{A}$ and DP$_\mathcal{A}$ programs, resp.,
 by just  exchanging $ \mathcal{A}$ for $ \mathcal{L} $ in Definition \ref{def:LP-rule-BASIC}. 
 We will refer to   NP$_\mathcal{A}$ and  NP$_\mathcal{L}$ 
rules/programs  as NP rules/programs. An NP 
 program   is said head-consistent if   
   its set of head  literals   $h \neq \bot$   is consistent.
\end{definition}
 



\begin{definition} \label{def:Model-BASIC}  
Let  $ P $ be   an NP program and   let  $r=h \leftarrow B \in P$. An interpretation  $  I $  is any    subset of   $  \mathcal{L}   $ which 
does not contain any complementary pair 
    $\{ a, \neg a \}$, for any $ a \in \mathcal{A} $.
  $I$ is a model of    $r  $,  
  denoted  $ I \models r $, if: 
$$ B^+  \not \subseteq I  \quad \mbox{or}  
 \quad  B^- \cap I \neq \emptyset \quad
 \mbox{{or}} \quad h  \in  I.$$


-- An interpretation $ I$  is a model of $P$, written $ I \models P $,  
  if  \ $ \forall r  \in P, \, I \models r $.
  
--  A model $ M $ of  $P$ is  minimal        if no  subset of $ M $ is a model of  $P$. 
 
-- When $P$  has a unique minimal model or least model, 
  it is  denoted by  $LM(P) $.

\end{definition}

\begin{definition} \label{def:inter-clsoed-BASIC}  
An  interpretation $I$
 is   closed under   an  NP program
  $ P $ if: 
    $$\forall (h \leftarrow B)  \in P: \quad h \in I  
\ \ \mbox{whenever}  \ \   B^+   \subseteq I    
\ \ \mbox{and}  \ \  B^- \cap I = \emptyset .$$ 
\end{definition}

\begin{definition} \label{def:inter-support-BASIC}  
An  interpretation $I$  is   supported by  an  NP program
 $ P $ if  $\forall \ell \in I $ we have:         
    $$    
  \ \exists (h \leftarrow B)  \in  P :  \quad    h=\ell,  \ \
    B^+   \subseteq I  \ \  \mbox{and} \ \  B^- \cap I = \emptyset .$$ 
  
\end{definition}

\begin{definition}  \label{def:CONSEQ-IMM-OPE} 
(See \cite{EmdenK76}) Let $ P $ be 
a  constraint-free
head-consistent NP  
program, ${B_P}$ the propositions occurring in $P$ (Herbrand universe)   and   
 $ I   $   an
interpretation.
 The   immediate-consequence-operator  $ T_P: 2^{B_P}\rightarrow 2^{B_P}$ 
   is defined by:
$$T_P (I)=\{h \, \vert \vert  \ \, h \leftarrow B \in P, \ \,
 B^+   \subseteq I,  \  \,  B^- \cap I = \emptyset \}.$$


-- $ T^i_P $ is defined as: 
$ T^0_P=\emptyset \ \mbox{and} \  T^{i+1}_P=T_P(T^i_P)  $. 

-- $ I $ is a fixpoint of 
$ P $ if $ T_P(I)=I $. \ \ 

-- $ I $ is  closed under
$ T_P $ if $ T_P (I) \subseteq I$.  

--    The  least fixpoint $ \mbox{\it lfp}(T_P) $ of  $ P $  
  is the smallest    $ I $ closed under $ T_P $.
\end{definition}

\begin{definition}  \label{def:ANSWER-I}
 (See   \cite{GelfondL88,GelfondL91}.)  
 Let  $P$ be an NP  program and $I $   an interpretation.   
 The    $I$-reduct $ P^I $ of $P$  is the {\em not}-free program
 defined as: 
$$P^I=\{h \leftarrow B^+  \ \vert \vert \   
h \leftarrow B \in P, 
\ \ B^- \cap I=\emptyset\} $$


--   $ I$ is  an answer set of $ P $ if  
$ I=LM(P^I) $. 
 
-- $ P $ is consistent if it has  an answer set and inconsistent otherwise. \qed
\end{definition}

 Althought there exist  several   definitions  of an answer set \cite{Lifschitz10},
we retain the previous one  as   it is the most suitable for our goal.
We now review some   
 properties of NP programs.
 
 \begin{theorem} \label{TH:Tarski55-Endem76} (See \cite{Tarski55,EmdenK76}.)
 For  any  {not-}free,  constraint-free and
 head-consistent NP program  $ P $, we have: 
$$ LM(P)=\bigcup_{0 \leq i} \,T^i_P= \mbox{\it lfp}(T_P).$$
\end{theorem}

\begin{proposition} \label{prop:BASIC-Properties-not-free-NP}  
If $  P$ is a {not-}free and constraint-free NP$_\mathcal{A}$ program and 
 $ Cn(P) $   denotes
  the smallest  interpretation  $ I $   closed under  $P$, then:

\vspace{.2cm}
(a)    $P$ is  head-consistent.

(b) $ Cn(P) $ is consistent and supported by $P$.

(c)    $LM(P) $ is the unique  answer set   of $P$ .
    
\end{proposition}  

\begin{proposition} \label{prop:BASIC-Properties-Extended-NP}  
If $  P$ is a {not-}free, constraint-free and 
head-consistent NP$_\mathcal{L}$ program, then:
$P$,  $Cn(P)$ and $LM(P)$ verify properties (b) and (c) in 
Proposition  \ref{prop:BASIC-Properties-not-free-NP}.  
%
%
\end{proposition}


\begin{proposition} \label{prop:BASIC-Propoerties-ANSWER-SET} 
Let   P be a constraint-free NP  program, S be an answer set of P
and  $\Sigma$ be a set of constraints. We have:

\vspace{.2cm}

(1) Any model of P  is closed under P.

(2) S is a minimal model of P.

(3)  S is  supported by P.

(4) S is an answer set of 
$P \cup \Sigma$ \ if S is a model of $\Sigma$.

\end{proposition}

\subsection{Nested ASP (NASP)} \label{subsec:BACKground-Nested-Expressions}

The  notions revisited here come from 
 \cite{LifschitzTT99}. 
In the sequel,  nested expressions  will be called just {\em expressions} for short. 
For  their  better readability, we  use: 
 (1)   a subindex $ k $ in the connectives  $\wedge_k$ and $\vee_k$ 
      to indicate  their  arity,  which  
      is omitted 
  when all their arguments $\varphi_1$ to $\varphi_k$ are  atomic; 
   (2)  two   expression    delimiters,     
 $\wedge_k  [\ldots ]$ and  $\vee_k  (\ldots )$,  to better 
   distinguish  sub-expressions
   inside expressions; and
   (3) prefix notation  
  as  it needs   one    connective 
      per expression (infix one needs
   $k-1$);  if  $ k=2 $, we   sometimes 
   use   infix notation. 



\begin{definition}  \label{def:NC-formulas} 
$\odot_k \langle  \varphi_1   \ldots  \varphi_k \rangle$  
   stands for  both
 $\vee_k   (\varphi_1   \ldots   \varphi_k ) \mbox{\ and }
\wedge_k   [\varphi_1    \ldots  \varphi_k ].$ 
        The class   of  expressions $\mathcal{E}_{\mathcal{A}}$  
is the smallest 
 set   verifying the inductive conditions below and
   $ \mathcal{E}_{\mathcal{L}}$ is constructed  
   as  $  \mathcal{E}_{\mathcal{A}}$ by just
 replacing $\mathcal{A} $ with  $\mathcal{L}$:
\begin{itemize}

\vspace{-.1cm}
\item   $   \mathcal{E}_\mathcal{A}^1   
\, \subset \, \mathcal{E}_{\mathcal{A}}$.


\vspace{-.15cm}
\item  If  \, $\varphi_1, \ldots, \varphi_k  \in  \mathcal{E}_{\mathcal{A}}$  \,then    
\,$\odot_k \langle  \varphi_1   \ldots  \varphi_k \rangle \in   \mathcal{E}_{\mathcal{A}}$.
\end{itemize}
\end{definition}

\begin{example} \label{exsec1:varisexamples}  
 We have that $\varphi \in \mathcal{E}_\mathcal{L}$, 
 while 
  $\neg {not} \neg a \notin \mathcal{E}_\mathcal{L}$ and    
$ \neg (\vee \ b \ a) \notin \mathcal{E}_\mathcal{A} $.
$$ \ \, \varphi= 
\bigvee_3 \, ( \wedge   [ \, \neg {a}  \ \ \top  ] \quad \
  \bigwedge_2 \ [ \, \vee  (\,\mbox{\em not} \, \neg {a}  \ \ c   \,) \ \ 
   \bigwedge_2 \, [\,b  \ \ \vee (a  \ \ \neg {d}  \,) \,] \ ] \ 
  \quad \ \wedge [\bot \ \  not \, b]\ ).$$  


\end{example}

\begin{definition} \label{def:sub-for} 
The unique sub-expression of an  
elementary  expression is  itself.  
The sub-expressions of 
 $\varphi=  \odot_k \langle  \varphi_1  \ldots \varphi_k \rangle$ 
  are $\varphi$ itself plus  the sub-expressions of  the     $\varphi_i$'s. 
  \end{definition}
  
  \begin{definition} Expressions are represented   by {trees} as follows: 
1) elementary expressions  are  {\em   leaves};  2)  
   occurrences of connectives $ \wedge /\vee $ 
     are   {\em  inner nodes}; and 3)    
any sub-expression $  \odot_k \langle \varphi_1   \ldots \,\varphi_k \rangle$    
 is represented by a {\em $k$-ary hyper-arc}
from  the node of  $\odot_k$      
  to, for every $i$, the node of $\varphi_i$  
  if $\varphi_i$  is a leaf and otherwise
  to that of its connective.
  \end{definition}



 \begin{definition} \label{def:NASP-RULE} (See \cite{LifschitzTT99}.) 
In  an (resp. extended)  NASP rule $ H \leftarrow B $,    one has 
(resp. $ B, H \in \mathcal{E}_{\mathcal{L}}$) $ B, H \in \mathcal{E}_{\mathcal{A}}$.
 A rule  $ \bot \leftarrow B $ is called 
 constraint.
 An (resp. extended) NASP program is a  
 set   of   $n \geq 1$ (resp. extended)  NASP  rules. 
\end{definition}

 \begin{definition} \label{def:interpretation-Model-BASIC} 
 (See \cite{LifschitzTT99}.)
    Let  $\Theta $ be an expression $\varphi_i \in 
    \mathcal{E}_\mathcal{L} $, 
    nested rule $H \leftarrow B$ or nested program $P$.
 An interpretation $ I $   is a model 
of  $\Theta$, in symbols  $ I \models \Theta $, 
 if $ I $  fulfills  the   
 inductive conditions  below, where  
     $ \ell \in \mathcal{L} $. A minimal model and the least model of  $P$
  are  defined in the same way as  in Definition  \ref{def:Model-BASIC}. 
\begin{itemize}
\vspace{-.05cm}
\item    
$I \models \top  $. \  \ $ I \models \wedge[\ ] $.  \ \ $I \not \models \, \bot $.  \ \  $ I  \not \models \, \vee(\,) $.

\vspace{-.15cm}
\item    $ I \models \ell $  \  \ if \ \ $ \ell \in I $.  \hspace{4.8cm} 

\vspace{-.15cm}  
\item   $ I   \models \mbox{\em not} \,  \ell$ \  \ if \ \   $ \ell \notin I $.
\hspace{4.2cm}

\vspace{-.15cm}
\item $ I \models \wedge [\varphi_1 \ldots \varphi_{i}  \ldots   \varphi_k] $ 
\ if \    $\forall i $, $  I \models  \varphi_{i} $.      

\vspace{-.15cm}
\item $ I \models \vee (\varphi_1 \ldots \varphi_{i}  \ldots  \varphi_k)  $  
\ if \ 
     $\exists i $, $  I \models  \varphi_{i} $. 

\vspace{-.15cm}
\item  $ I \models  H \leftarrow B $ \quad \,if  \quad 
$ I \not \models B $ \ or \ $ I   \models H $. 

\vspace{-.15cm}
\item $ I \models P $ \quad if \quad 
$\forall r\in P $, \  $ I \models  r $.    
\end{itemize}

\end{definition}

  \begin{definition} 
  If $ \varphi, \varphi' \in \mathcal{E}_\mathcal{L} $  are
 {\em not}-free expressions, then:
  
  \vspace{.2cm}
 -- $\varphi$ and $\varphi' $ are   logically  equivalent,   
 denoted  $\varphi \equiv \varphi'$,  \ if  \  
    $\forall I$,  $I \models \varphi$ iff $I \models \varphi'$.
   
   \vspace{.1cm} 
 -- $\varphi'$ is a logical consequence of $\varphi$,  
 denoted $\varphi \models \varphi'$, \  if  \
    $\forall I$,   $I \models \varphi$ entails $I \models \varphi'$.
  \end{definition}

  \begin{definition}  \label{def:REDUCT_BASIC} (See \cite{LifschitzTT99}.)  
  Let $ P $ be  a nested program  and 
    $ I  $ be any interpretation. 
   The $ I $-reduct $ P^I $ of $ P $ 
   is the {\em not}-free program    below, 
   wherein $ \ell \in \mathcal{L} $,  
$ \varphi_i \in \mathcal{E}_{\mathcal{L}}$ and $ \odot \in \{\wedge, \vee \}$.
   $ I $  is closed 
  under $ P^I $ if  $\forall H \leftarrow B \in    P^I $:
    $ I \models H $
  whenever $ I \models B $.
$ I $ is an answer set for $ P  $ if 
  $ I $ is   minimal among 
  the closed interpretations of  $ P^I $. 
  \begin{itemize}

  \vspace{-.05cm}
  \item $x^I=x $ \ \ if \ \ $ x \in   \{\bot, \top\} \cup \mathcal{L} $.
  
 \vspace{-.05cm}
  \item $(\mbox{\em not} \, \ell)^I=\bot $ 
  (resp. $\top $) \ \ if \ \ $ \ell \in I $ (resp. $ \ell \notin I $).
  
 \vspace{-.05cm}
  \item $ \odot_k  \langle   \varphi_1   \ldots   \varphi_k\rangle^I = \odot_k \langle \varphi_1^I \ldots    \varphi_k^I\rangle$.

  \vspace{-.05cm}
  \item $ P^I=\{H^I \leftarrow B^I: H \leftarrow B \in P\} $.
    
  \end{itemize}

  \end{definition}

  \begin{definition}  \label{def:BACKGROUND-STRONG-EQUIV} (See \cite{LifschitzPV01}.) Two programs  $ P_1 $ and $ P_2 $ are strongly equivalent,
  denoted by $ P_1 \Leftrightarrow P_2 $,
if for any program $P$,   $ P_1 \cup P$ and $ P_2 \cup P$ have
the same answer sets. 
\end{definition}

\begin{proposition} \label{prop:BACKGROUND-STRONG-EQUIV} 
(See \cite{Ferraris05a,FerrarisL05b}.)      Let  $ P_1 $ and $ P_2 $
 be NASP programs. We have $P_1 \Leftrightarrow P_2 $ 
 if for each interpretation $I$:  $P_1^I \equiv P_2^I$, 
or equally, if for all  interpretations $I,J$:
   $$\forall I, J: \quad J \models P_1^I \quad \mbox{iff} 
   \quad  J \models P_2^I.$$
\end{proposition}

\begin{proposition}   \label{pro:STRONG-PRESERVED} (See \cite{LifschitzTT99}.)
 Relation  $\Leftrightarrow  $ 
is preserved under the
transformations   on expressions and programs below, where 
  $E, F, G \in \mathcal{E}_\mathcal{L}$, $ \odot \in  \{\vee, \wedge \}$ and $\overline{\odot}=\{\vee, \wedge \}/\odot$:
\begin{enumerate}

\item $F \odot G \quad \Longleftrightarrow \quad G \odot F $.

\vspace{-.15cm}
\item $F \odot (E \,\overline{\odot} \,G) \quad \Longleftrightarrow \quad (F  \odot E) \,\overline{\odot} \, (F  \odot G) $.

\vspace{-.15cm}
\item $ E \wedge F \leftarrow G \quad \ \, \Longleftrightarrow  
\quad \{E  \leftarrow G, \  F \leftarrow G\}.$

\vspace{-.15cm}
\item $E \leftarrow F \vee G  \quad \ \, \Longleftrightarrow  
\quad \{E  \leftarrow F, \ E \leftarrow G\}.$


\end{enumerate}
\end{proposition}

   
   \noindent {\bf Note.}     
As $\top$ $\Leftrightarrow \wedge [] $ and $\bot  $ $\Leftrightarrow \vee () $  can be   removed from  expressions $ \varphi  $ using the equivalences below, henceforth, 
       {\em only $ \bot $-occurrences 
      in heads modeling constraints} 
      will be considered.
\vspace{-.2cm}
 $$\neg \top \Leftrightarrow \bot, \ \ \neg \bot \Leftrightarrow  \top,
  \ \, [\wedge \  \bot \  \varphi] \Leftrightarrow  
\bot ,  \ \, 
   (\vee \  \top \  \varphi) \Leftrightarrow  \top , \ \,
   [\wedge \ \top \  \varphi ] \Leftrightarrow  \varphi   
  \ \mbox{and} \ (\vee \  \bot \  \varphi)     \Leftrightarrow  \varphi.$$
  



 \section{ The Proposal: Principle} \label{sec:General-Presentation}

This section gives   some   intuitive ideas
 and  formal notions  underlying
 our proposal  and   considers   successively
normal ASP, flat NASP  and NASP programs.  
It is limited to  extended programs since
  the    introduced concepts applied to  non-extended ones 
  are  easily deduced  from those   (below) for extended programs.
 Table 1 displays all    program types 
treated within this  work  along with their associated  
employed acronyms.

\begin{table}
\begin{center} 
\begin{tabular}{ |p{3.5cm}||p{.8cm}|p{2.8cm}|p{1.9cm}| p{1.1cm}| }
 \hline
 &   ASP& Semantical ASP & Flat NASP & {\bf NASP} \\
 \hline
  Normal   & NP$_\mathcal{A}$    & \ \ \ \ \ \ \ \ SN$_\mathcal{A}$   
  & \ \    NFNP$_\mathcal{A}$ & {\bf NNP}$_\mathcal{A}$ \\
   Extended Normal & NP$_\mathcal{L}$  & \ \ \ \ \ \ \ \ SN$_\mathcal{L}$ & 
   \ \  NFNP$_\mathcal{L}$ & {\bf NNP}$_\mathcal{L}$\\
  Disjunctive &  DP$_\mathcal{A}$  & \ \ \ \ \ \ \ \ SD$_\mathcal{A}$   & 
  \ \ DFNP$_\mathcal{A}$ & {\bf DNP}$_\mathcal{A}$\\
 Extend. Disjunctive  & DP$_\mathcal{L}$ & \ \ \ \ \ \ \ \ 
 SD$_\mathcal{L}$ & \ \ DFNP$_\mathcal{L}$ & {\bf DNP}$_\mathcal{L}$\\
 \hline
\end{tabular}
\caption{\em Type of programs and  their acronyms.}
\end{center}
\end{table}
 

\subsection{ASP Framework} \label{subsec:GENREAL-APPRAOCh-ASP}
 Syntactically, we can view  rules 
 $r=h \leftarrow \wedge 
[l_1 \ldots l_k \mbox{\em not} \, \ell_1 \ldots \mbox{\em not} \, \ell_n ]$ 
 under a slightly different syntax by shifting literals from the body to the head,
i.e. having the syntax: 
\begin{equation} \label{eq:FIRST}
 \vee   (h \  D )  
 \longleftarrow \wedge \,[l_{1} \ldots    l_{i-1} \, 
 \mbox{\em not} \, l_1  \ldots \mbox{\em not} \, l_{j-1}], \quad  
 D=\vee (\overline{\l_{i}} \ldots \overline{\l_k} \,   
 \overline{\mbox{\em not} \, \ell_{j}}   \ldots  \overline{\mbox{\em not} \, \ell_{n}}) 
\end{equation}

Then $h$ should be  deduced as usual  when  $\wedge [l_1 \ldots l_k \mbox{\em not} \, \ell_1 \ldots \mbox{\em not} \, \ell_n ]$ 
is satisfied, or   equivalently,  
  when $\wedge \,[l_{1} \ldots    l_{i-1} \,
 \mbox{\em not} \, l_1  \ldots \mbox{\em not} \, l_{j-1}]$ 
 is satisfied  and   ${D}$ is falsified 
 (which holds exactly
 when all its overlined   elementary expressions are falsified).
  Thus, that an interpretation $I$ falsifies an expression, for instance $D$, will be denoted   by $ I \fmodels \, D$.
 
 \vspace{.05cm}
If one  shifted   $\vee (\ell_i  \ldots \mbox{\em not} \, {\ell_n}) $ 
instead of  $\wedge [l_{i} \ldots   \mbox{\em not} \, \ell_{n}]$, 
  then   expression $D$  in the rule head of  (\ref{eq:FIRST})  would be
  $D=\wedge [\overline{\ell_i}  \ldots \overline{\mbox{\em not} \, \ell_n}]  $. 
  In the general case, we require to substitute  an arbitrary  expression  
  conjunctively linked to the rule body  $F_B$ by  another   expression  
   disjunctively linked to the rule head ${F}$, 
   more formally:
   $$ H \leftarrow B \wedge F_B \quad \mbox{by}\quad H \vee {F} \leftarrow B,   $$
   
   where $F_B, {F}$ verify 
   $ I \models \, F_B \ \mbox{iff} \ I \fmodels  \, {F}$, for 
   any   interpretation $I$. So we define both ${F}$  from $F_B$ and   
   $F_B$ from ${F} $  (towards shifting  ${F} $ 
   from the head  to the body) as follows:

   \begin{definition} \label{def:DEFINITION-of-OVERLINE-F} 
    Let  $F_B \in \mathcal{E}_{\mathcal{L}}$. 
      ${F} $   results from
     exchanging   $\wedge$ and $\vee$ in $F_B$
    for $\vee$ and $\wedge$, respect.,
   and by overlining  all its elementary expressions. 
   Conversely,  $F_B$ results  from   
     exchanging   $\wedge$ and $\vee$ in ${F} $
    for $\vee$ and $\wedge$, respect.,   and by  eliminating  
    overlines.
   \end{definition}

   %
   






The aforesaid falsification relation applied to overlined expressions is  defined as:

\begin{definition} \label{def:LITERAL-HEAD-SEMAN} Let   $I$ be  
any  interpretation  and ${F}$  an   expression all of whose   
elementary expressions are overlined.
The falsification relation $I \fmodels \, {F}$ is given below: 
 \begin{itemize}
\vspace{-.05cm}
\item    
 $I  \fmodels \, \overline{\bot} $.  \ \  $ I   \fmodels \, \overline{\vee(\,)} $. \ \  $I \not \fmodels \, \overline{\top} $.
 \ \ $I \not \fmodels \, \overline{\wedge [\,]} $.

\vspace{-.15cm}
\item   
$ I \fmodels \, \overline{\ell} $  \  \ if \ \ $  \ell \in I $.

\vspace{-.15cm}  
\item    $ I   \fmodels \, \overline{{\mbox{\em not} \,  \ell}}$ \  \ if \ \   $ \ell \notin I $.

\vspace{-.15cm}
\item  $ I \fmodels \, \wedge [\varphi_1 \ldots \varphi_{i}  \ldots  \varphi_k]  $  
\ if \ 
     $\exists i $, $  I \fmodels  \varphi_{i} $.

\vspace{-.15cm}
\item  $ I \fmodels \, \vee (\varphi_1 \ldots \varphi_{i}  \ldots  \varphi_k)  $  
\ if \     $\forall i $, $  I \fmodels  \varphi_{i} $. \qed

\end{itemize}
\end{definition} 

  \begin{example} \label{ex:ormla-F-and-overline-F}   
 We have that $ {F}_B=\vee (\, { \mbox{\em not} \, \neg e} \ \  
\wedge [ {\neg f} \ \  {a}] \,)$  if and only if
${F}=\wedge [\,\overline{{ \mbox{\em not} \, \neg e}} \ \  
\vee ( \overline{\neg f} \ \ \overline{{a}})]$.   
If we take $I=\{\neg f\} $ then we obtain:
  $$\{\neg f\} \fmodels \, \overline{\neg f},  
\qquad \{\neg f\}\fmodels \, \overline{{ \mbox{\em not} \, \neg e}},  
\qquad \{\neg f\} \not \fmodels \, \overline{{a}}, \qquad  \{\neg f\} \fmodels \, {F} 
\qquad \mbox{and} \qquad \{\neg f\} \models \,{F}_B.$$
\end{example}

 \begin{corollary}  \label{cor:DE-Morgan} If $\odot \in \{\wedge, \vee\} $,  
 $\overline{\odot} =\{\wedge, \vee\} / \odot $  and    $F',G'$ are expressions
 obtained   by applying De Morgan's laws to  $F,G$. Then:

\vspace{-.15cm}
$$ I \fmodels \,{   F  \,  \odot \, G  }  \qquad  \mbox{\em iff} \qquad
       I \fmodels \, {F}'  \ \overline{\odot} \  {G}'  \qquad \mbox{(De Morgan)}.$$
\end{corollary}


\begin{definition} \label{def:LITERAL-HEAD-REDUCTT} We establish 
the $I$-reduct of overlined 
elementary expressions and, by extension, that of a general expression ${F} $ 
with overlined elements as follows: 

\vspace{.25cm}
$\bullet$  $(\overline{x})^I=\overline{x}$  \ if  $x \in \mathcal{L} \cup \{\top, \bot\}$. 

\vspace{.2cm}
$\bullet$ $(\overline{{not} \, \ell})^I =\overline{({{not} \, \ell})^I} $. 

\vspace{.2cm}
$\bullet$ $ \odot_k  \langle   \varphi_1   \ldots   \varphi_k\rangle^I = \odot_k \langle \varphi_1^I \ldots    \varphi_k^I\rangle$. 
\end{definition}

\begin{proposition} \label{prop:I-REDUCTS}
Rules $R=H \leftarrow B \wedge F_B $ and $R'=H \vee {F} \leftarrow B  $, 
wherein $F_B$ and ${F}$ are defined as specified in Definition \ref{def:DEFINITION-of-OVERLINE-F}, 
are strongly-equivalent  $ R \Leftrightarrow R'$.
\end{proposition}

\begin{niceproof} (Sketch) 
From    Definitions \ref{def:DEFINITION-of-OVERLINE-F}  and \ref{def:LITERAL-HEAD-REDUCTT},
one can prove: for any interpretation $I$,  
the $I$-reducts of the rules verify  $R^I \equiv R'^I $. Hence, by 
Proposition \ref{prop:BACKGROUND-STRONG-EQUIV},  $ R \Leftrightarrow R'$. 
\end{niceproof}    


      \begin{example} \label{ex:POLARIZED-NFNP} 
Let us consider programs $P$ and  $P'$, where  $P'$ is obtained from $P$ by just
shifting some literal  from   bodies to  heads of rules:
$$ P=\{a \leftarrow \neg b, \ \  b \leftarrow not \, \neg a \}. 
\qquad P'= \{(a \ \vee \  \overline{\neg b}) \leftarrow \top, \quad  
(b \ \vee \ \overline{not \, \neg a} )  \leftarrow \top \}.$$

We take interpretation   $I=\{\neg a\} $ and we have:  
$$P^{\{{\neg a}\}}= \{a \leftarrow \neg b\} 
\equiv  \{(a \ \vee \  \overline{\neg b}) \leftarrow \top  \}=P'^{\{{\neg a}\}} .$$

Since  one can check that for all $I$,
 $P^I \equiv P'^I$,  we have
     $ P\Leftrightarrow P'$. 
\qed
\end{example}

Clearly, {Definitions \ref{def:DEFINITION-of-OVERLINE-F}, 
\ref{def:LITERAL-HEAD-SEMAN} and \ref{def:LITERAL-HEAD-REDUCTT} 
  confer the same   semantical   role     on both  $F_B$  in the body and  
    ${F} $  in the head of a rule $r$, 
 that is,   the    models of  $r$ 
 remain unaltered.

\vspace{.25cm}
{\underline{\bf Syntax.}}
We return to  equation (\ref{eq:FIRST}). 
If $h$ is considered as positive and the overlined literals   
as negative, (\ref{eq:FIRST}) has the form: $h \vee   {D} 
 \leftarrow   B$,  wherein $B$ is a usual ASP body, ${D}$ is a negative clause 
and $h \vee   {D}$  is a positive-Horn clause, i.e. 
  Horn with a positive disjunct $h$.

  \vspace{.05cm}
Correspondingly, a  semantically (or truly) disjunctive ASP rule 
has the syntax (\ref{eq:SECOND}),
where  $B,{D}$ are as above and the $ h_i $'s   as previous  $h$, i.e.
  its head is positive non-Horn.
 \begin{equation} \label{eq:SECOND}
 \vee (h_1 h_2 \ldots h_m \ {D})   
 \longleftarrow  B, \ \mbox{with} \ m > 1.  
\end{equation}


  \begin{definition} \label{def:POLARITY}   
$a]$ Any $x \in \mathcal{L} \cup \{\bot\}$      is   positive.  
Any overlined elementary expression   $\overline{x}, x \in \mathcal{E}^1_\mathcal{L}$  
        is    negative. 
 $b]$ A     positive Horn (resp.  non-Horn) clause  
  has $n \geq 0$ negative disjuncts and one (resp. $ k > 1$) positive.
   $c]$ A   CNF is  positive-Horn  if 
 so are  all its clauses. $d]$ A  positive non-Horn CNF   has 
 $n \geq 0$ positive-Horn and $k \geq 1$ positive non-Horn clauses.
\end{definition}

\begin{example} \label{ex:INTUITION-Exam}
Rule   $ r  $ (resp. $r'$)  below is extended normal (resp. disjunctive):
  $$r=  \vee \, (\,\neg d  \ \
 \overline{ \mbox{\em not} \, \neg e} \ \  \overline{\neg f}) 
\longleftarrow \wedge \, [\neg b   \ \ \mbox{\em not} \, c]. \qquad
r'=  \vee \, ( \, \bot \ \overline{d} \ \  \overline{f} \ \ d) 
\longleftarrow  \wedge \, [ b   \ \ \mbox{\em not} \, \neg c].$$ 

\vspace{-.85cm}
\qed
\end{example} 

{\bf Note.} Negative clauses, i.e.  with only  negative elements, are useless  
 because by assumption, ASP  rules always contain  a positive disjunct, i.e. 
 $h \in \mathcal{L} \cup \{\bot\}$.  \qed
        
\vspace{.1cm}
In a nutshell,   ASP  rules whose head  
     contains   positive   and negative disjuncts 
     (overlined to   be distinguished  from positive ones)
     and with the form
     \begin{tcolorbox}  
  $$H_{C}^+ \leftarrow  B, \quad \mbox{} 
 \  \  \mbox{\em  $H^+_{C}$ is a   positive-Horn  clause,}$$ 

\end{tcolorbox} 
 can be thought of as ``syntactic sugar" for   traditional  normal ASP rules.

\vspace{.2cm}
{\underline{\bf Discussion.}}
For our ultimate goal, the advantage  of the  "clausal-head" form 
in equations  (\ref{eq:FIRST}) and (\ref{eq:SECOND}) 
versus the classical form is that it 
      {\bf uniformly}  views both normal (\ref{eq:FIRST})
      and disjunctive (\ref{eq:SECOND}) rules because indeed, under this form,    
    {\em   heads are always clauses} 
   and this feature     resembles that encountered in NASP     
  in which  {\em  heads are always   expressions.} 
  
  \vspace{.05cm}
 And having both NASP and ASP rules under an analogous form makes it reasonable 
   employing the same policy  within both of them to  divide  rules into  
    normal and disjunctive ones. 
     This means that we can project onto NASP the existing ASP policy 
     supporting normal and disjunctive     rules  by
     positive Horn  and positive non-Horn clauses, respect., 
     and consequently, propose positive-Horn  and  positive-non-Horn expressions 
     to support normal and disjunctive nested rules,   respectively. Below, we 
      elaborate upon such a division.

    \subsection{Flat NASP and NASP} 

  We consider  CNFs and DNFs as the only existing  {\em  flat   expressions.}  The head of  flat NASP (FNASP) rules is a CNF  
       and their body is  a {\em  flat   expression}. 
      A positive-Horn (resp. positive non-Horn) CNF
is a set of positive-Horn (resp. and positive non-Horn) clauses.

\vspace{.05cm}
Along the  lines of Subsection \ref{subsec:GENREAL-APPRAOCh-ASP},  FNASP  rules whose head is   a  
positive-Horn (resp.  positive non-Horn) CNF are normal  (resp.    disjunctive). 
  Actually, any normal FNASP rule   results from merging 
  (by  Proposition \ref{pro:STRONG-PRESERVED},  line 3)  a set of normal ASP   rules    
   and it has the following form, where
    $B$ is a flat expression:
$$\tcboxmath{ H^+_{CNF}  \leftarrow  B,  \quad \mbox{\em} 
 \ H^+_{CNF} \  \mbox{\em is a  positive-Horn CNF}. }$$
  
   \vspace{.0cm}   
 Normal FNASP is tightly connected to     
 normal ASP and normal NASP as the former (resp. the latter)
  is a specific  case (resp. a generalization)  of normal FNASP.
 We highlight also 
    the following coincidences  (to be proven) between normal FNASP and normal NASP: 

  \vspace{.25cm}

 {\em    \noindent \ \ \ \  -- Normal NASP  semantic notions\footnote{The ones in Subsection \ref{subsec:Background-LP}:
 model,   closedness, supportedness, immediate consequence operator, etc. } are defined as Normal FNASP ones. 

\vspace{.05cm}
\noindent \ \ \  --  Their NASP components  
are just nested versions   of  FNASP ones.}

\vspace{.25cm} 
Altogether,  normal FNASP plays a  bridge role  
between normal  ASP and  NASP,  and so, it
will greatly  assists us to lift 
   semantical notions from  ASP   to  NASP.

\vspace{.1cm}
As   positive-Horn   clauses and 
positive-Horn CNFs  support normal ASP  and normal FNASP, respect., 
  enriching  normal programs to full NASP
  demands   climbing a further level up the scale of  Horn-like expressiveness. 
  Therefore, we postulate  that 
  {\em positive-Horn expressions}
  should underpin     normal NASP. 
Accordingly, NASP rules 
with a positive non-Horn expression as head  
are  disjunctive. 
 In a nutshell, we propose that  normal NASP  rules should present the following  form, wherein
   $  B \in \mathcal{E}_{\mathcal{L}}$ ($\mathcal{E}_{\mathcal{L}}$ is in Definition \ref{def:NC-formulas}):
$$ \tcboxmath{  H^+_{NNF}  \leftarrow B, \quad \mbox{} 
 \  \  \mbox{\em   
 $ H^+_{NNF}$  is a  positive-Horn NNF expression.} }$$

   We will demonstrate that a positive-Horn   expression  is a succinct form of a set of  
    positive-Horn clauses and CNFs, which allows   normal NASP programs, underpinned by  
    positive-Horn expressions,
    to be a succinct form of a set of normal ASP and FNASP programs, underpinned
    by   positive Horn clauses and positive Horn CNFs, respectively.

\section{Semantically-Normal Programs} \label{sec:NDs}

 We call semantically-normal (SN) rules  those under
 the clause-head form  
 discussed in Section \ref{sec:General-Presentation}.  
 This section   defines  the syntax of 
 SNs and then establishes  for them,
 the same semantical notions  
 given in Subsection \ref{subsec:Background-LP} for NPs.
 Also, as SN programs are  elementary  NNPs, 
 such notions
 form a   starting point to set  up  later  
 those corresponding to  NNPs.

\begin{definition} \label{def:syntact-ND-A} 
Let   $r= h \vee  {D}    \leftarrow B,$   $B$ being an
 NP$_\mathcal{A}$ body.       
  $ r $ is  normal or SN$_\mathcal{A} $ 
(resp. disjunctive or SD$_\mathcal{A} $) 
if $h \vee  {D}$  is  a positive- Horn (resp. non-Horn) clause.
An  SN$_\mathcal{A} $ (resp. SD$_\mathcal{A} $) program  
is a set of $\geq 1$ SN$_\mathcal{A} $  (resp. $\geq 0$ SN$_\mathcal{A} $  and $\geq 1$  SD$_\mathcal{A} $) rules. 
An SN$_\mathcal{A} $ rule:

\vspace{-.15cm} 
 \begin{enumerate}
 
\vspace{-.0cm}
\item [] \quad  -- $r$ is a fact, if $B=\top$,  ${D}  =\bot$ and $h \in \mathcal{A}$; 

\vspace{-.2cm}
\item [] \quad -- $r$ is {\em not}-free, if $ B=B^\mathcal{A} $ and 
${D}  ={D}  ^\mathcal{A}$; and 
 
\vspace{-.2cm}
\item [] \quad -- $r$ is a  constraint, if $ h=\bot $. \qed
  \end{enumerate}  
\end{definition}

 $\vee  (  \bot \ \ \overline{b} \ \ \overline{not \, c})
  \leftarrow 
 \top $ is a constraint.

\begin{definition} \label{def:syntac-sug-NP}    
Extended semantically- normal (SN$_\mathcal{L} $) 
and  -disjunctive (SD$_\mathcal{L} $)  
programs are defined as SN$_\mathcal{A} $  and DN$_\mathcal{A} $ ones, respect.,  
   by substituting  in Definition \ref{def:syntact-ND-A}, $ \mathcal{A} $ by $ \mathcal{L} $. 
We will refer to  SN$_\mathcal{A} $ and SN$_\mathcal{L} $ rules/programs as SN rules/programs.
Any SN  program
 is head-consistent 
if the set of  its  head literals $h \neq \bot$  is consistent. 
\end{definition}

 See {Example   \ref{ex:INTUITION-Exam}} for a  case of SN and SD rules.
%
Below, we deal with {\bf semantic aspects}.  
As usual, NP semantics  was based 
 on set operations in Subsection \ref{subsec:Background-LP}. Instead and from now on, 
 we will  employ  definitions  in Subsection \ref{subsec:BACKground-Nested-Expressions} based on logic concepts, e.g.,   to check whether   $I$ verifies a body  $B$, 
    we will  check whether 
  \,$I   \models    B$ according to  Definition \ref{def:interpretation-Model-BASIC}.

 \begin{definition} \label{def:model-Horn-rule} Let $P$ be    an SN  program
    and 
    $  I    $ be an interpretation.  
$ I $ is a model of    $ P $ 
{when}  for  all $ ( h \vee {D} \leftarrow B) \in P$,  the conditions
below hold.
A minimal model  and the least model $LM(P)$ of $P$
 are defined as  
those in Definition \ref{def:Model-BASIC}:
$$ I  \not \models    B
 \quad  \mbox{  {or}} \quad I \not \fmodels  \,  {D}
 \quad  \mbox{ {or}} \quad h \in I  . 
$$ 
\end{definition}

\begin{definition} \label{cor:LP-CLOSED-SUPPORTED} 
 An interpretation $I $ 
 is closed under an SN  program  $ P $ if: 
     $$\forall (h \vee {D} \leftarrow B)   \in P: \qquad  h \in I   \quad  
\mbox{whenever} \quad  
I \models {B} \quad   \mbox{{and}} \quad  
I    \fmodels {D}.$$ 
\end{definition}

\begin{definition} \label{cor:LP-CLOSED-SUPPORTED} 
An interpretation $I  $ 
 is supported by an SN  program  $ P $ if \,$ \forall \ell \in I  $: 
$$  \exists (h \vee {D} \leftarrow B) \in P \ \ \mbox{such that:} \qquad \ell=h, 
\quad  I \models {B} \quad   \mbox{{and}} \quad  
I   \fmodels {D} .$$ 
\end{definition}

\begin{definition}   Let  $P$ be any 
 constraint-free,   head-consistent SN program 
    and  $ I   $
 be an  interpretation.
 The immediate consequence operator $ ST_P (I) $ for SN programs
    is defined  below and  $ ST^i_P $ and  
$ \mbox{\it lfp}(ST_P) $ are defined as those  in Definition \ref{def:CONSEQ-IMM-OPE}. 
$$ST_P (I)=\{h \ \vert \vert \  h \vee {D} \leftarrow B  \in P, \, 
\  I   \models   B, \      
I   \fmodels  {D}  \}.$$ 
 
\end{definition}

Along the lines of the comment made above Definition \ref{def:model-Horn-rule}, in the sequel,
reducts  $F^I$  of  expressions $F$ will be  obtained by applying Definition \ref{def:REDUCT_BASIC}.

\begin{definition}  Let    $ P $ be an SN  program  and $ I  $   an interpretation.  
 The    $I$-reduct of $P$ is the {\em not}-free program  $ P^I $ 
defined below. Answer sets and consistency of $P$   are
 defined as those in Definition \ref{def:ANSWER-I}.
$$P^I=\{ \, h  \vee    {D}^I \leftarrow B^I \ \, 
\vert \vert \ \, h  \vee    {D} \leftarrow B \in P,  
\ \,    B^I \not \equiv \bot, \ \,  {D}^I \not \equiv \top\}.$$
 
\end{definition}

 Next we convert 
an SN  program   
into an NP one as follows:

\begin{definition} \label{def:TR1}  
Let  $P$ be an SN program  and    $r= h  \vee {D}  \longleftarrow  B   \in P$.
 We associate  to $P$ the  NP program $SN(P)$   below, where $D_B$
 is obtained from ${D}$ by    Definition \ref{def:DEFINITION-of-OVERLINE-F}:
$$ S(r)= h \leftarrow B  \wedge   D_B, \qquad
 \ \mbox{\bf } \qquad 
SN(P)=\{  S(r) \parallel r \in P\}.$$ 
\end{definition} 
 
Clearly if $P$ is an SN program,  
then $ SN(P) \ \ \mbox{is NP}$.

\vspace{.1cm}
Below, we state that claims from  Theorem \ref{TH:Tarski55-Endem76} to  
Proposition    \ref{prop:BASIC-Propoerties-ANSWER-SET} 
verified by NP programs are also verified by SN ones, 
 but first,
we need  Propositions \ref{Prop:EQUIV-GNP-NP} and \ref{prop:SN-3-4-5},  manifesting the tight 
 relationships between  NPs   
and  SNs:
 
\begin{proposition} \label{Prop:EQUIV-GNP-NP} 
Let $P$ be an  SN program and $I$ be  an interpretation.  
Then  we have: 
 \begin{itemize}
 
 \item [{\em (1)}] 
  $P$ is head-consistent iff 
 so is $SN(P)$.
 
 \vspace{-.2cm}
 \item [{\em (2)}]
 $I $ is closed under $P$ iff 
   $I $ is closed  under $SN(P)$.

 \vspace{-.2cm}
 \item [{\em (3)}] $I $ is supported by $P$ iff 
   $I $ is supported  by $SN(P)$.
 
 \end{itemize} 
\end{proposition}

The proofs 
 follow immediately 
from  the involved definitions applied to NP programs
(in Subsection \ref{subsec:Background-LP}) 
 and to SN programs (given above) and from Definition \ref{def:TR1}   of $SN(P)$.

 \begin{proposition} \label{prop:SN-3-4-5}  Let $P$ be any   not-free,
 constraint-free and   head-consistent SN program and denote by 
     $Cn(P)$  the smallest interpretation $I$  closed under $P$. We have:
  \begin{itemize} 
 \item [{\em (4)}] $ Cn(P)=Cn(SN(P)) $.
 
 \vspace{-.2cm}
 \item [{\em (5)}] $ST_P(I) = T_{SN(P)}(I)$.
 
 \vspace{-.2cm}
 \item [{\em (6)}] $P \Leftrightarrow  SN(P)$.
 
 \end{itemize}
 \end{proposition}

The proofs of (4) and (5) are immediate 
from  the involved definitions.
Next we sketch the proof of (6).
Any arbitrary interpretation $I$ and 
  SN rule  $r$ produce the  reducts  $r^I $
and  $SN(r)^I$ and it is easy to check that $ r^I \equiv SN(r)^I $.  Hence,
for  any interpretations $I,J$, we have: $ J \models r^I $  iff  $ J \models SN(r)^I$, which entails $P \Leftrightarrow  SN(P)$ 
by Proposition \ref{prop:BACKGROUND-STRONG-EQUIV}.

\begin{proposition} \label{TH:Tarski55-Endem76-GNP} 
 For   {not}-free,   constraint-free and
    head-consistent SN  programs  $ P $, 
  we have:
  $$LM(P)=\bigcup_{0 \leq i} \,ST^i_P=\mbox{lfp}(ST_P) .$$
\end{proposition}

\vspace{-.2cm}
The proof follows from  Proposition 
\ref{prop:SN-3-4-5},   (4) and (5), and  from the fact 
that the same equalities
are verified by their NP counterparts as claimed in 
Theorem \ref{TH:Tarski55-Endem76}.

\begin{proposition} \label{prop:BASIC-Properties-not-free-GNP}  
If $  P$ is a {not-}free and constraint-free SN$_{\mathcal{A}}$ 
program, then $P$, $ Cn(P) $ and $ LM(P) $ fulfill  properties (a) to (c) in Proposition \ref{prop:BASIC-Properties-not-free-NP}.
\end{proposition}  

The proof
   follows    from Proposition \ref{Prop:EQUIV-GNP-NP}, claims (1) to (3), 
  Proposition \ref{prop:SN-3-4-5},  claim (4), 
    and from the fact that  NP$_{\mathcal{A}}$  programs verify such properties  (a) to (c) by
  Proposition \ref{prop:BASIC-Properties-not-free-NP}.

 \begin{proposition}
  \label{prop:BASIC-Propertiesprograms-not-free-ENP}  
If $  P$ is a {not-}free,  constraint-free and
 head-consistent  SN$_{\mathcal{L}}$ program, then  
$P$, $ Cn(P) $ and $ LM(P) $  verify properties  (b) and (c)
in Proposition \ref{prop:BASIC-Properties-not-free-NP}. 
\end{proposition} 

The proof
is similar to that of Proposition 
\ref{prop:BASIC-Properties-not-free-GNP}, by using now Proposition \ref{prop:BASIC-Properties-Extended-NP}.

\begin{proposition} \label{prop:BASIC-Propoerties-ANSWER-SET-ND} 
Let  P be a constraint-free SN program, S    an answer set of P
and  $\Sigma$   a set of constraints. Then P, S and $\Sigma$
verify   (1) to (4) in Proposition \ref{prop:BASIC-Propoerties-ANSWER-SET}. 
\end{proposition}

   The proof follows from 
Proposition \ref{prop:SN-3-4-5}, 
claim  (6),
and the fact that NP  programs also fulfill
 such relationships  (1) to (4) by Proposition \ref{prop:BASIC-Propoerties-ANSWER-SET}.

\vspace{.1cm}
 \begin{tcolorbox}
The above claims entail that 
 no semantical differences  between NP  and SN programs  exist
and so that   SN
is indeed just a \underline{\bf syntactical alternative}   to NP.
\end{tcolorbox}

  \section{ Normal FNASP Programs} \label{sec:FLAT-programs}
  
   As already said, CNFs  and DNFs are considered as 
 flat  expressions. Subsection \ref{subsec:SYNTAX-NFNP} 
   and  \ref{subsec:SEMANTIC-NFNP}
    define the syntax and semantics, respect., 
    of normal flat NASP  (NFNP) programs and analyze related issues.
It is shown that  any NFNP rule   syntactically relies on a positive-Horn CNF and 
semantically  is  strongly equivalent to a set of SN rules. 

 \subsection{Syntax of NFNP  } \label{subsec:SYNTAX-NFNP}


\begin{definition}   \label{def:FNASP-rule}
 Rule  $r=H\leftarrow B $, with $H,B  \in \mathcal{E}_\mathcal{A}$, is   FNASP$_\mathcal{A} $  
if    $B$ is 
   flat  and   
    $ H=\wedge_k \ [\, (h_1    \vee  D_1)  \ldots 
  (h_k    \vee D_k) \,] $  is a CNF.  
  $r$ 
 is  normal or NFNP$_\mathcal{A}$  
 (resp. disjunctive or  DFNP$_\mathcal{A}$) 
 if $H$ is   positive-  (resp. non-) Horn.  An NFNP$_\mathcal{A} $  (resp. FDNP$_\mathcal{A} $) program is a 
 set of $\geq 1 $  NFNP$_\mathcal{A} $ (resp. $\geq 0 $  NFNP$_\mathcal{A} $
 and $\geq 1 $  DFNP$_\mathcal{A} $) rules.
 An NFNP$_\mathcal{A}$ rule:

\vspace{.25cm}

-- $ r $   contains (resp. is) 
 a fact   if ${B}=\top $
 and  $\exists i$ (resp. $\forall i$): \ ${D}_i=\bot $ and $ h_i \in \mathcal{A} $.
 
 \vspace{.1cm}  
  -- $ r $   is partially (resp. is)  
  {\em not}-free if $B={B}^\mathcal{A}  $
  and $\exists i$ (resp. $\forall i $): \ ${D}_i={D}_i^\mathcal{A}$.

 \vspace{.1cm}  
-- $ r $   contains (resp. is)   a constraint if     $\exists i$ (resp. $\forall i$): \ 
$h_i=\bot$. \qed
\end{definition}

\begin{example} \label{ex:NFNP-rules} 
$r_1$  is NFNP$_\mathcal{A} $ and contains a constraint and a fact $b$.  
 $$ r_1= \bigwedge_3 \, [ \, \vee \, (\,c   \ \ \overline{b}  \ \overline{g})  \ \ \
  b  \ \ \ \vee \, (\bot \   \overline{\mbox{\em not}  \, g})  \, ]   
  \longleftarrow     \bigvee \, (\,a  \ \mbox{\em not} \,  e   \ m).  $$
%
 %
  
\end{example}

\begin{definition}     Extended  normal FNASP$_\mathcal{L} $
 (NFNP$_\mathcal{L} $) and    disjunctive (DFNP$_\mathcal{L} $) programs
 are  defined as NFNP$_\mathcal{A} $ and DFNP$_\mathcal{A} $ ones    by replacing   
  $ \mathcal{A} $ with $ \mathcal{L} $ in Definition \ref{def:FNASP-rule}. 
  We refer   to   NFNP$_\mathcal{A} $ and NFNP$_\mathcal{L} $ rules/programs as 
NFNP rules/programs. An NFNP program   is head-consistent
if the set of head literals    $h \neq \bot$ existing in all its rule heads  is
  consistent.
\end{definition}

\begin{example} \label{ex:FXNP-program} 
 Program $P=\{r,r'\}$, 
whose rules are below, is NFNP$_\mathcal{L} $. $P$ is not
head-consistent as its set
of head literals  is $ \{\neg d, d, b, \neg c\} $.
 $$ r= \bigwedge_2 \, [  \vee \, (\,  \neg d   \ \ \overline{{not} \, c}  )  \ \ \
   d     \, ]   \longleftarrow  
   \,  \vee \, (\, a  \ \ \mbox{\em not} \,  \neg e   \ \ \neg m) $$

\vspace{-.25cm}
$$ r'= \bigwedge_3 \, [ \, \vee \, (\,  \overline{\neg  b}   \ \ \neg c  )  \ \ \
   b \  \ \ (\vee \ \bot \ \overline{d})  \, ]   \longleftarrow    
  \vee \, (\, a \ \ \mbox{\em not} \,  e \  \ m).  $$ 
  
  \vspace{-.85cm}
   \qed
\end{example}

 
 \vspace{-.35cm}
 \begin{example} \label{ex:Literature-1} We show a non-ground 
   DFNP$_\mathcal{L} $ program  from a  data-integration   
  context \cite{BriaFL08,BriaFL09,BriaFL09b}.
   Consider a global relation 
   $ p(ID,name,surname,age) $ 
   for persons with a key-constraint 
   on the first attribute $ID$. To perform
  consistent query-answering \cite{Bertossi06}, when two 
  tuples share the same key, the relation person is
  "repaired" by intentionally deleting one of them.
This is obtained by the 
DP$_\mathcal{L}$ program  below, where  $ \neg {p} $
  stands for delete tuples
  and $ p' $ is 
  the resulting consistent relation on which  query 
  answers are computed:
  
  \vspace{.2cm}
  \hspace{1.cm} $ \neg {p}(i,n,s,a) \ \vee \ \neg {p}(i,m,t,b) 
  \  \longleftarrow \ \wedge \, [\, p(i,n,s,a) \ \ p(i,m,t,b) 
  \ \ n \neq m \,]$
  
  \vspace{.15cm}
 \hspace{1.cm} $ \neg {p}(i,n,s,a) \ \vee \ \neg {p}(i,m,t,b) 
  \  \longleftarrow \ \wedge \, [\, p(i,n,s,a) \ \ p(i,m,t,b) 
  \ \ s \neq t \,]$
  
  \vspace{.15cm}
 \hspace{1.cm} $ \neg {p}(i,n,s,a) \ \vee \ \neg {p}(i,m,t,b) 
  \  \longleftarrow \ \wedge \, [\, p(i,n,s,a) \ \ p(i,m,t,b) 
  \ \ a \neq b \, ]$
  
  \vspace{.15cm}
\hspace{1.6cm} $\qquad \quad  \ \ R=p'(i,n,s,a) \ \,\longleftarrow \ 
  \wedge \, [\, p(i,n,s,a) \ \ \mbox{\em not} 
   \, \neg {p}(i,n,s,a) \, ]$
  
  \vspace{.2cm}
The  top three  rules are more  succinctly and    naturally  
 encoded into a DFNP$_\mathcal{L}$ rule: 

 \vspace{.25cm}
  $ \neg {p}(i,n,s,a)  \vee  \neg {p}(i,m,t,b)  
   \longleftarrow  \bigwedge_3 \, [\, p(i,n,s,a) \ \, p(i,m,t,b)  \ \,   
  \vee \, (\, n \neq m \ \, s \neq t \ \, a \neq b) \, ]$ \qed
  
\end{example}

By Proposition \ref{pro:STRONG-PRESERVED}, line 3,    an NFNP program is  
  strongly-equivalent to  an NP  one: 

\begin{definition}  \label{cor:SECT-APPROACH:pos}
   Let   $P$ be an   NFNP  
       program and $ r=H \leftarrow B \in P $.   
   We associate to $P$ the NP program  
     $FN(P)$
       specified below, wherein $dnf(B) $  denotes the   translation 
of $B $ into DNF,  
   $ C  $ is a conjunct in  $   dnf(B)$ and $C\ell$ is a clause in $H$. 
 $$F(r) = \{C\ell  \leftarrow  C   \ \vert \vert \ 
C\ell \in H , \   C  \in  dnf(B)  \} \ \ \qquad FN(P)= \{SN(r') \parallel r \in P, \,r' \in F(r)\}  . $$
    
\end{definition}

\begin{example}
One can check that 
 rule $r_1$ in Example \ref{ex:NFNP-rules}   compacts 9 NP$_\mathcal{A} $ rules, 
  $ c \leftarrow   \wedge \,[ b \ g \  \mbox{\em not} \,  e ] $ 
being one of them.
\noindent  On the other hand, it is also easy to check  that $P$ in 
Example \ref{ex:FXNP-program} compacts 15 NP$_\mathcal{L} $ rules as follows:
\begin{enumerate}

\vspace{-.cm} 
\item $r$ compacts 6 NP$_\mathcal{L} $ rules, 
 $ d \leftarrow a$ \ and \  
$ \neg d \leftarrow 
\wedge \,[\mbox{\em not} \,  \neg e \ \ \ {not} \, c]$ being two of them. 

\vspace{-.2cm} 
\item $r'$ compacts 9 NP$_\mathcal{L} $ rules,  
$ \bot     \leftarrow   
  \wedge \,[\mbox{\em not} \,  e  \, \ d ] $ being  one of them.
 \qed

\end{enumerate}
\end{example}

\begin{proposition} \label{prop:NFNP-and-SN} If $P$ is an NFNP program, then 
 $   FN(P)  \ \mbox{is NP}$.
\end{proposition}

  The proof   is immediate from the  definitions of NFNP program 
and from the facts: 

\vspace{.1cm}
(a) $F(r)$ is SN if $r$ is NFNP; and 

(b)   $SN(P) $ was shown to be NP if 
$P$ is SN.

\subsection{ Semantics of NFNP} \label{subsec:SEMANTIC-NFNP}

\begin{definition} \label{cor:LP-CLOSED-SUPPORTED} 
 Let  $P$   be an NFNP  program and
     $ I $  be   an interpretation.
 $I $ is a {model}  of $ P $, 
 denoted by $ I \models P $, when 
 for all $  (H \leftarrow B) \in P$,  the conditions below hold. 
A minimal model and the least model $LM(P)$ of $P$ are defined as 
in Definition \ref{def:Model-BASIC}.
$$ I  \not \models    B  \quad 
\mbox{{or}} \quad \forall \,  {(h \vee {D})} \in H:
\quad  I \not \fmodels   \, {D}   \quad \mbox{{or}} \quad 
 h \in I.$$ 
\end{definition}

\begin{example}  \label{ex:Model-NFNP} We take 
  $r_1$  from Example  \ref{ex:NFNP-rules} and $r$ from 
  Example \ref{ex:FXNP-program}.
%
   (1)  below is an NFNP$_\mathcal{A}$ case, and   (2) and (3)
  are NFNP$_\mathcal{L}$ cases:
\begin{enumerate}

\item [(1)]  $\{e,m, b,   c,g\}$   is a model of  $r_1$  and
     its   minimal  models are $ \{b,  c,g\} $ and $ \{e\} $. 
  
\item  [(2)]   $\{\neg e,\neg  m,  d,  c\}$ is a model
  of $r$ and     its minimal  models are  $\{ d,c\} $ and $\{\neg e\} $.

 \item [(3)] Let  $P=\{r_1,r\}$ be the NFNP$_\mathcal{L}$ program  including both previous rules: 
 \begin{enumerate}
 
 \item $\{ b,c, d,g\}$, $\{  e,    d, c\}$ and $\{  \neg e, b,c,g  \}$
  are its minimal models. 
  
 \item $P$ is 
 not head-consistent because its set of head literals  is $\{c,b, \neg d, d\} $.
   \end{enumerate} 
   \end{enumerate} 
  \end{example}

\begin{definition} \label{cor:NFNP-CLOSED} 
  An interpretation $ I  $  is  closed under  an NFNP  program $ P $ if 
  \mbox{ $  \forall H\leftarrow B \in P$:}
     $$
     \forall  \, {(h \vee {D})} \in H: 
     \quad  h \in I \quad 
\mbox{whenever} \quad   
I \models B 
\quad \mbox{{and}} 
\quad I   \fmodels \, {D}.  $$
\end{definition}

\noindent {\bf Remark.}   It is simply verified 
that both Definition \ref{cor:NFNP-CLOSED} 
and   \ref{def:REDUCT_BASIC} (taking from
  \cite{LifschitzTT99})  
  particularized to {\em not}-free   and
    NFNP programs, respect., coincide.


\begin{definition} \label{cor:NFNP-SUPPORTED} An interpretation $I$ 
is  supported by  an NFNP  program $ P $ if 
$   \forall \ell \in I $: 
$$    \exists (H \leftarrow B) \in P, \quad \exists 
\,{(h \vee {D})} \in H: \qquad \mbox{} h =\ell, \quad 
 I \models  B
  \quad \mbox{{and}} \quad
I   \fmodels \, {D}.$$ 
\end{definition}

\begin{example} One can easily confirm  that any model 
of program $P$ from Example  \ref{ex:Model-NFNP} 
 is closed under $P$ and that its minimal models  
are supported by $P$. 
\end{example}

\begin{definition} Let    $ P $ be a  constraint-free,
 head-consistent NFNP  program  and    $ I  $ be   an interpretation. 
 The  immediate consequence operator $ FT_P (I) $  for 
 NFNP programs is defined below. 
    $ FT^i_P $ and  $ \mbox{\it lfp}(FT_P) $  are defined in Definition \ref{def:CONSEQ-IMM-OPE}.
$$FT_P (I)=\{h \ \vert \vert \ H \leftarrow B \in P, 
\ \,  (h  \vee   {D}) \in H,  \ \,  
 I   \models    B,   \ \,  I   \fmodels \, {D}  \}.$$ 
\end{definition}

\begin{example} \label{ex:operator-pTP} 
We consider the {\em not}-free NFNP$_\mathcal{A}$  program $ P $ containing
the   rules:
$$   f \wedge g  \leftarrow \top. 
 \qquad    [\wedge \ (\vee \ b \ \overline{c}) \ \ e \ \
(\vee \ \overline{a} \   h) \, ] \ 
\longleftarrow \    f \wedge g.  \qquad
[\wedge \  \ a \ \ \
(\vee \ \overline{h} \   c) \, ] \ 
\longleftarrow    \ f \wedge e  .$$

  Below we give the powers  $FT^{k}_P$ for 
$ k \in \{1,2,3,6\} $ and $ k > 6 $:
\begin{gather*}
  FT^{1}_P =\{f,g\},  \quad
   \ FT_P^{2}=\{f,g,e\}, \quad
  \ FT_P^{3}=\{f,g,e,a\}, \\  
  \vspace{.3cm}
     FT_P^{4}=\{f,g,e,a,h\}, \quad  
   FT_P^{5}=\{f,g,e,a,h,c\}, \quad
    FT_P^{6}=\{f,g,e,a,h,c,b\},\\  
  FT_P^{k}= FT_P^{6} \  
 \mbox{for}  \  k > 6.   
\end{gather*}
 
 \vspace{-.7cm}
 \qed 
\end{example}

\begin{definition}   Let    $ P $ be an 
NFNP   program,    
 $ I  $ be   an interpretation and let $ H \leftarrow B \in P $.  
 The    $I$-reduct of $P$, denoted by $ P^I $,  is
defined below  (with a slight abuse of notation). Answer set and consistency of $P$  
  are
 defined as in Definition \ref{def:ANSWER-I}.
$$H^I=   \bigwedge [\, h  \vee {D}^I \parallel h \vee {D}  \in H, 
\  \ {D}^I  \not \equiv \top\, ]. $$
$$P^I=\{H^I \leftarrow B^I
\ \ \vert \vert \ \ H \leftarrow B \in P, 
\  \ B^I \not \equiv \bot, \ \ H^I \neq \wedge []  \}.$$
\end{definition}

We now  show that analogous claims to those within
 Theorem \ref{TH:Tarski55-Endem76} 
to  Proposition \ref{prop:BASIC-Propoerties-ANSWER-SET}  
are also valid for NFNP programs.  Previously,
  Propositions \ref{Prop:EQUIV-NFNP-ND} and \ref{Prop:EQUIV-NFNP-ND-2}  state
the relationships making evident the tight semantical connections between  NFNPs 
and  SNs. 

\vspace{.05cm}
We omit  the formal proofs of the claims    below as:
$(i)$ they  follow 
from Propositions \ref{Prop:EQUIV-NFNP-ND} and  \ref{Prop:EQUIV-NFNP-ND-2}  and the fact that SN programs
 verify such claims (Section \ref{sec:NDs});
 and  $(ii)$  they will be later on   proven  for NNP programs
  that are a generalization of NFNP ones.

\begin{proposition} \label{Prop:EQUIV-NFNP-ND} 
Let  $P$ be an NFNP  program and $I$ be an interpretation. We have:
  \begin{itemize}
 
 \vspace{-.05cm}
 \item [{\em (1)}] 
  $P$ is head-consistent iff 
 so is $FN(P)$.
 
 \vspace{-.2cm}
 \item [{\em (2)}]
 $I$ is closed under $P$ iff 
 $I$ is closed under $FN(P)$.

 \vspace{-.2cm}
 \item [{\em (3)}] $I$ is supported by $P$ iff 
 $I$ is supported   by $FN(P)$.
 
 \end{itemize} 
\end{proposition}
 

\begin{proposition}   \label{Prop:EQUIV-NFNP-ND-2} Let $P$ be   any   not-free,
 constraint-free and  head-consistent NFNP program $P$ and denote the smallest interpretation $I$  closed under $P$
 by $Cn(P)$. We have:
 \begin{itemize}
 
 \item [{\em (4)}] $ Cn(P)=Cn(FN(P)) $.
  
 \vspace{-.2cm}
 \item [{\em (5)}] $FT_P(I) = T_{FN(P)}(I)$.
 
 \vspace{-.2cm}
 \item [{\em (6)}] $P \Leftrightarrow  FN(P)$.
 
 \end{itemize} 
\end{proposition}

\begin{proposition} \label{TH:Tarski55-Endem76-GNP} 
 For   {not}-free,  constraint-free and head-consistent   NFNP programs  $ P $:
  $$LM(P)=\bigcup_{0 \leq i} \,FT^i_P=\mbox{lfp}(FT_P) .$$
\end{proposition}

\vspace{-.5cm}
\begin{proposition} \label{prop:BASIC-Properties-not-free-NFNP}  
If $  P$ is a {not-}free and constraint-free  NFNP$_\mathcal{A} $ program, then $P$,  
  $ Cn(P) $ and $LM(P)$ verify the properties (a) to (c)
   in Proposition \ref{prop:BASIC-Properties-not-free-NP}.
%


\end{proposition}  


 \begin{proposition} \label{prop:BASIC-Propertiesprograms-not-free-FXNP}  
	If $  P$ is a {not-}free, constraint-free and head-consistent 
NFNP$_\mathcal{L} $ program, then $P$, $Cn(P) $ and $LM(P)$ verify
 properties (b) and (c) in Proposition  \ref{prop:BASIC-Properties-not-free-NP}.
%
%
%
\end{proposition} 



\begin{proposition} \label{prop:BASIC-Propoerties-ANSWER-SET-NFNP} 
Let   P be an  NFNP program, S  be  an answer set of P
and  $\Sigma$ be  a set of constraints. Then P, S and $\Sigma$
verify the properties expressed in Proposition \ref{prop:BASIC-Propoerties-ANSWER-SET}.
\end{proposition}


\section{Positive-Horn  Expressions} \label{sec:Positive-Horn-Nested-Form}

 Subsection  \ref{subsec:Horn-nested}  quickly  recalls  
  the  Horn expressions as given in 
  \cite{Imaz2021horn,Imaz2023a,Imaz2023b}. 
  We denote by $\mathcal{H}_\mathcal{L} $ and $\mathcal{H}_\mathcal{A} $ the classes 
  of Horn expressions built from literals and atoms, respectively.
  
\vspace{.1cm}
   Subsection \ref{subsec:POSITIVE-HORN}  presents for the first time and in this work  
the classes $\mathcal{H}^+_\mathcal{L} \subset \mathcal{H}_\mathcal{L}$ and 
$\mathcal{H}^+_\mathcal{A} \subset \mathcal{H}_\mathcal{A}$ of positive-Horn expressions
(see Figure 1). 
Subsection \ref{subsec:Posit-Non-Horn} specifies the classes 
 $\mathcal{H}^{2+}_\mathcal{L} $ and $\mathcal{H}^{2+}_\mathcal{A} $ of positive non-Horn expressions.
Subsection \ref{sec:PROTERTIES-HORN-NEXPRESS} demonstrates a relevant property regarding our objectives 
of positive-Horn expressions. 

\vspace{.1cm}
Since $\mathcal{H}_\mathcal{A} $ and $\mathcal{H}^+_\mathcal{A} $ are easily obtainable from 
$\mathcal{H}_\mathcal{L} $ and $\mathcal{H}^+_\mathcal{L} $, 
we will  only mention $\mathcal{H}_\mathcal{L} $ and $\mathcal{H}^+_\mathcal{L} $ within the discussions 
included in the following subsections.

\subsection{Horn Expressions $\mathcal{H}_\mathcal{L} $  } \label{subsec:Horn-nested}

We recall that by Definition \ref{def:POLARITY},  $ \mathcal{L}\cup \{\bot\}$ elements
are {\bf positive}
and that
     overlined  elementary expressions $\{\overline{x} \parallel x \in  \mathcal{E}_\mathcal{L}^1\} $  
are {\bf negative}.  
Since head expressions can contain positive and negative  
elements, the class of expressions  allocatable to NASP heads is:

\begin{definition}  \label{def:NC-HEAD-formulas} 
The class   of head   expressions $\mathcal{HE}_{\mathcal{L}}$
is the smallest 
 set    verifying the inductive conditions below.
 $ \mathcal{HE}_{\mathcal{A}}$ is constructed
   as  $  \mathcal{HE}_{\mathcal{L}}$ but substituting  $\mathcal{L} $ with  $\mathcal{A}$.
\begin{itemize}

\vspace{-.05cm}
\item   $ \mathcal{L}\cup \{\bot\} \cup  \{\overline{x} \parallel x \in  \mathcal{E}_\mathcal{L}^1\}
\, \subset \, \mathcal{HE}_{\mathcal{L}}$.


\vspace{-.15cm}
\item  If  \, $\varphi_1, \ldots, \varphi_k  \in  \mathcal{HE}_{\mathcal{L}}$  \,then    
\,$\odot_k \langle  \varphi_1   \ldots  \varphi_k \rangle \in   \mathcal{HE}_{\mathcal{L}}$. \qed
\end{itemize}
\end{definition}

 Thus   we first define the (fully) negative 
   expressions that are both,  
  essential members of   Horn expressions and
  the counterparts of  
   negative literals from the unnesting setting.

  \begin{definition} \label{def:negative-formula}   
  Any    $ \varphi \in \mathcal{HE}_{\mathcal{L}}$ (resp. $ \varphi \in \mathcal{HE}_{\mathcal{A}}$)
  is     fully negative 
   if all its elementary expressions are overlined. 
  $\mathcal{N}_\mathcal{L} $ (resp. $\mathcal{N}_\mathcal{A} $) denotes the set 
 of  negative expressions constructed from $\mathcal{L} $ (resp.  $\mathcal{A} $). 
\end{definition}

\begin{example} The  expression  $ \bigvee_2  \, (\wedge \, [\,\overline{\mbox{\em not} \, b}  
\ \ \overline{\mbox{\em not} \, \neg \, b}  ] \  \  
\wedge  [\,    \overline{c}   \   \   \overline{a}  \,] \,) \in \mathcal{N}_\mathcal{L}$  is fully negative.
 \qed
\end{example}

    We now proceed with the definition of the class $\mathcal{H}_\mathcal{L} $, 
 first informally  and then formally. Thus, {informally,}  
a head expression {\em   $ \varphi \in \mathcal{HE}_\mathcal{L}  $   is  Horn, namely 
$ \varphi \in \mathcal{H}_\mathcal{L} $,  
whenever:} 

\vspace{.25cm}    

 \qquad {\em {\em $\bullet$}   either  $ \varphi$ does not contain disjunctions; or}

\vspace{.1cm}
 \qquad {\em {\em $\bullet$}  all  its disjunctions  have  at most one non-negative  disjunct.} 



\begin{example} \label{exam:Horn-nested-formulas-1}  
Let us analyze   formulas
$H_1$ to  $H_3$ below. 
The  unique disjunction within $H _1$   is the proper $H_1$ and as
$H _1$   has  only one non-negative  disjunct,  then $H_1 \in  \mathcal{H}_\mathcal{A} $.  
  $H _2$  has two non-negative  disjuncts,
  and hence, it is trivial  that   $H_2 \notin  \mathcal{H}_\mathcal{A} $.  
 \begin{itemize}
 
\item  [] \hspace{3.cm} $H_1 = 
 \bigvee_2  \, ( \,  \wedge \, [ \overline{\mbox{\em not} \, b}  \ \ \ \overline{d} \, ] \  \  
 \wedge [c   \   \   a   \,] \,).$  

\item [] \hspace{3.cm} $H_2 = 
\bigvee_2 \, ( \,  \wedge \, [\overline{b} \ \ d  \, ] \ \ \
 \wedge [\bot \ \ \overline{a}  \,] \ ).$     

\item [] \hspace{3.cm} 
$ H_3=\bigvee_2 \, (\, H_1 \ \  \bigvee_2 ( \overline{a} \ \
 \wedge [\overline{\mbox{\em not} \, b} \ \ \overline{c}\,] \,) \ )$

\end{itemize}

The disjunctions of $ H_3 $ 
are: all disjunctions inside its leftmost disjunct $ H_1 $, 
its rightmost disjunct and $ H_3 $  itself. The rightmost disjunct 
is fully negative and  we already checked that  $ H_1 $  
complies with the requirements of a Horn expression. 
  $ H_3 $  
is a Horn expression $ H_3 \in \mathcal{H}_\mathcal{A} $ as 
the proper  $ H_3 $  and all its disjunctions have at most one non-negative disjunct. 
   \qed
\end{example}

 \begin{example} \label{exam:Horn-nested-formulas-2}   $H_2$ below   results
 from  $H_1$ by only switching   its left-most literal $\overline{a}  $  for $a   $. 
 Since all  the three   disjunctions in $H _1$,
 including the whole expression  $H _1$,
  have   exactly one  
\begin{itemize}
\item [] \hspace{2.cm} $ H_1= \bigvee_2 \, (\overline{a}  \quad  
 \bigwedge_2 \, [\,    \vee \,(\overline{e}   \ \ c  )  \quad
\bigwedge_2 \,  [ b   \quad    \vee (a \ \ \overline{\mbox{\em not} \, d}) \ ] \ ] \ )$

\item [] \hspace{2.cm} $H_2= \bigvee_2 \, ({a}  \quad  
 \bigwedge_2 \, [\,    \vee \,(\overline{e}   \ \ c  )  \quad
\bigwedge_2 \,  [ b   \quad    \vee (a \ \ \overline{\mbox{\em not} \, d}) \ ] \ ] \ )$ 

\end{itemize}

 non-negative disjunct, then $H _1 \in \mathcal{H}_\mathcal{A}$. 
 Concerning $ H_2 $,
 one can check that $ H_2 $  
has the form $H_2 =\vee (a   \ \phi_3)$ whose two disjuncts  are non-negative, 
and  hence $H _2 \notin \mathcal{H}_\mathcal{A}$.  \qed
\end{example}

In Example \ref{exam:Horn-nested-formulas-3}  below, we analyze 
  $ H_1$ and $ H_2$  
($ H_2$ comes from the Introduction, Example \ref{ex:INTRO-rule}) 
 having unspecified  sub-expressions represented by the $\phi $'s.  So,   
 whether or not $ H_1$ and $ H_2$ are  Horn 
obviously depends on  the  concrete  features of the $\phi $'s. 
We determine the conditions to be held by such sub-expressions in order for $H_1$
and $H_2$ to be Horn.

\begin{example} \label{exam:Horn-nested-formulas-3}  Consider  
$ H_1 $ and $ H_2 $ below, with   
$ \phi_1 $ to $ \phi_3 $ being unspecified  expressions. 
$$H_1=\bigwedge_3 \, [\, \phi_1  \ \ \bigvee_2 \ 
( \overline{a} \ \ \bigwedge_3 \  
[ \, \vee  \ ( \overline{a} \  \ \overline{c} \,) 
\quad \bigvee_2 (\phi_2 \ \   \wedge  [{b}  \ \ 
{\textcolor{black} {{a}}} \,] \, ) \ \ \bot \, ] \ ) \quad {\textcolor{black} a} \,].$$
 $$\mbox{\it H}_2=\bigwedge_3  \, [ f   \ \ \  \bigvee_2 \, ( \overline{ g}  
 \ \ \ \bigwedge_3 \ [ \
\vee \, ( \overline{f}  \ \ \overline{ not g}  \ \ c \,) \ 
\ \bigvee_2 \,  (\phi_1 \ \ \wedge \,  
[\phi_2  \ \  \overline{not  f }\,] \, )
 \ \ \ g \,] \ \ ) \ \  \ {\phi_3} \, ] $$
 
 The different syntactical options\footnote{For space saving purposes,
  we  omit for  now how these conditions  are determined but
facilitate this kind of details  when examining whether $H_1$ and $H_2$ are positive-Horn,
which is more relevant as positive-Horn  expressions are the ones allocated to NNP heads.}
  for $ H_1$ and $ H_2$ to be Horn expressions are:

\vspace{.25cm}
-- $ H_1 \in \mathcal{H}_{\mathcal{L}}$ \  \ iff \ \ 
both \ $ \phi_1 \in \mathcal{H}_{\mathcal{L}}$  \ \,{\bf and} \  
$ \phi_2 \in \mathcal{N}_{\mathcal{L}}$.

\vspace{.25cm}
-- $ H_2 \in \mathcal{H}_{\mathcal{L}}$ \ \ 
iff \ \ $ \phi_1,   \phi_2, \phi_3 \in \mathcal{H}_{\mathcal{L}}$  \   {\bf and} \  
at least one of $ \phi_1 $ or $ \phi_2 $  is in $ \mathcal{N}_{\mathcal{L}} $.  
 \qed
\end{example}

We now tackle the {\bf formal definition} 
of $\mathcal{H}_{\mathcal{L}}$. 
 For that purpose, we  require an inductive 
formulation of what a Horn expression is. Thus
 the  prior informal definition of a Horn expression
 can be equivalently reformulated  inductively as:

\vspace{.25cm}
\noindent {\em (1) A conjunction is Horn if its conjuncts are Horn.
 
 \vspace*{.1cm}
\noindent (2) A disjunction  is Horn 
if it
has any number  of negative
disjuncts and  one Horn disjunct.}

\vspace{.25cm}
This recursive feature is the base of the formal 
Definition \ref{def:Horn-Nested}  of a Horn-expression, whose   lines (1) to (3) impose
 the following restrictions on each sub-expression: 

\vspace{.15cm}
1) The elementary Horn-expressions are just:  
$\mathcal{L} \cup  \{\bot\} \cup \mathcal{N}_\mathcal{L}$.

\vspace{.15cm}
2)  All conjuncts  of  a 
conjunctive Horn-expression must be Horn.

\vspace{.15cm}
3)  There must be $  \geq 0$ negative and one Horn  disjuncts in a disjunctive Horn-expression.

\begin{tcolorbox}
\begin{definition} \label{def:Horn-Nested} 
    The set  $\mathcal{H}_{\mathcal{L}}$ of   Horn expressions 
      is   the smallest set  verifying the inductive conditions below.
       $\mathcal{H}_{\mathcal{A}}$ is defined as $\mathcal{H}_{\mathcal{L}}$ by just substituting
      $\mathcal{L}$ with $\mathcal{A}$.
    \begin{itemize}

\item [1)] \, $ \mathcal{L}  
\, \cup  \, \{\bot\} \, \cup \, \mathcal{N}_{\mathcal{L}}  \subset \mathcal{H}_{\mathcal{L}}.$    \hspace{10.05cm}

\item [2)]  \,  If  
\,$\varphi_1, \ldots, \varphi_k \in \mathcal{H}_{\mathcal{L}} 
 \ \mbox{then} \ 
        \bigwedge  \, [ \varphi_1   \,\ldots\,  \varphi_k \,] 
   \in \mathcal{H}_{\mathcal{L}}.$ \hspace{3.27cm}  

\item  [3)]    If \,$\varphi_i \in \mathcal{H}_{\mathcal{L}} $ 
\,and \,$ 1 \leq j \neq  i \leq k, \  \varphi_j  \in \mathcal{N}_{\mathcal{L}} \ \, \mbox{then}
\  \bigvee \, (\varphi_1 \ldots \varphi_i \ldots \varphi_j \ldots  \varphi_k \, ) 
 \in \mathcal{H}_{\mathcal{L}}.$      
 
\end{itemize}
\end{definition}

\end{tcolorbox}
\noindent A fast analysis of 
the above recursive function indicates, among other things, that:  

 \vspace{.2cm}
--  Classical Horn formulas are subsumed by $\mathcal{H}_{\mathcal{L}} $.

 \vspace{.1cm}
--   Sub-expressions  of $\mathcal{H}_{\mathcal{L}}$ members are either 
  (fully) negative or non-negative Horn.

  \vspace{.2cm}
  Additional
 properties enjoyed by  $\mathcal{H}_{\mathcal{L}} $ members are  proved in
   \cite{Imaz2021horn,Imaz2023a,Imaz2023b}, for instance,
   that the nested-Horn recognition problem, namely,
    to know whether any arbitrary expression 
     $\varphi \in \mathcal{HE}_\mathcal{A}$ is Horn, or  
     $\varphi \in \mathcal{H}_\mathcal{A}$, is solved in  strictly linear time 
  \cite{Imaz2023b}.

\subsection{Positive-Horn Expressions 
$\mathcal{H}^+_{\mathcal{L}}$ } \label{subsec:POSITIVE-HORN}

\noindent
Positive-Horn expressions  are the nested counterpart of   positive-Horn CNFs. 
  We  first  informally define them,  and  after that, 
  to define them formally, we 
 will  add  to lines 1 to 3 of Definition \ref{def:Horn-Nested} 
   only    expressions that besides being Horn  are also positive.

 \vspace{.15cm}
      \underline{Informally,}  positive-Horn expressions 
  are  defined as follows:
\begin{itemize}

\vspace{-.05cm}
\item   The  elementary     positive-Horn expressions are: $\mathcal{L} \cup \{\bot\} $.
 
 \vspace{-.15cm}
 \item  All conjuncts in  a conjunctive positive-Horn expression
  must be positive-Horn.
\vspace{-.15cm}     
\item A disjunctive Horn expression is positive  if it has  
 one positive-Horn  disjunct. 
 \end{itemize} 
 
\noindent {\bf Example \ref{exam:Horn-nested-formulas-1}.} (Continued). 
Checking  whether any $H_1$ to $H_3$ is positive-Horn can
be done by just using the informal definition given above.
 \begin{itemize}

\item  
$H _1 \in \mathcal{H}_{\mathcal{A}}$ and it is a disjunction with  
one positive-Horn disjunct, so 
$H _1 \in \mathcal{H}_{\mathcal{A}}^+$.
  
  \vspace{-.1cm} 
  \item   Since $ H_2 \notin \mathcal{H}_{\mathcal{A}} $ 
  and $  \mathcal{H}^+_{\mathcal{A}} \subset  \mathcal{H}_{\mathcal{A}} $  then 
 $ H_2 \notin \mathcal{H}_{\mathcal{A}}^+$. 

\vspace{-.1cm}
\item     $ H_3 \in \mathcal{H}_{\mathcal{A}} $ is a conjunction with one   
negative conjunct. Hence  
$ H_3 \notin \mathcal{H}_{\mathcal{A}}^+ $.  \qed 

\end{itemize}

 By looking at the \underline{informal} definition above, one can deduce
 very simple conditions  to reject or to accept expressions as being positive-Horn.  
%
  The following simple features  rule out a
  positive-Horn status for an expression $\varphi$:
\begin{enumerate}

\vspace{-.05cm}
\item $\varphi$ is not Horn.

\vspace{-.1cm}
\item  $\varphi$ is a Horn conjunction with at least one fully negative conjunct.

\vspace{-.1cm}
\item  $\varphi$ is a Horn disjunction without any non-negative disjunct.

\end{enumerate} 


\vspace{.1cm}
 In the opposite sense, a {sufficient (yet not necessary) condition} (SC)  
for a Horn-expression $\varphi$  to be positive  is: 

\vspace{.25cm}
{\em (a) all literals in a conjunction of $\varphi$ are positive.}

\vspace{.1cm}
{\em (b) all disjunctions of literals of $\varphi$ are positive-Horn.}


\begin{example}  
 SC rightly \underline{excludes} the expression $\vee (\overline{a} \ \ \wedge [\overline{b} \ \ c])$
from $\mathcal{H}_{\mathcal{A}}^+$ because it does not fulfill (a). On the other hand,
 one can easily verify that $H_1$ from Example  \ref{exam:Horn-nested-formulas-2}
fulfills  {\em SC}, and then,   rightly again, SC \underline{includes} 
  $H_1$ into $\mathcal{H}_{\mathcal{A}}^+$. 
\end{example}

Thus, {\em SC} is   a sufficient, but clearly,  a     unnecessary  condition to recognize
positive-Horn expressions as
the latter can indeed have
fully negative sub-expressions  other than  literals, and so, do
 not verify conditions (a) and (b)  of  {\em SC} above.

\vspace{.1cm}
  Definition \ref{def:Positive-Horn-Nested} 
  is the formal definition of  $\mathcal{H}_{\mathcal{L}}^+$. Items (1) to (3)  
 account for  the conditions in the \underline{informal} definition  above and
   they are applied to  all sub-expressions $ \varphi_i$  
by taking into account
  whether $ \varphi_i$ is
 basic (1), conjunctive (2) or 
disjunctive (3): 
  
\begin{tcolorbox}
\begin{definition} \label{def:Positive-Horn-Nested} 
  The class of positive-Horn expressions  $ \mathcal{H}_{\mathcal{L}}^{+}$ is  
 the smallest set verifying  the  conditions below. 
  $ \mathcal{H}_{\mathcal{A}}^{+}$ is defined as $ \mathcal{H}_{\mathcal{L}}^{+}$ by replacing 
   $ \mathcal{L}$ with  $ \mathcal{A}$.
    \begin{itemize}

\item [(1)] \, $\mathcal{L} \cup \{\bot\}  \subset \mathcal{H}_{\mathcal{L}}^+.$    \hspace{10.05cm}   

\item [(2)]  \,  If  
\  $\varphi_1, \ldots,  \varphi_k \in \mathcal{H}_{\mathcal{L}}^+
 \, \ \mbox{then}   \  \
       \bigwedge  \, [  \varphi_1  \ldots  \varphi_k \,] 
   \in \mathcal{H}_{\mathcal{L}}^+.$ \hspace{3.27cm}  

\item  [(3)]  If  
 $\varphi_i \in \mathcal{H}_{\mathcal{L}}^+ $    and   
  $1 \leq j \neq i \leq k,  \varphi_j \in \mathcal{N}_{\mathcal{L}}  
   \, \mbox{then}  \,  \bigvee   (\varphi_1  \ldots \varphi_i \ldots  \varphi_j \ldots \varphi_k )  \in \mathcal{H}_{\mathcal{L}}^+.$      
 
\end{itemize}

\vspace{-.3cm}
 \qed 
\end{definition}
\end{tcolorbox}


We now illustrate the   morphology  of $\mathcal{H}_{\mathcal{L}}^+$ by
 checking which    expressions from  Subsection \ref{subsec:Horn-nested}, besides  being Horn, 
   are also positive. Thus we have: 
  
  \vspace{.2cm} 
  --    Example \ref{exam:Horn-nested-formulas-1} 
  was checked without difficulty with the informal definition of  $ \mathcal{H}^+_{\mathcal{L}}$;
  
  \vspace{.05cm} 
 --  Example \ref{exam:Horn-nested-formulas-2},       $H_1$
 was easily checked through criterion {\em SC}; and 
 
 \vspace{.05cm} 
 --   Example \ref{exam:Horn-nested-formulas-2}, $H_2$  
is not positive-Horn as   was shown not being Horn.

\vspace{.2cm}
Hence, there is  only Example \ref{exam:Horn-nested-formulas-3} left to be
checked for positive-Horn compliance:

\vspace{.25cm}
\noindent {\bf Example \ref{exam:Horn-nested-formulas-3}.} (Continued). 
We verify whether any  characteristics 
of sub-expressions $\phi_1$ and $\phi_2$ make it possible to include
 $ H_1 $ into $\mathcal{H}_{\mathcal{L}}^+$.  We  do so by
  applying step-by-step  Definition \ref{def:Positive-Horn-Nested} in  
a bottom-up way, i.e. from sub-expressions   to   expressions:
 \begin{itemize}

 \vspace{-.05cm}
\item   Sub-expression $\psi_1$  within $H_1$ given below belongs 
to $\mathcal{H}_{\mathcal{L}}$ if $ \phi_2 \in \mathcal{N}_{\mathcal{L}} $. 
Yet, whichever the features of  $\phi_2$  are,  
$\psi_1$  cannot verify  (2)  since
 $\vee (\overline{a} \ \ \overline{c} \,) \notin \mathcal{H}_{\mathcal{L}}^+ $. 
 Hence $ \psi_1 \notin \mathcal{H}_{\mathcal{L}}^+ $. 
 $$\psi_1=  \bigwedge_3 \ [\, \vee (\overline{a} \ \ \overline{c} \,) 
\quad  
 \bigvee_2 ( \phi_2 \ \  \wedge   [ {b}  \  
{\textcolor{black} {{a}}} \,] \, ) 
\quad  \bot \, ].$$

 \vspace{-.45cm}
\item Consequently,    sub-expression
$\psi_2= (\vee \  \overline{a} \  \psi_1)$ containing $ \psi_1 $ 
 does not belong to $ \mathcal{H}_{\mathcal{L}}^+ $ either, 
because it does not fulfill condition (3) given that
$ \psi_1 \notin  \mathcal{H}_{\mathcal{L}}^+$. 

 \vspace{-.15cm}
\item    Thus 
 $ H_1 =[\wedge \ \ \phi_1 \ \ \psi_2 \ \ a]$ has one  
 conjunct $ \psi_2 $ such that $ \psi_2 \notin \mathcal{H}_{\mathcal{L}}^+ $;  
 so  whichever  the features of $\phi_1$ are, $ H_1$ cannot verify 
 condition (2), and so  $ H_1 \notin \mathcal{H}_{\mathcal{L}}^+$.  \qed
  \end{itemize}

  \vspace{.25cm}
\noindent {\bf Example \ref{exam:Horn-nested-formulas-3}.} (Continued) 
Let us consider $ H_2 $   and check the characteristics required by its sub-expressions
$ \phi_1 $ to $ \phi_3 $  for  $ H_2 $  to be positive-Horn. Contrary  
to the previous example,   we now carry out a top-down checking, 
namely from expressions to  sub-expressions: 
 \begin{itemize}
 
\item  [--]  
For $H_2= \wedge [a \ \Psi_1 \ \phi_3] $ to be in $\mathcal{H}_{\mathcal{L}}^+ $, 
we need $ \Psi_1   \in \mathcal{H}_{\mathcal{L}}^+ $ and 
$  \phi_3  \in \mathcal{H}_{\mathcal{L}}^+ $.
 
 \vspace{-.1cm}
 \item  [--]  For $ \Psi_1= \vee (\overline{b} \ \Psi_2)$ to be in $\mathcal{H}_{\mathcal{L}}^+ $, 
 one needs  $ \Psi_2   \in \mathcal{H}_{\mathcal{L}}^+ $.
 
  \vspace{-.1cm}
 \item  [--]   For $ \Psi_2= \bigwedge_3 \,  
 [ \, \vee (\overline{a}  \  \ \overline{b}  \  \ c \,) \  \Psi_3 \  b \,] $ 
 to be in $\mathcal{H}_{\mathcal{L}}^+ $, 
 we need $ \Psi_3   \in \mathcal{H}_{\mathcal{L}}^+ $.
 
  \vspace{-.1cm}
\item [--]   
For $\Psi_3 = \bigvee_2 \,  ( \, \phi_1 \   \wedge  [\,  \phi_2  \   \overline{a} ] ) $ to be in $\mathcal{H}_{\mathcal{L}}^+ $, we need 
 $ \phi_2 \in \mathcal{N}_{\mathcal{L}}$ 
and $ \phi_1 \in \mathcal{H}_{\mathcal{L}}^+$.

 \vspace{-.1cm}
\item  [--] Summing up, for  $ H_2 $ to be in $\mathcal{H}_{\mathcal{L}}^+ $,  we need
 $ \phi_1, \phi_3 \in \mathcal{H}_{\mathcal{L}}^+$  and 
 $ \phi_2 \in \mathcal{N}_{\mathcal{L}}$.
  \qed
 \end{itemize}
 
\noindent  {\bf Remark.}      {Example \ref{exam:Horn-nested-formulas-3}} clearly shows 
that the  conditions  
 for an expression to be  positive-Horn 
 are   stronger than those to be just  Horn.
Regarding $H_1$, under certain characteristics  of  $\phi_1$ to  $\phi_3$, $H_1 $  can belong to $\mathcal{H}_{\mathcal{L}}$,
  whereas no characteristics  of $\phi_1$ to  $\phi_3$ make it 
  possible for $H_1$ to belong to  $\mathcal{H}^+_{\mathcal{L}}$. 
Regarding $H_2$, the  conditions  
 for $H_2$ to be  positive-Horn,  rewritten in (1) below,
 are clearly  stronger than those  to be just  Horn, 
 recalled in (2) below too.
  $$ (1) \quad \phi_1, \phi_3 \in \mathcal{H}_{\mathcal{L}}^+  \quad   \mbox{\bf and}  \quad
   \phi_2 \in \mathcal{N}_{\mathcal{L}}.$$
   $$(2) \quad  \phi_1,   \phi_2, \phi_3 \in \mathcal{H}_{\mathcal{L}}  \quad   
  \mbox{{\bf and} \  at least one of} \    \phi_1  \  \mbox{or} \  \phi_2 \ \mbox{is in} \  \mathcal{N}_{\mathcal{L}} .$$  
 
 \vspace{-.6cm} 
  \qed
 
 \vspace{.15cm}
  All previous examples contained positive-Horn expressions constructed from atoms, namely
 belonging to $\mathcal{H}^+_{\mathcal{A}}.$ Next we provide two 
 positive-Horn expressions constructed from literals and so belonging to $\mathcal{H}^+_{\mathcal{L}}.$
  
\begin{example} \label{ex:POSIT-POLAR-Horn}
We supply 
 two  positive-Horn expressions for   extended  NNP heads
and whose   positive occurrences  
are  $ \{\neg y, z, \neg x\} $. Actually, one can check that
both positive-Horn expressions verify the aforementioned criterion {\em SC}.
   $$ H'=\bigvee_2 \, (\overline{x}  \quad  
 \bigwedge_2 \, [\   \vee \,(\overline{\neg z}    \ \ \neg y  )  \quad
\bigwedge_2 \,  [\, z  \quad    \vee (\neg x \ \ \overline{\mbox{\em not} \, \neg v}) \ ] \ ] \ )
\in \mathcal{H}^+_{\mathcal{L}}.$$
$$ H''=\bigvee_2 \, (\overline{\mbox{\em not} \, x}  \quad  
 \bigwedge_2 \, [\    \vee \,(\overline{\mbox{\em not} \, \neg y}     \ \ \neg y  )  \quad
\bigwedge_2 \,  [ \, \neg y   \quad    \vee (\neg x \ \ \overline{v} ) \ ] \ ] \ )
\in \mathcal{H}^+_{\mathcal{L}}.$$

\vspace{-.7cm}
\qed
\end{example}

\subsection{Positive Non-Horn Expressions} \label{subsec:Posit-Non-Horn}

For the sake of  completeness,  we provide the formal definition
of the class $\mathcal{H}^{2+}_{\mathcal{L}}$ of positive non-Horn expressions  
allocated to DNP  heads but  do not elaborate upon their analysis 
 as they are not relevant for this article 
 (are needed in Definition \ref{def:NNP-rule} only). 

\vspace{.2cm}
 Definition \ref{def:XXXX} is the formal definition of  $\mathcal{H}^{2+}_{\mathcal{L}}$, whose lines (1) to (3)
are such that:
\begin{itemize}
\item   (1) includes all   clauses containing  
exactly two  positive and different elementary expressions, 
and so excluding negative and positive-Horn clauses. To the latter clauses,
negative, positive-Horn and positive non-Horn clauses can be added in line (3);

\item   (2) is very similar to that of previous definitions of Horn classes; and 

\item   (3) includes disjunctive expressions with \underline{at least one} 
positive non-Horn disjunct and potentially having other  positive-Horn and positive-non-Horn disjuncts.
\end{itemize}

\vspace{.05cm}
 Although we do not prove here, one can check that excluded from $\mathcal{H}^{2+}_{\mathcal{L}}$ 
are  expressions whose translation to CNF  either  is a positive-Horn CNF or 
contains a fully negative clause.
On the other hand, the function below is not  the unique existing  one that
  formally defines  $ \mathcal{H}^{2+}_{\mathcal{L}}$ and other different alternative functions 
  could  be proposed.

\begin{definition} \label{def:XXXX} 
    $ \mathcal{H}^{2+}_{\mathcal{L}}$ is  
 the smallest set verifying  the following inductive conditions below. 
 $ \mathcal{H}^{2+}_{\mathcal{A}}$  is defined as $ \mathcal{H}^{2+}_{\mathcal{L}}$  by replacing 
 $\mathcal{L} $ with $ {\mathcal{A}}$:
    \begin{itemize}

\item [(1)] \, If $ x, x' \in \mathcal{L}  \cup \{\bot\},  x \neq x', \ \mbox{and}  \ {y_i} \in  \mathcal{E}^1_{\mathcal{L}}$ \, then \, 
$\vee (\overline{y_1} \ldots x \ldots \overline{y_i} \ldots x' \ldots \overline{y_k}) \in 
  \mathcal{H}^{2+}_{\mathcal{L}}$.  

\item [(2)]  \,  If  
\  $\varphi_1, \ldots,  \varphi_k \in \mathcal{H}^{2+}_{\mathcal{L}}$ \, then
 \,    $\bigwedge  \, [  \varphi_1  \ldots  \varphi_k \,] 
   \in \mathcal{H}^{2+}_{\mathcal{L}}.$ \hspace{3.27cm}  

 
 \item  [(3)]    If  $ \varphi_i  \in \mathcal{H}^{2+}_{\mathcal{L}} $  
  and    $    \, \varphi_j \in  \mathcal{N}_{\mathcal{L}}  \cup \mathcal{H}_{\mathcal{L}}^+ \cup \mathcal{H}^{2+}_{\mathcal{L}}, j \neq i,$ then
    $\vee   (\varphi_1   \ldots \varphi_i \ldots \varphi_j \ldots  \varphi_k  ) 
 \in \mathcal{H}_{\mathcal{L}}^{2+}.$ 
 
\end{itemize}
\end{definition}

\subsection{Nested-Horn: A Property} \label{sec:PROTERTIES-HORN-NEXPRESS}

This section  describes a fundamental property from  our objective viewpoint of  Horn  expressions.
As we saw, positive-Horn CNFs   have the  syntax:
 \begin{equation} \label{eq:fourth}
H = \wedge_k [\,   (h_1 \vee {D_1}) \ \ldots \
  (h_k \vee {D_k}) \, ], 
\end{equation} 

wherein  $ h_i \in \mathcal{L} \cup \{\bot\}$  is positive and  
  ${D_i}$ is a  negative clause and
   both  constitute the pillar  of    NFNP semantic concepts as  shown
    in Subsection \ref{subsec:SEMANTIC-NFNP}. 
%
This readable and computationally comfortable 
  pattern seems,   at first glance, inexistent within  Horn  expressions 
      since positive  and negative 
   sub-expressions  are apparently
      dissolved inside the  structure of 
      the global Horn expression (see  examples in  Section \ref{sec:Positive-Horn-Nested-Form}). 
   
   \vspace{.1cm} 
  Interestingly,  one of our main results  is proving that   
 Horn expressions   have such a structurally-simple pattern, but 
  it is presented in an {\em implicit  and  generalized nested way}. 
  This implicit pattern
eases the leverage of semantic notions in order to  
      extend  many  properties   
      from NFNP  
      to NNP (their heads are  a positive Horn-CNF and a positive Horn-expression, respect.)
     and so indirectly also from NP to NNP.
 
  \vspace{.1cm}
    More precisely, we  prove that if 
 $ h_1, h_2   \ldots, h_k \in  \mathcal{L} \cup \{\bot\}$ 
 are  the positive   occurrences  in $H$,
 then there  exists an expression  ${H}_\Delta$   strongly-equivalent  to $H$ such that:
\begin{tcolorbox}
  $$H_\Delta= \wedge_k [\,   (h_1 \vee \Delta_1) \ \ldots \ 
  (h_k \vee \Delta_k) \, ],$$  
\end{tcolorbox}
 
 wherein each $\Delta_i $ is a fully    negative 
 sub-expression\footnote{Although as  will be shown later they are   not sub-expressions in the sense 
of Definition \ref{def:sub-for}.} of $H$.
   $H_\Delta$ is obviously a generalization of  (\ref{eq:fourth}) resulting 
   from simply replacing negative clauses $D_i$
 with negative expressions $\Delta_i$.    
Subsections \ref{subsec:NEGATIVE-expressions} and  
 \ref{subsec:Positive-Negative-Pattern}  explain
how to find  any specific  $\Delta_i $ 
 and   $H_\Delta$, respectively.
 
 \begin{tcolorbox}
{\em {\bf \em Notation.} From a formal viewpoint, we should  write 
 $\Delta(H,h)$  but for the sake of  simplicity,  
 will omit both  $H$ and $h$, or only $H$,   when they are clear from the context.}
\end{tcolorbox}

\subsubsection{Obtaining Negative $\Delta$'s } 
\label{subsec:NEGATIVE-expressions}


 Let $H$  be a positive-Horn expression and ${\textcolor{black} {h}}$
be a  positive
\underline{occurrence}  of an elementary expression $\mathcal{L} \cup \{\bot\}$
in $H$.
Our  goal is   finding a sub-expression
 $\Delta$ of $H$ verifying: 
 $$ H \models    h \vee \Delta, $$ 
 or, in other words,  finding $\Delta$ 
 fulfilling the following logical relationship:  
  $$ \mbox{If $I \models H$  \ and \ $I \fmodels \, \Delta$ 
\ then \ $h \in I$.}$$ 

Observe that this logical relationship  also exists between a 
positive-Horn CNF $H$  and  its clauses $( h  \vee D )$.  
We  formally define    
$\Delta$ and  then
  provide some illustrative  examples.
\begin{tcolorbox} 
 \begin{definition} \label{def:DELTA} 
 Let  $ H  \in  \mathcal{H}_{\mathcal{L}}^+$      
  and denote by $H(  {\textcolor{black} {h}} )$ that $h \in \mathcal{L} \cup \{\bot\}$ is a specific positive occurrence 
  in   $ H $.  
      ${\Delta}(H,{\textcolor{black} {h}})$   
   and  inductively defined as:       
      $$\Delta(H,{\textcolor{black} {h}})= \left\{
\begin{array}{l l l}

\vspace{.25cm}

  \bot   & \quad   H({\textcolor{black} {h}})= {\textcolor{black} {h}}.   \\ 

\vspace{.25cm}
 {\Delta}(H_i,{\textcolor{black} {h}})   
& \quad  H({\textcolor{black} {h}})=\bigwedge \, [ \, H_1  \ldots  \,    
H_{i}({\textcolor{black} {h}})  \, \ldots  H_k \,]   . \\

\bigvee (\, H_1 \ldots \, 
  \Delta(H_i,{\textcolor{black} {h}})
   \, \ldots H_k )   & 
   \quad  H({\textcolor{black} {h}})=
   \bigvee \, (\, H_1 \ldots \,  
    H_{i}( {\textcolor{black} {h}} ) 
   \, \ldots H_k ).
\end{array} \right.$$ 
\end{definition}
\end{tcolorbox}


\begin{example} Let us consider expression $H$ below and its positive occurrence $a$. 
If one applies Definition \ref{def:DELTA} to $H$, one then obtains 
with no difficulty that $\Delta(a)=\phi_2 $.
$$H=\bigwedge_2 \, [\, {\textcolor{black} {\overline{b}}} \quad 
  \bigvee_2 \, ( \,\phi_2 \quad  {a}) \, ].$$
\end{example}

\begin{example} \label{ex:Syntact-Dp-1}
   Let us  focus  on $H$ below, and, inside it, on the  occurrence  of  
   ${\textcolor{black} {a}}$:
   
   \vspace{-.05cm}
   $$H({\textcolor{black} {a}})  =   \bigwedge_2 \, [\, \varphi_4 \quad 
   \bigvee_2 \, ( \, \varphi_1 \quad  
   \bigvee_2 \, (  \,    \bigwedge_2 \, [\, {\textcolor{black} {a}} \quad 
  \bigvee_2 \, ( \,\phi_2 \quad \overline{a}) \, ] \ \varphi_2) \,) \, ]$$ 
 
 \vspace{-.1cm}
\noindent  We upwards from ${\textcolor{black} {a}} $ 
apply Definition \ref{def:DELTA}  to $H({\textcolor{black} {a}}) $  
 and  obtain
   $\Delta(H({\textcolor{black} {a}})) $ as follows:
 $$\Delta(H,{\textcolor{black} {a}})= \left\{
\begin{array}{l l l}

\vspace{.25cm}
 \bot & \quad H_0({\textcolor{black} {a}})={\textcolor{black} {a}}.   \\ 

\vspace{.25cm}
 \bot  & \quad H_1({\textcolor{black} {a}}) =   \bigwedge \,  [\,
  H_{0 }( {\textcolor{black} {a}} ) \ \
 {\vee  (\, \phi_2 \ \overline{a})} \, ] . \\
   
 \varphi_2 \vspace{.25cm}
   & \quad  H_2 ({\textcolor{black} {a}})=
   \bigvee \, ( \,  H_{1}( {\textcolor{black} {a}} ) 
 \ \ \varphi_2). \\
   
     \vspace{.25cm}
 \vee  ( \, \varphi_1 \ \varphi_2)   & \quad  H_3 ({\textcolor{black} {a}})= 
  \bigvee (\, \varphi_1 \ \  
  H_{2}( {\textcolor{black} {a}} )   \,) . \\
  
 \vee  ( \, \varphi_1 \ \varphi_2) & \quad  H ({\textcolor{black} {a}})=
    \bigwedge \, [ \, \varphi_4 \ \  
   H_{3} ( {\textcolor{black} {a}} )\,]   .
\end{array} \right.$$ 

\vspace{-.6cm}  
     \qed
   
\end{example}

\begin{example}  \label{ex:syntac-D(p)-2} 
Consider now $H'({\textcolor{black} {a}})  $ below, wherein 
 the $ \phi $'s and 
$\Phi$'s are unspecified  sub-expressions and
$H ( {\textcolor{black} {a}})$ is assumed to be
the one from  Example \ref{ex:Syntact-Dp-1}. 
Besides, we focus on the same 
 atom occurrence $ {\textcolor{black} {a}} $ within $H ( {\textcolor{black} {a}})$ 
 as in Example \ref{ex:Syntact-Dp-1}:
$$ \mathrm{H}'( {\textcolor{black} {a}})=
   \bigwedge_2 \ \ [  \bigvee_2 \ \ ( \bigwedge_2  \ \ [\bigvee_2 \ \ ( \, \bigwedge_2  
 \ \ [ \, \bigvee_2 \, (\mathrm{H}_{}( {\textcolor{black} {a}}) \quad \phi_1) \quad \Phi_1 ] 
\quad \phi_2) \quad \Phi_2 ] \quad \phi_3) \quad \Phi_3 ]. $$

  It is not hard to obtain with  Definition \ref{def:DELTA} 
  and formula $ \mathrm{H}'( {\textcolor{black} {a}})$:
$$\Delta(H',{{\textcolor{black} {a}}})=\bigvee \ (\,\Delta(H,{{\textcolor{black} {a}}}) \quad \phi_1 \quad \phi_2 \quad \phi_3).$$

By Example  \ref{ex:Syntact-Dp-1} we have that
$\Delta(H,{{\textcolor{black} {a}}})=
 \vee \, ( \, \varphi_1 \ \varphi_2)$, and then:
$$\Delta(H',{{\textcolor{black} {a}}})
= \bigvee \, ( \, \varphi_1 \ \ \varphi_2 \ \ \phi_1 \ \ \phi_2 \ \ \phi_3).$$  

\vspace{-.75cm}
\qed
\end{example}

We now turn our attention  to the clausal context. 
Assume that  a positive-Horn CNF
  $ H$ contains 
a clause  $    {\textcolor{black} {h}} \vee D$, where $h$ is positive. By  applying
  Definition \ref{def:DELTA}     to  
  $ H  $, one 
   obtains  that 
 $\Delta$ is $D $, and hence that for CNFs,
 $\Delta$  
 has indeed the intended meaning. 

 \begin{corollary}  \label{cor: Horn-CLASAL-D}
If $ H $ is a positive-Horn CNF  and    $ {\textcolor{black} {h}} \vee D \in H$, then: \  
  $\Delta= D .$
\end{corollary}

\begin{niceproof} It  is immediate from
Definition \ref{def:DELTA} of $\Delta$.
\end{niceproof}


\subsubsection{\bf Obtaining Expression  $H_\Delta$}
\label{subsec:Positive-Negative-Pattern}

Once we know how to obtain sub-expressions $\Delta$'s, we can proceed to determine $H_\Delta$.

\begin{theorem} \label{lem:HyNegD=a} If   $H  \in  \mathcal{H}_{\mathcal{L}}^+$, 
 ${\textcolor{black} {h}}$  
  is any  occurrence  of    $\mathcal{L} \cup \{\bot\} $  in $H$ and  $\Delta $ is its associated sub-expression given 
  by Definition \ref{def:DELTA},
  then we have:  
    $$\forall I, \,   H^I \models  \, (h \vee {\Delta(H,h))^I}.$$ 
\end{theorem} 

\begin{niceproof} 
See Appendix.
\end{niceproof}

Next,  we supply two examples in which we show that 
if an interpretation $I$  satisfies a positive-Horn expression $H$ 
and falsifies its sub-expression  $ \Delta$ 
 then necessarily $h \in I$.

\vspace{.3cm}
\noindent  {\bf Example \ref{ex:Syntact-Dp-1}.}   ({\em Continued}\,).   
 We check step-by-step Theorem \ref{lem:HyNegD=a}.  
 We already determined that: $\Delta(a) 
 =(\vee \ \varphi_1 \ \varphi_2) $.
 $$I \models H^I \quad \mbox{and} \quad I \fmodels \, (\vee \ \varphi_1 \ \varphi_2)^I
 \quad  \mbox{entails} \quad
   I \models (\bigwedge_2 [\, \varphi_4 \ \  
        \bigwedge_2 [\, {\textcolor{black} {a}} \ \ 
  \bigvee_2 (\phi_2 \quad \neg {a}) \, ] \  \, ])^I.$$ 
 
 Hence,   the entailment written above  leads to 
 $  a \in I$. \quad
 \qed

\vspace{.35cm}
\noindent {\bf Example \ref{ex:syntac-D(p)-2}} ({\em Continued}\,).  
We already obtained that:
$\Delta({{\textcolor{black} {a}})}
=\vee (\varphi_1 \  \varphi_2 \  \phi_1 \  \phi_2 \  \phi_3) .$ 
 Thus, when $I \models H^I$ and $I \fmodels  \Delta({{\textcolor{black} {a}})^I}$,  $I$ verifies the expression below,
  and hence clearly $   a \in I$:
 $$ I \models   (\bigwedge_4 \  [\bigwedge_2 \ {\textcolor{black} {a}} \ 
 (\vee \ \phi_2 \ \neg {a})  ]   \ \Phi_1   \ \Phi_2 \ \Phi_3 ])^I.$$
  
\vspace{-.8cm} 
\qed

\vspace{.3cm}


\noindent From Theorem \ref{lem:HyNegD=a}, one  can deduce  the   generalized Horn-CNF pattern $ {H}_\Delta $ of $ {H} $: 

\begin{tcolorbox} 
\begin{theorem} \label{THE:CLAUSAL-PATTERN}  
If $H  \in  \mathcal{H}_{\mathcal{L}}^+$, $ h_1,   \ldots, h_k \in \mathcal{L} \cup \{\bot\}$ 
are all the  positive occurrences   in $H$ and 
 $ \Delta_1,   \ldots, \Delta_k $ are by Definition \ref{def:DELTA} their associated sub-expressions, then:
$$  {H}  \Leftrightarrow   H_\Delta=
\wedge_k \, [ \, {  (h_1 \vee \,  \Delta_1(H,h_1))}  
\ \ldots \
   { (h_k \vee \,  \Delta_k(H,h_k)}) \, ].$$
\end{theorem}
\end{tcolorbox}

\begin{niceproof} See Appendix.
\end{niceproof}

\begin{example} \label{ex:FORMULA-TILDE-H} We check 
  positive-Horn $H_1$ with positive occurrences $\{c,b,\bot\} $:
$$ H_1=\bigvee_2 \, (\overline{a}  \quad 
\bigwedge_2 \, [ \ \ \bigvee_2  ( \wedge \, [ \overline{b} \ \ \ \overline{g}]  \ \ \ c  )  \quad
\bigwedge_2 \, [ b   \ \ \  \vee (\bot \ \ \overline{\mbox{\em not} \, d}) \, ] \ \, ] \ \, ).$$

 -- Applying Definition \ref{def:DELTA} leads to the following  $\Delta $ formulas:
$$\Delta(b)=\overline{a} .$$ 
$$\Delta(c)=\bigvee_2 (\overline{a} \ \ \wedge [\overline{b} \ \ \ \overline{g}] \ ) .$$ 
$$\Delta(\bot)=\vee (\overline{a} \ \ \overline{\mbox{\em not} \, d}) .$$


\vspace{.25cm}
-- Hence expression  $ {H}_\Delta$ that is strongly-equivalent to $H$ is:
$${H}_\Delta=\bigwedge_3 \, [ \,  \vee \, (b \ \overline{a}) \ \ 
\bigvee_3 \, ( c \ \   \overline{a} \ \ \wedge [\overline{b} \ \overline{g}] ) \ \ 
\vee   (\bot \ \ \overline{a} \ \overline{\mbox{\em not} \, d}) \ ] $$

\vspace{-.7cm}
\qed
\end{example}

\begin{example}  \label{ex:MANY-NOTs} Now we take  $H''$ in Example \ref{ex:POSIT-POLAR-Horn}.
 The positive  literals in $H''$ are $\neg y$ and $\neg x$, and 
regarding the two occurrences of  $\neg y$,   we will differentiate them 
  by denoting the leftmost (resp. rightmost) one by $\neg y^1$  (resp. $\neg y^2$).
  So $H''$ can be rewritten as:
   $$ H''=\bigvee_2 \, (\overline{\mbox{\em not} \, x}  \quad  
 \bigwedge_2 \, [\    \vee \,(\overline{\mbox{\em not} \, \neg y}     \ \ \neg y^1  )  \quad
\bigwedge_2 \,  [ \, \neg y^2   \quad    \vee (\neg x \ \ \overline{v} ) \ ] \ ] \ )
\in \mathcal{H}^+_{\mathcal{L}}.$$
    Definition \ref{def:DELTA} leads to the following   formulas:
$$\Delta(\neg y^2)=\overline{\mbox{\em not} \, x}$$ 
$$\Delta(\neg y^1)=\vee (\overline{\mbox{\em not} \, x} \ \ \overline{\mbox{\em not} \, \neg y})$$ 
$$\Delta(\neg x)=\vee (\overline{\mbox{\em not} \, x} \ \ \overline{v})$$

With these values, one  obtains that the   expression ${H''}_\Delta$ is the following:
$$\bigwedge_3 \, [ \, \vee (\neg y  \ \ \overline{\mbox{\em not} \, x}) 
\quad \vee (\neg y \ \ \overline{\mbox{\em not} \, x} \ \ \overline{\mbox{\em not} \, \neg y}) \
\quad \vee (\neg x \ \ \overline{\mbox{\em not} \, x} \ \ \overline{v}  ) \ ] $$

 \vspace{-.75cm}
 \qed
\end{example}

We now switch to the clausal scenario in order to  check that within 
it, the  meaning of ${H}_\Delta$
is indeed the expected one. Thus,  if we apply Theorem \ref{THE:CLAUSAL-PATTERN}  to   
a positive-Horn CNF
 $ H $, then we obtain that ${H}_\Delta$ is the proper  $ H $, and so  formally:
 

\begin{corollary} If   $ H $ is a positive-Horn CNF, 
then: \  $ {H}_\Delta=H $.
\end{corollary}

\begin{niceproof} 
It follows from Theorem \ref{THE:CLAUSAL-PATTERN}  and Corollary \ref{cor: Horn-CLASAL-D}.
\end{niceproof}

\section{Syntax of Normal NASP} \label{sec:DefiniteNLP-Syntax}

In this section, we
   define  the syntax of normal NASP (NNP) rules,
  and thereby, trace the syntactical boundary between   normal and disjunctive   NASP programs. 
We also show that an NNP rule is strongly-equivalent to a set of NFNP rules but it is up to an exponential factor 
 more succinct than  the latter set.
 
 \vspace{.05cm}
 Subsection \ref{subsec:SYNTAX-NNP-Definition} furnishes the syntax of NNP rules. Subsection \ref{subsec:SPLITING} specifies informally
 how to split any arbitrary (disjunctive) NASP program into a set of NNP programs. Subsection \ref{subsec:EXPONEN-SUCCINCT}
 discusses the exponential succinctness of NNP w.r.t. classical NP. The last subsection
 formalizes and gives examples of translations between NNP and  NP programs.

\subsection{Definition of the  Syntax} \label{subsec:SYNTAX-NNP-Definition}
  
 \begin{definition}   \label{def:NNP-rule}
An NASP rule  $H \leftarrow B $, with 
$B \in \mathcal{E}_\mathcal{A}$, 
 is normal or NNP$_\mathcal{A} $    
 (resp.   disjunctive or  DNP$_\mathcal{A}$) 
 if $H \in \mathcal{H}^+_\mathcal{A}$ (resp. $H \in \mathcal{H}^{2+}_\mathcal{A}$). 
 An  NNP$_\mathcal{A} $ (resp.  DNP$_\mathcal{A} $) program is a 
 set of $ n \geq 1  $ NNP$_\mathcal{A} $ 
 (resp. $\geq 0$ NNP$_\mathcal{A} $   and  $\geq 1$ DNP$_\mathcal{A} $) rules.
  For    NNP$_\mathcal{A} $ rules 
  $r=H \leftarrow B$,   
  we have $H \Leftrightarrow   {H}_\Delta= \wedge_k [(h_1    \vee  \Delta_1) \ \ldots 
\ (h_k    \vee \Delta_k) \,]$ (see Theorem \ref{THE:CLAUSAL-PATTERN}), 
and 
   thus:
\begin{enumerate}

 \vspace{-.cm}
\item $ r $  contains  (resp. is)
 a fact   if $ B = \top $
 and  $\exists i$ (resp. $\forall i$): \ $\Delta_i = \bot$.
 
 \vspace{-.2cm}
 \item $ r $  is partially (resp. is) {\em not}-free if 
 $ B = B^\mathcal{A}  $ 
  and $\exists i$ (resp. $\forall i$): $ \Delta_i = \Delta_i^\mathcal{A}$.
  
\vspace{-.2cm}
\item  $ r $ contains (resp. is)    a constraint if  $\exists i$   (resp. $\forall i$):  
$h_i = \bot$.   \qed
\end{enumerate} 
\end{definition}

\noindent {\bf Remark 1.} According to Definition  \ref{def:NNP-rule}, we have that:

\vspace{.2cm}
-- a rule $H \leftarrow B$ is NNP$_\mathcal{A} $ if $H \in \mathcal{H}^+_\mathcal{A}$ (is positive-Horn); and

\vspace{.1cm}
-- a rule $H \leftarrow B$ is DNP$_\mathcal{A} $ if 
$H \in \mathcal{H}^{2+}_\mathcal{A}$ (is positive-non-Horn).

  \vspace{.2cm}
  So non-positive Horn expressions included in the set
  $\mathcal{H}_\mathcal{A} - \mathcal{H}^+_\mathcal{A} \cup \mathcal{H}^{2+}_\mathcal{A}$ 
  are useless for NASP.  But this is similarly met within the
  unnesting framework, wherein negative Horn clauses  are not employed 
  as was suggested after Example \ref{ex:INTUITION-Exam}, in Section \ref{sec:General-Presentation}.
  

\begin{example} \label{ex:first-NNP-Rule-SYNTAX}
Rule $r$ below  is  NNP$_\mathcal{A} $ as its head expression is  positive-Horn.
Besides, this NNP$_\mathcal{A} $ rule is {\em not}-free and it contains no fact and one constraint.

\vspace{-.2cm}
$$ r=  \bigvee_2  ( \wedge \, [   \overline{b}  \  \ \ \overline{d} ] \quad    
\bigwedge_3 [\,c  \  \ a  \ \ \vee (\overline{g} \ \ \bot) \,]\,)  \leftarrow f.$$  

\vspace{-.2cm}
The positive occurrences  in $H$ are $ \{c,a,\bot\} $, 
and then, we can obtain without difficulty:

\vspace{-.2cm}
$$\Delta=   \wedge \, [   \overline{b}  \  \ \ \overline{d} ] \ \mbox{\ for  $a,c$ }
\qquad \mbox{} \qquad 
\Delta=\bigvee_2 (\,  \wedge \, [   \overline{b}  \  \ \ \overline{d} ]  \quad \overline{g}) \ \mbox{\ for  $\bot$ }.$$ 

\vspace{-.85cm}
 \qed
\end{example}

\begin{example} The positive occurrences in the head  $H$ of the rule below are
$\{\bot,b,g\}$. In this particular case, one has:
$ H = H_\Delta$ and the  sub-expressions $\Delta$'s are easily identifiable. 
This rule is partially {\em not}-free and it contains a fact $b$ and a constraint.
$$  \bigwedge_3 \ [ \ \vee (\overline{b}   \ \ \bot )  \quad \
  b \ \quad  \bigvee_2 \, ( g   \ \ \ \wedge  
  [ \overline{d} \ \ \overline{\mbox{\em not} \, f} \, ] \ )  \ \ ]   \longleftarrow   \  \top.$$ 
 
  \vspace{-.8cm}
 \qed 
 
\end{example}

 Example \ref{ex:literature-2} gives a  non-ground, 
 {\em not}-free,  constraint-free  NASP  
 program,   borrowed from  
  \cite{BriaFL08,BriaFL09,Bria09c}, which
    makes evident the easiness  offered by NASP
   to   skip  ASP redundancies.

\begin{example} \label{ex:literature-2}  The  program 
co-CERT3COL (left column, Table 2) -- is an extension of graph 3-uncolorability -- due to I. Stewart \cite{Stewart91a}:
Given a graph $ G $, whose edges $ (x,y) $ are labeled with non-empty sets of variables
$ v_1, \ldots, v_n $, find a truth assignment 
to $ v_1, \ldots, v_n $
 s.t. the subgraph $ G' $ of $ G $, containing all edges
$ e $ with at least one literal in the label of $e$ is satisfied, is not 3-colorable.
Let  the labeled edge of $ G $ 
represented by predicates $ p(x,y,v) $ and 
 $ n(x,y,v) $  indicate that the edge   $(x , y) $
 has as a positive or negative label $ v $, respect.
 
 \vspace{.1cm}
 Table 2, on the left (resp. right), 
  reports an ASP (resp. equivalent NASP) encoding from \cite{EiterIbarakiMakino97}
  (resp.  \cite{BriaFL08,BriaFL09,Bria09c}).
  $ rb $ to $ rd $  belong to both encodings.  $Ra$ to $Rc$ are 
  obtained by applying    steps 3-4,   Proposition  \ref{pro:STRONG-PRESERVED}, 
  to   ASP rules.
Observe that program in Table 2, on the right,  
is FDNP$_\mathcal{A} $ (Definit. \ref{def:NNP-rule}) and that it saves 7 rules and 
the intermediate predicate $ bool $.
 A more succinct NASP program,
 not provided in the literature,
 is given next.
 
\noindent
\begin{table}[h!]
  \begin{center}
    \label{tab:table1}
    \begin{tabular}{|l|r|}
     \hline
      $r1 : v(x) \leftarrow  p(x,y,v)$ & 
         \multirow{4}{*}{$Ra : v(x) \wedge v(y) \leftarrow  p(x,y,v) \vee n(x,y,v)$} \\
      $r2 : v(y) \leftarrow   p(x,y,v)$ &    \\
      $r3 : v(x) \leftarrow  n(x,y,v)$ &    \\
      $r4 : v(y) \leftarrow  n(x,y,v)$ &    \\
      \hline
       $r5 : bool(v) \leftarrow p(x,y,v)$ &  
       \multirow{3}{*}{$ Rb : t(v)  \vee f(v) \leftarrow p(x,y,v) \vee n(x,y,v)$} \\
      $ r6 : bool(v) \leftarrow n(x,y,v)$ &  \\ 
      $ r7 : t(v) \vee f(v) \leftarrow bool(v)$ &   \\ 
      \hline
      $r8 : c(x,r) \leftarrow w, v(x) $ & 
      \multirow{3}{*}{$Rc : [\wedge \ c(x,r) \  c(x,g) \  c(x,b) ] 
      \leftarrow w,  v(x) $} \\
      $ r9 : c(x,g) \leftarrow w \wedge v(x) $ & \\
       $ ra : c(x,b) \leftarrow w \wedge v(x) $ &  \\
       \hline
       $rb : (\vee \ c(x,r) \   c(x,g) \   c(x,b)) \leftarrow v(x)$ 
       &   $\mbox{ \ rule} \ rb$  \hspace{2.6cm}  \\ 
       $rc : w \leftarrow   \wedge [ p(x,y,v) \ t(v) \ c(x,A)  
       \ c(y,A) ]$ 
       &        $\mbox{ \ rule} \ rc$  \hspace{2.6cm}   \\
       $rd : w \leftarrow   \wedge[ n(x,y,v) \ f(v) \ c(x,A) 
       \ c(y,A) ]$ 
       &      $\mbox{ \ rule} \ rd$ \hspace{2.55cm}    \\
       \hline       
    \end{tabular}
  \end{center} 
    \vspace{-.4cm}
\caption{\em ASP (left) and  FNASP (right) programs   \cite{BriaFL08,BriaFL09,Bria09c}.}
\end{table}

\vspace{-.4cm}
Rules $Ra$ and $Rb$ can be merged into 
  $ R_{AB} $ in
  Table 3.
 Rules $rc$ and $rd$  share 
the head literal $w$ and 
the disjunction  $ \vee (B(rc) \ \ B(rd)) $ 
of their body expressions, after  factorizing, gives rise to
 $\Psi$ in  
Table 3. Now, integrating $\Psi$ into  $Rc$ leads to 
the   rule:  
$$R{c}'=     \wedge  [c(x,r) \ \ c(x,g) \ \ c(x,b) \, ]   \longleftarrow  \Psi \wedge v(x) $$

 And now, the original  rule $rb$ 
 shares the body literal $v(x)$ with  $Rc'$.
\begin{table}
\begin{center}
\begin{tabular}{|l|}
  \hline
  \\
\hspace{1.6cm} $R_{AB}=  \bigwedge_3  [\,v(x) \ \ v(y) \ \ 
    (\,t(v)  \vee  f(v) \, ) \, ] 
\longleftarrow  p(x,y,v) \vee n(x,y,v)$  \\
\\
 \hspace{0.8cm}$\Psi=   \bigwedge_3 [ \, c(x,a) \  \ \ c(y,a) \quad 
 \bigvee_2 (  [\,p(x,y,v) \wedge  t(v) \,] 
\ \    [\,n(x,y,v) \wedge f(v) \,] \, ) \   \, ]$
\\ 
\\ 
 \hspace{0.cm}$R=   \bigwedge_2 [\,  \vee (c(x,r) \  c(x,g) \   c(x,b)  ) \ \  
 \bigvee_2  
 ( \wedge  [c(x,r) \  c(x,g) \  c(x,b) ] \ \ \overline{\Psi} ) \ ]
\leftarrow   v(x)$ 
\\
\\
\hline
\end{tabular}
\end{center}

\vspace{-.4cm}
\caption{\em Definitive NASP  equivalent to the ASP in Table 1.}
\end{table}
In order to shift expression $\Psi $ to the head of $Rc'$, 
we must obtain the   NNF expression $\overline{\Psi} $ strongly-equivalent to  $\Psi $ 
by employing Definition \ref{def:DEFINITION-of-OVERLINE-F}, which is  a kind
of application of De Morgan law's
(supported by Corollary \ref{cor:DE-Morgan}). 
Then placing the obtained NNF expression $\overline{\Psi }$
   in the head of $R{c}'$ and merging 
 the resulting rule 
with  rule $ rb $  leads to  $ R $ in  Table 3. \qed
\end{example}

 \begin{tcolorbox}
{\em  
 The initial ASP program,
  Table 2,  left column, 
 is compacted into an NASP one  having only  2  rules ($ R_{AB} $ and $R$)
  and 19 literals,   contrasting
 with the initial 38 literals 
 and even the 27 literals in the FNASP program,  Table 2,   right column.}
 \end{tcolorbox}

  \begin{definition}   \label{def:FNASP-EXTENDED-rule} 
    Extended normal (or  NNP$_\mathcal{L} $)   
    and disjunctive (or  DNP$_\mathcal{L} $) nested
programs   are defined   as NNP$_\mathcal{A} $ and DNP$_\mathcal{A} $  ones, respect., by replacing, 
in Definition \ref{def:NNP-rule}, $\mathcal{A}$ with  $\mathcal{L}$.
We will refer  to   NNP$_\mathcal{A} $
and NNP$_\mathcal{L} $ rules/programs as NNP rules/programs. An NNP   program  is head-consistent
 if the set  constituted by all its head literals  $ h \neq \bot$
 is consistent.
 
\end{definition}



  \noindent {\bf Example \ref{ex:INTRO-compact}.} 
(Continued) The nested rule  is indeed NNP$_\mathcal{L}$ because one can check that
its head is in $\mathcal{H}^+_\mathcal{L}$ and   its body 
in $\mathcal{E}_\mathcal{L}$.  In this  case, we have $ H =H_\Delta$ and so its $\Delta$'s are easily 
identifiable. This rule is partially {\em not}-free and
 contains neither facts nor constraints.

\begin{example} A rule $H \leftarrow B$, with $B \in \mathcal{E_\mathcal{L}}$, is   NNP$_\mathcal{L} $  if  
$H$ is any of the positive-Horn expressions $H'$ and $H''$ provided in Example \ref{ex:POSIT-POLAR-Horn}. 
\end{example}

\subsection{Splitting Disjunctive NASPs} \label{subsec:SPLITING}

We have chosen the two disjunctive NASP (DNP) programs in 
Examples \ref{ex:Literature-1} and  \ref{ex:literature-2} that are borrowed from the state of the art, 
and we employ them  as illustrative examples to informally show how to split any arbitrary NASP program
into a set of NNP programs.

\vspace{.25cm}
\noindent {\bf Example \ref{ex:Literature-1}.} {\em (Continued)}   
The  specified  DFNP$_\mathcal{L}$ rule at the end of the example   can   be easily split
  into the  two NFNP$_\mathcal{L}$ rules  below  whose head is just a single
  literal:
  
  \vspace{.25cm}
  \hspace{.5cm} $r_1=\neg {p}(i,n,s,a) \   
   \longleftarrow  \bigwedge_3 \, [\, p(i,n,s,a) \ \ p(i,m,t,b)  \ \ \  
  \bigvee_3 (\, n \neq m \ \ s \neq t \ \ a \neq b) \ ]$
  
   \vspace{.25cm}
 \hspace{.5cm}
 $r_2= \neg {p}(i,m,t,b)  \ 
   \longleftarrow  \bigwedge_3 \, [ \, p(i,n,s,a) \ \ p(i,m,t,b)  \ \ \  
  \bigvee_3 \, (\, n \neq m \ \ s \neq t \ \ a \neq b) \ ]$ 
  
  \vspace{.35cm}
  So the initial DP$_\mathcal{L}$ program can be represented 
  by the two NFNP$_\mathcal{L} $ programs: 
  $P_1=\{r_1,R\} $  and  $P_2=\{r_2,R\} $, 
  where $R$ is the bottom rule of Example \ref{ex:Literature-1}, and thus, more appropriately modeled by 
  either one DFNP$_\mathcal{L} $ program or two 
  NFNP$_\mathcal{L} $ programs.
  By  shifting inequalities from  body to  head,
$r_1$ and $r_2$  can also be  
  represented by the following NASP  rules, wherein, for the sake of notational simplicity, 
we have written as $=$ the relation ${\neq} $ being overlined $\overline{\neq} $, and so, wherein literals $x=y$  
   are in fact negative ones:
   $$\vee_2 (\neg {p}(i,n,s,a) \ 
    \ \wedge [n=m \ \ s = t  
 \ \ a=b] \, )
   \ \longleftarrow  \ p(i,n,s,a) \wedge p(i,m,t,b)$$ 
   $$\vee_2 (\neg {p}(i,m,t,b) \ \ 
   \wedge [n=m \ \ s = t \ \ a=b] \, )
   \ \longleftarrow  \ p(i,n,s,a) \wedge p(i,m,t,b)$$

 We next check that both rules are NNP$_\mathcal{L}$ and  not NFNP$_\mathcal{L}$, 
 that is,  shifting expressions   can oblige us
  to jump from  
 NFNP to  NNP.
Literals $\neg  {p}(\ldots)$ correspond 
to    literals   $h$ in our  definitions and so  
are positive. Altogether, 
such heads    have the following syntax: 
   $$\vee_2 (h \ \ \
   \wedge [\overline{\ell_1} \ \ \overline{\ell_2}  \ \ \,\overline{\ell_3}] \ ), \ \mbox{where} \ \ell_i \in \mathcal{L},$$  
  which fits into the definition of a positive-Horn expression, and more precisely, 
 it  is a positive-Horn DNF.
 So  both rules  are NNP$_\mathcal{L}$. On the other hand,  
the  program is head-consistent given that the set  $ \{\neg {p}(i,n,s,a), \ \neg {p}(i,m,t,b)\} $ 
is  consistent.

\vspace{.2cm}
Finally, it is easily obtained that the negative sub-expression
$\Delta$ associated to both literals $\neg {p}(i,n,s,a)$ and $\neg {p}(i,m,t,b)$
is $\Delta=\wedge [n=m \ \ s = t \ \ a=b] $. \qed

 \vspace{.3cm}
\noindent {\bf Example \ref{ex:literature-2}.} {\em (Continued)} 
 $Rab$ and $R$   are   DNP$_\mathcal{A}$, but their heads have only
  one non-Horn disjunctive sub-expression,  concretely  
$ \vee (t(v)  \ f(v))$ within   $ Rab$ and 
$  \vee (  c(x,r) \  c(x,g) \   c(x,b)) $ within $ R $, since
the disjunctive expression in $R$: 
$$ \bigvee_2 ( \wedge [c(x,r) \  c(x,g)  \ c(x,b) \,] \ \ \   \overline{\Psi} \,)$$ 

\vspace{-.2cm}
 is indeed positive-Horn given that  expression  $\overline{\Psi}$
is fully negative. 
Thus   $Rab$ and $ R $ can be split into 2 and 3 NNP$_{\mathcal{A}}$ rules, respect.,
and more concretely,  the initial DNP$_\mathcal{A}$ program can be decomposed into   6 NNP$_{\mathcal{A}}$ programs 
 wherein each one of them results from choosing a literal   
from  $ \{t(v),  \ f(v)\} $ and   another one
from  $\{c(x,r), \  c(x,g), \   c(x,b)\} $. 
One can  check that one of such six NNP$_{\mathcal{A}}$ programs
  is: 
$$ \wedge [ t(v)  \ \  v(x) \ \ v(y) \, ] 
\longleftarrow  p(x,y,v) \vee  n(x,y,v).$$
$$  \bigwedge_2 \ [\, c(x,b) \ \ \  
 \bigvee_2 \ ( \ \wedge [ c(x,r) \  c(x,g) \  c(x,b) \, ] \ \ \ \overline{\Psi}) 
\  ]  
\longleftarrow   v(x).$$

 It is not hard to verify that  the  six NNP$_{\mathcal{A}}$ programs 
 can be   alternatively obtained:
 
 \vspace{.2cm}
--  firstly by  \underline{splitting} the input DP$_{\mathcal{A}}$ rules, Table 2, on the left, into 
 NP$_{\mathcal{A}}$ rules; and  
 
 \vspace{.05cm}
 --  then by  \underline{compacting} each of the six  NP$_{\mathcal{A}}$ programs 
 into an NNP$_{\mathcal{A}}$ program. 
 
 \vspace{.2cm}
 It is simply obtained that  $\Delta $ for all the  three positive literals in the head 
 of the first rule is $\bot.$  Similarly, for the leftmost positive literal $c(x,b)$
 in the head of the second rule, its associated $\Delta$ is also $\bot$. Finally, 
the associated $\Delta$ with
 literals $c(x,r),  c(x,g)$ and  $c(x,b)$ in the head of the second rule  is  $\Delta= \overline{\Psi}$.
\qed

\vspace{.25cm}
As expected, such a  decomposition leads to, in the worst case,  
an exponential number of NNP programs,
just as  happens in the  unnested framework.
However, since  a DNP program 
can be exponentially more succinct than 
 disjunctive ASP (DP)  programs, 
the number of disjunctions
within  DNP programs  can also be exponentially smaller than 
the number of disjunctive rules of their  DP counterparts.

\vspace{.1cm}
 \begin{tcolorbox}
This means that  we can have exponentially less
NNP programs than NP ones, and moreover, that the size of every
NNP program can be exponentially smaller than that of its equivalent NP programs.
\end{tcolorbox}


\subsection{Exponential Succinctness} \label{subsec:EXPONEN-SUCCINCT}
We start this subsection by discussing  two NASP program subclasses handled in the literature. 
 In the first one, e.g. \cite{LloydT84,ErdemL01,Turner03,YouYM03}, alluded to in the introduction,
the head of   rules   
 is a single literal, 
 concretely, these rules have the  form:
$$   h   \ \leftarrow \ B,  \ \ \mbox{where} \  \ h \in \mathcal{L} \cup \{\bot\}
  \ \ \mbox{and} 
\ \ B \in \mathcal{E}_\mathcal{L}.$$

These rules  are 
technically,  yet not truly, NNP  as
 their head   is unnested.
  They are named  non-disjunctive nested rules by many authors and,    
 to the best of our knowledge, they are {\bf the unique  NASP rules considered 
 up to now  as normal.}

\vspace{.1cm}
The second NASP  program subclass to be discussed here
is also tackled by numerous authors,  e.g. in \cite{BriaFL08,BriaFL09},
 and  the   syntax of its rules is 
concretely the following:
 $$r= \bigvee_n \,(C_1 \ldots C_i \ldots C_n) \ \longleftarrow \ 
 \bigwedge_k \,[ D_1 \ldots D_j \ldots D_k]
  $$ 
 
 where the $C_i$'s are conjunctions and the $D_j$'s are  disjunctions 
   of elementary expressions  in both cases.  These rules are {\bf not} FNASP because their head is not a CNF but a DNF.
 
 \begin{example} 
 Rules in {Example \ref{ex:Literature-1}} and specified in Subsection \ref{subsec:SPLITING}
 are a sub-case of this simple type of  programs,
 specifically, the  $C_i$'s   and  $D_j $'s of $r_1$
  are as follows:  
 $$C_1=\neg {p}(i,n,s,a), 
 \quad C_2=[ n=m \ \ s = t   \ \ a=b]$$  $$D_1=p(i,n,s,a), \quad D_2=p(i,m,t,b).$$
 
 \vspace{-.75cm}
 \qed
\end{example}

According to the  definitions   introduced above,
 nested rules of this type  are normal or NNP$_\mathcal{L}$ when the following condition 
is fulfilled:

\begin{corollary} \label{Prop:SUB-CLASS-NNP} 
Let  $r= \vee \, (C_1 \ldots   C_{n}) \leftarrow B$, where each $C_i$ is a conjunction of literals. 
 $r$ is   NNP  if $n-1$ conjuncts  are
  fully negative  and   one     
  is fully positive (with $\mathcal{L} \cup \{\bot\} $ elements). 
    
 \end{corollary}

\begin{niceproof} It follows from  Definitions  \ref{def:NNP-rule} 
and   (and also   \ref{def:FNASP-EXTENDED-rule}) of an NNP program 
and the definition of
a positive-Horn expression.
\end{niceproof}

\begin{corollary} \label{cor:DNF-of-NNF} Let  
$r= \vee \, (C_1 \ldots   C_{k}) \leftarrow B$ be an NNP rule, 
all of whose $C_i$'s are formed by literals
 and  such that $\forall i, \vert C_i \vert=n$.  
$r$ is strongly-equivalent to $ n^k $  NP rules.
\end{corollary}

\begin{niceproof} It is a consequence of  applying
 distributivity to convert a DNF into a CNF. 
 \end{niceproof}

So {\em even simple NNP programs can
be exponentially more succinct than    NP ones}.

\vspace{.1cm}
 We now proceed with the presentation and analysis of  NNP programs
 that are also exponentially more compact
 than   NP ones and, besides, are highly nested.

\begin{example} \label{ex:EXPONEN}  Let us consider 
the following   NNP$_{\mathcal{A}}$ rule $H \leftarrow \top$, wherein:
$$H=\bigvee_2 ( B  \ \  \bigwedge_2 [ a_1 \ \
\bigvee_2 (C  \ \ \bigwedge_2 [ a_2 \ \ \bigvee_2  (D \ \ a_3) \ ] \ ) \ ] \ ), 
\quad \mbox{where:}$$
$$B=\wedge  [ \overline{b_1} \ \ \overline{b_2} \ \ \overline{b_3} ], 
\quad C=\wedge [\overline{c_1} \ \ \overline{c_2} \ \ \overline{c_3} ]  \ \ \mbox{and}
\ \ D=\wedge [\overline{d_1} \ \ \overline{d_2} \ \ \overline{d_3}].$$  

 It can be easily  obtained  that  the above NNP$_{\mathcal{A}}$ rule is strongly-equivalent to $3 + 3^2 + 3^3$ 
classical NP$_{\mathcal{A}}$ rules,
which,  formulated  in a compact way,   are  exactly
the following:
$$ \{a_1 \leftarrow b \parallel \quad   b \in \{b_1,b_2,b_3\}\},$$
$$ \{ a_2 \leftarrow \wedge [b \  c]  \parallel
\ \ b \in \{b_1,b_2,b_3\}, \ \ c \in \{c_1,c_2,c_3\}  \},$$ 
$$ \{ a_3 \leftarrow \wedge \,  [b \  c  \  d] \parallel
 \ \ b \in \{b_1,b_2,b_3\}, \ \ c \in \{c_1,c_2,c_3\}, 
 \ \ d \in \{d_1,d_2,d_3\}  \}.$$
 
 \vspace{-.7cm}
 \qed
\end{example}

Now    we  both formalize  and generalize  the pattern of the previous example:

\begin{proposition}  Let   $r= H \leftarrow \top $ 
be an NNP  rule 
 whose head $ H$ has the pattern:
$$ H=  \bigvee_2 ({C}_1 \  \ \bigwedge_2 [A_1  
    \ \ \bigvee_2   ({C}_2 \ldots \ \   
    \bigvee_2 ({C}_{k-1} \ \   \bigwedge_2 [ A_{k-1}  
    \ \ \bigvee_2 ({C}_k \ \ A_k )]) \ldots )])$$

\noindent where    $ A_j $ (resp. $  {C}_{i} $) is a  conjunction of 
  positive (resp. negative)  elementary expressions and  
$\vert C_i \vert= \vert A_j \vert=n$.   
$r$ is strongly-equivalent to  $O(n^{k})$
NP  rules.
\end{proposition}

\begin{niceproof} (Sketch) Looking at  Example \ref{ex:EXPONEN}, it is not difficult 
to obtain by induction that the NP  rules obtained 
from any rule $r$ of    level $1 \leq j \leq k$ are:
$$ l_j \leftarrow \wedge \,[\ell_1 \ \ell_2 \ldots 
\ell_{j} ], \ 
\ \mbox{where} \ l_j \in A_j, \ \mbox{and} \   \ 1 \leq i \leq j, \ 
 \ell_i \in C_i . $$

Since by hypothesis $\vert C_i \vert= \vert A_j \vert=n$,   for such 
a level $j$, the number of rules is clearly $ n^{j+1} $. 
Hence, the total number of rules for all levels is  
$ n^2 + n^{3} + \ldots  + n^{k+1}$.
\end{niceproof}

\subsection{Translating NNP into NP } \label{subsec:TRANSL-NNP-into-NP}
Next, we show two fashions  to convert an NNP program  into a strongly-equivalent NP 
one, but previously, 
we need the following property,  saying that 
 repeatedly applying  distributivity to a positive-Horn expression 
  leads to a positive-Horn CNF:

\begin{proposition} \label{pro:HORN-Exp-HORN-CNF}  Applying distributivity to 
 $ \varphi \in  \mathcal{H}^+_{\mathcal{L}}$
 leads to a positive-Horn  CNF.
\end{proposition}

\begin{niceproof} See Appendix.
\end{niceproof}

\begin{definition} \label{def:LP-of-NLP} Let  
$ P $ be an    NNP program and let $ r=( H \leftarrow B )\in P $.  
 We associate to  $P$  the  NP   program $NN_1(P)$  below, 
 where   dnf$(\varphi)$  
 and cnf$(\varphi)$ stand for the  DNF and CNF, respect., resulting 
 from applying distributivity to  $ \varphi$ and $FN(P)$ is  in Definition \ref{cor:SECT-APPROACH:pos}:
  $$ N_1(r)  =\mbox{cnf}(H) \leftarrow \mbox{dnf}(B),  
   \qquad \mbox{\em NN}_1(P)=\bigcup_{r \in P} \mbox{\em FN}(N_1(r)).$$ 
\end{definition}

\vspace{-.4cm}
\begin{proposition} \label{prop:P=FP(P)} We have that:

\vspace{.25cm}
-- If $r$ is an NNP rule then $N_1(r)$ is an NFNP rule.

\vspace{.05cm}
-- If  $ P $ is an NNP program   then 
$ \mbox{NN}_1(P)$ is an NP program.
\end{proposition}

\begin{niceproof} 
 The first statement is an immediate  consequence of Proposition \ref{pro:HORN-Exp-HORN-CNF}. 
From this fact and  by Proposition \ref{prop:NFNP-and-SN},    one 
can establish the second claim. 
\end{niceproof}

\begin{example} \label{def:NNP-to-NP-1}  
Assume the NNP$_\mathcal{A}$ program $ P=\{r\} $  whose rule is:
$$ r=\bigvee_2     \,  (\wedge \, [  \,   \overline{b}  \  \   \overline{d}  ] \  \  
\wedge    [\,  c  \   a   \,] \,) \longleftarrow 
\bigwedge_2 \, [ m \ \bigvee_2 \,(n \quad \wedge [not \, g_1 \ \ g_2 ]) ].$$
 
 Rule $ \mbox{\em N}_1(r) $ obtained by applying distributivity laws is the following:
$$\bigwedge_4 [\, (\overline{b} \vee c) \ ( \overline{b} \vee a) \ ( \overline{d} \vee c) \ ( \overline{d} \vee a) \, ] 
\longleftarrow  \bigvee_2 (\wedge [m \ n ] \ \ \wedge [ m \ \, not \, g_1 \,  \ g_2] ).$$

\vspace{-.1cm}
Observe that the head expression is a positive-Horn CNF, and so in concordance  with Proposition \ref{pro:HORN-Exp-HORN-CNF}. Then,
one can obtain that  
the program $\mbox{\em NN}_1(P)=FN(N_1(r)) $ is  NP$_\mathcal{A}$ and 
concretely it is formed by the following eight rules:
\begin{gather*}
  a \leftarrow  \wedge [ d \ m \  not \, g_1   \ g_2] \qquad \qquad  a \leftarrow  \wedge [ d \ m \  n]. \\
     c \leftarrow  \wedge [ d \ m \  not \, g_1   \ g_2] \qquad \qquad c \leftarrow  \wedge [ d \ m \ n]. \\
   a \leftarrow  \wedge [ b \ m \  not \, g_1   \ g_2] \qquad \qquad a \leftarrow  \wedge [ b \ m \  n]. \\
   c \leftarrow  \wedge [ b \ m \  not \, g_1   \ g_2] \qquad \qquad  c \leftarrow  \wedge [ b \ m \  n]. 
\end{gather*} 

\vspace{-.7cm}
\qed
\end{example}

\vspace{.1cm}
\noindent {\bf  Example \ref{ex:INTRO-compact}}. \underline{(Continued)}. 
Let  $r$ be its nested rule.  By applying distributivity to its   head $H$,
 one obtains $cnf(H)$  below 
 which is a positive-Horn CNF:
  $$cnf(H)=  \bigwedge_4 \, [ \ \vee \ ( \overline{b}   \ \ \neg c  )  \quad 
  b  \quad  ( g  \ \ \ \overline{d}) \quad  
  (g \ \ \overline{\mbox{\em not} \, f} \,  )  \, ] $$
  
    \vspace{-.1cm} 
  With no difficulty, one can obtain $N_1(r)$  and check that it is an
  NFNP$_\mathcal{L}$ rule. Moreover, 
   we can apply Definition \ref{def:LP-of-NLP} and check that 
   $NN_1(P)= \mbox{\em FN}(\mbox{\em N}_1(r))$ is exactly the  
   NP$_\mathcal{L}$ program given in Example  \ref{ex:INTRO-compact}
   in the Introduction section. \qed

\begin{tcolorbox}
We  give next a {\bf second way}  to obtain  an  NP program  strongly equivalent to an NNP one,
which is based on the expression $H_\Delta$ associated to $H$. This second way is the one we will retain
in the rest of the paper.
\end{tcolorbox}

\begin{definition} \label{pro:ALTERNTIVE-FP-PROGRAM}  
Let  $ H \leftarrow B$ be an NNP rule  and denote by $ {\Delta}_B$
 the expression obtained  by applying Definition \ref{def:DEFINITION-of-OVERLINE-F} 
to  $\Delta$ within $H_\Delta$. Thus we define:
$$ N( H \leftarrow B)=
\{h \leftarrow C \ \vert \vert \ \  h \vee \Delta  \in H_\Delta, \ \  
C \in dnf(B \wedge {\Delta}_B)  \,   \}. \quad NN(P)=\bigcup_{r \in P} N(r).$$
\end{definition}

\begin{proposition}  \label{prop:NN-1=NN}  Let  $P$ be an NNP program. We have $NN_1(P) \Leftrightarrow NN(P) $.
\end{proposition}

\begin{niceproof} See Appendix.
\end{niceproof}

\noindent {\bf Example \ref{def:NNP-to-NP-1}} (Continued) In order  to obtain
$H_\Delta$ associated to   the head $H$  of the rule, we obtain:
$\Delta=\wedge \, [  \,   \overline{b}  \   \ \   \overline{d}  \,] $ for $c,a$, and then we have:
$$H_\Delta=\bigwedge_2 [\, \bigvee_2 (c \ \ \wedge  [  \,   \overline{b}  \   \ \   \overline{d}  \,]) 
\ \ \ \bigvee_2 (a \ \ \wedge  [   \,   \overline{b}  \   \ \   \overline{d}  \,]) \ ]$$ 

 By applying Definition \ref{def:DEFINITION-of-OVERLINE-F}, we 
obtain ${\Delta}_B= \vee (b \ \ d)$ and Proposition \ref{pro:ALTERNTIVE-FP-PROGRAM} leads  to:
 $$ B \wedge {\Delta}_B= 
\bigwedge_2 [\ \ \bigvee_2 (\wedge [m \ n ] \ \ 
\wedge [ m \ \, not \, g_1 \,  \ g_2] ) \ \ \ \vee (b \ \ d) \ ].$$

Now, we have that  $dnf(B \wedge {\Delta}_B)$ has exactly four conjuncts.
Concretely,   obtaining  $N(H\leftarrow B)$ in Definition
\ref{pro:ALTERNTIVE-FP-PROGRAM} leads to the eight NP rules 
 in Example  \ref{def:NNP-to-NP-1} above. So 
Proposition \ref{prop:NN-1=NN}  is checked for this particular example. \qed

\vspace{.3cm}
\noindent {\bf   Example \ref{ex:INTRO-compact}}. (\underline{Continued}). 
In this particular case,  expression $H_\Delta$  is exactly $H$,
and  on the other hand, we have 
$B=\vee ( a \ \, \mbox{\em not} \,  e  \ \,  m )$. Hence, we can easily obtain:
$${\Delta(\neg c)}= {\overline{b}} \qquad \qquad \Delta_B(\neg c)=b $$  
$${\Delta(b)}=\overline{\bot} \qquad \qquad {\Delta(b)}_B=  \top$$  
$${\Delta(g)}= 
{\wedge [ \overline{d} \ \ \overline{\mbox{\em not} \, f} \, ]} \qquad \qquad
{\Delta(g)}_B=\vee (d \ \   \mbox{\em not} \, f) \qquad \qquad $$
$$B\wedge {\Delta(\neg c)}_B=B \wedge b=\vee_3 (a \wedge b \quad \mbox{\em not} \,  e \wedge b \quad m \wedge b) $$
$$B\wedge {\Delta(b)}_B=B \wedge  \top=B $$
$$B\wedge {\Delta(g)}_B=
B \wedge   (d \ \vee  \ \mbox{\em not} \, f)  =
\bigwedge_2 [ \   (a  \quad \mbox{\em not} \,  e  \quad m ) ]
 \quad  \vee (d \ \ \mbox{\em not} \, f) \ ]$$
 
 Now, determining the rules indicated by $N(H \leftarrow B)$, 
 in Definition  \ref{pro:ALTERNTIVE-FP-PROGRAM}, with the values above leads to the NP rules
 initially given in Example \ref{ex:INTRO-compact} in the introduction. \qed

\begin{example} Assume the  NNP$_\mathcal{A}$ program $P=\{r\}$,  
where $r=H \leftarrow \wedge [\, e \ f ]$ and
$H$ is  the positive-Horn expression in 
Example \ref{ex:FORMULA-TILDE-H}.    
%
Then   obtaining the right side of 
$N(H \leftarrow B)$ in Definition  \ref{pro:ALTERNTIVE-FP-PROGRAM} leads to:
$$ b \leftarrow \wedge \,[ e \ f \ a ] \qquad \qquad \ c \leftarrow \wedge \,[ e \ f \ a \ \ b]$$  
$$ \bot \leftarrow \wedge \, [e \ f \ \mbox{\em not} \, d  \ \, a ] \qquad  c \leftarrow \wedge \, [e \ f \ a \ \ g].$$

One can easily verify that this NP$_\mathcal{A}$ program
is exactly the same as the one   obtained by applying Definition \ref{def:LP-of-NLP} of $\mbox{\em NN}_1(P)$.
\end{example}

\section{Semantics of NNP} \label{sec:DefiniteNLP-Semantics}

In this section we recover the semantical notions tackled in prior sections 
such as model, answer set,   closed and supported model, etc.  
in order to specify  them
for NNPs.

\begin{definition} \label{cor:NNP-CLOSED-SUPPORTED} 
 Let  $P$   be an NNP  program and
     $ I   $  be   an interpretation.
 $I $ is a {model}  of $ P $, 
 denoted by $ I \models P $, if   
 for all $  (H \leftarrow B) \in P$,  the conditions below hold. 
 A minimal model and the least model $LM(P)$ of $P$ are defined as in Definition \ref{def:Model-BASIC}.
$$ I  \not \models    B  \quad 
\mbox{\underline {or}} \quad \forall \,  {(h \vee  \Delta)} \in H_\Delta:
\quad  I \not \fmodels  \,  \Delta   \quad \mbox{\bf \underline {or}} \quad 
 h \in I.$$ 
\end{definition}

\noindent {\bf Example
\ref{ex:first-NNP-Rule-SYNTAX}.} (\underline{Continued}) Let us consider  rule $r$.
\begin{itemize}

\item $ I=\{f,b,c\} \not \models r$ because   
 $ I  \models B  $, \ \ $ I  \fmodels  \, \Delta(a)$  \ and \ $   a \notin I $.

\item $ I =\{f,d,c,a\}  \models r$ 
because   $  a,c \in I, \  \ \mbox{and}  
\  \ I \not \fmodels \, \Delta(\bot)$ (as $I \not \fmodels  \, g$). \qed

\end{itemize}

\vspace{.0cm}
\begin{example} \label{ex:INTERPRE-DN-asp-2} Let us consider  
 the {\em not}-free NNP$_\mathcal{A}$ rule $r= (H \leftarrow \top)$, where $ H  $ is: 
$$ H= \bigvee_2 (\,\wedge [\,\overline{a} \  \ \overline{e} 
 \ \ \overline{ m}] \quad \quad
\bigwedge_3  [\,\vee (\overline{b}   \ \ c  )  \quad 
 b   \quad   \bigvee_2 (g   \ \ \wedge [\overline{d} \ \ \overline{f} ] \,) \, ] \ \ )  $$

 The positive elementary expressions in $ H $ are $  \{c,b,g\} $ and we have the following $\Delta$'s:  
\begin{itemize}

\item $   \Delta(c) =\bigvee_2 (\wedge [\,\overline{a} \  \ \overline{e} 
 \ \ \overline{ m}] \quad \overline{b})$ 

\item  $\Delta(b) =  \wedge [\,\overline{a} \  \ \overline{e} 
 \ \ \overline{ m}].$

\item $\Delta(g) =\bigvee_2 
(\wedge  [\,\overline{a} \  \ \overline{e} 
 \ \ \overline{ m}] \quad \wedge [\overline{d} \ \ \overline{f} ] \ ).$
\end{itemize}

\noindent Below, we take three interpretations and check whether they satisfy $ r $: 
\begin{itemize}

\item   $ I=\{a, b,c\}   \models r$ \ \ because   \ \ $ {c}, b \in I$, 
     \ and \ $ I  \not \fmodels   \, \Delta(g) $.
   

\item $ I=\{a,b,c,d \} \not \models r $  \ \ because \ \ $I \models B$, 
\ \ $ I   \fmodels    \, \Delta(g)   $ \ and \
   $   {g} \notin I$.

\item $ I=\{a,b, g,c,d \} \models r $ \ \ because \ \ $\{{c}, {b}, {g} \} \subset I.$ 
   \qed

\end{itemize}
\end{example}

\begin{example} \label{ex:INTERPRE-DN-asp-3}  We take the  NNP$_\mathcal{A} $ rule $ H \leftarrow \top $, 
where $ H $ is specified below.  
$\Psi$ is a sub-expression  inside $H$ and so the set of positive occurrences in $H$ is $\{m,c,\bot,a\}$:
 $$H=\bigwedge_3  [\,\vee (\overline{b} \ \ m) \quad \Psi \quad \bigvee_2 (\overline{d} \ \     
\wedge [c \  a \,]  \, ) \ ] $$ 
$$\Psi=\bigvee_2 (\, \overline{f}  \quad
\bigwedge_2 [\,\bigvee_2 (\overline{m} \quad \bigwedge_2 [c  \ \ \vee (\overline{g} \ \ \bot)] \,) 
  \quad  \ \vee (m \ \overline{e}) \, ] \ )$$ 

To differentiate between 
occurrences of a given  $h \in \mathcal{L} \cup \{\bot\}$, 
we  use a sub-index so that $ h^i $, $ h^j $ and $ i \neq j$ stand for distinct occurrences 
of  $h$. With this notation, the  positive occurrences  in $H$ are $\{m^1,c^1,\bot,m^2,c^2,a\}$ and by  reflecting this 
onto the original $H$:
$$H=\bigwedge_3  [\,\vee (\overline{b} \ \ m^1) 
\quad \Psi \quad \bigvee_2 (\overline{d} \ \     \wedge [c^2 \  a \,]  \, ) \ ] $$ 
$$\Psi=\bigvee_2 (\, \overline{f}  \quad
\bigwedge_2 [\,\bigvee_2 (\overline{m} \quad \bigwedge_2 [c^1  \ \ \vee (\overline{g} \ \ \bot)] \,) 
  \quad  \ \vee (m^2 \ \overline{e}) \, ] \ )$$ 
\noindent  
  
The  negative sub-formulas $ \Delta $'s for the positive occurrences are:
\begin{itemize}

\item $\Delta(m^1)   =  \overline{b} $  \hspace{4.4cm} 
$\bullet \ \Delta(m^2) =  \vee (\overline{f} \ \ \overline{m} \ \ \overline{e}) $   

\item   $\Delta(c^1) =\vee (\overline{f} \ \ \overline{m}) $   
    \hspace{3.4cm}
$\bullet \ \Delta(c^2)   =   \overline{d}    $
   
\item $  \Delta(\bot)   =\vee (\overline{f} \ \ \overline{m} \ \ \overline{g}  ) $  
\hspace{3.cm}
$\bullet \ \Delta(a) =  \overline{d}    $

\end{itemize}



\noindent We select some interpretations and check whether  
they satisfy $ r $ (omitting that $ I \models B$):
\begin{itemize}
\item $ I=\{f\}   \models r $ as  
$ I \not  \fmodels  \Delta({m_1}), \,  I \not  \fmodels  \Delta({c_1}),  \,  I \not  \fmodels   \Delta(\bot),
\,  I \not  \fmodels   \Delta({m_2}), \,   I \not  \fmodels  \Delta({c_2}), \,  I \not  \fmodels \Delta({a}).  $

\item $ I=\{f,m,d,a\} \not  \models r $ since \ \ $ I  \fmodels  \Delta({c_1})$  \ and \ 
  $ c \notin  I $  \   
   \  \   (or also $ I   \fmodels  \Delta({c_2})$   and $ c \notin  I $).

\item  $ I=\{f ,m,g,c\} \not \models r$ since \ \  
$ I  \fmodels \Delta(\bot)$.

\item  $ I=\{f ,m,b,c\}  \models r$  since \ \  $ I  \not \fmodels   \Delta(\bot),  \  I  \not \fmodels  \Delta({a})$   \ and \ ${m}, c \in I  $.

\end{itemize}
\end{example}

\begin{definition} \label{cor:NNP-CLOSED}  An interpretation  
$  I  $  is  {\bf closed} under  an NNP  program 
       $ P $ if: 
     $$ \forall H\leftarrow B \in P, \quad     \forall  \, {(h \vee \Delta)} \in H_\Delta: \quad  h \in I \quad 
\mbox{whenever} \quad   
I \models B 
\quad \mbox{{and}} 
\quad I  \fmodels \, \Delta.  $$
\end{definition}

As with classical NP programs, we have the  next property:    

\begin{proposition} \label{Prop:MODEl_NNP-CLOSED} Any model of an NNP program $P$ is closed under $P$. 
\end{proposition}

\begin{niceproof} See Appendix.
\end{niceproof}

This proposition can be checked
 by using any of  the above examples.

\begin{definition} \label{cor:NNP-SUPPORTED} 
An interpretation  $I  $ 
    is {\bf supported} by    an NNP  program $ P $ if $\forall \ell \in I$: 
$$    
\quad \exists (H \leftarrow B) \in P, \quad \exists 
\,{ (h \vee \Delta)} \in H_\Delta: \qquad \mbox{} h =\ell, \quad 
 I \models  B 
  \quad \mbox{{and}} \quad
I  \fmodels \,\Delta.$$ 
\end{definition}

\begin{proposition} \label{Prop:MINIMAL_MODEL_SUPPORTED_NNP} A  minimal model of an NNP program $P$ is supported by $P$.
\end{proposition}

\begin{niceproof} See Appendix.
\end{niceproof}

As a consequence of the previous claim,  $LM(P)$ is the unique supported model 
 of a {\em not}-free NNP program $P$ 
 (Proposition \ref{prop:BASIC-Propertiesprograms-not-free-FXNP},
 claim (c)). One can verify both statements 
by using the previous examples. 
We now add an example containing  defaults:

\begin{example} \label{ex:NNP-SUPPORTED}  Let us consider the  NNP$_\mathcal{A}$ program $ P $ with the   rules below.
One can check that  the least model of $P$ is $\{m,g_2,c,a\}$ which is supported by $P$.
$$ r=\bigvee_2     \,  (\wedge \, [  \,   \overline{m}  \   \  \overline{d} ] \  \  
\wedge    [\,  c  \  \ a   \,] \,) \longleftarrow 
\bigwedge_2 \, [ m \ \ \bigvee_2 \,(n \quad \wedge [not \, g_1 \ \ \ g_2 ]) ].$$
$$  m \wedge g_2\longleftarrow \top. $$

\vspace{-.75cm}
\qed
\end{example}

\begin{definition}  \label{def:OPERATOR:NT} Let    $ P $ be a {\em not}-free, constraint-free
 and head-consistent NNP   program 
 and    $ I  $ be   an interpretation. 
 The  immediate consequence operator $ NT_P (I) $  for 
 NNP programs is defined below. 
    $ NT^i_P $ and  $ \mbox{\it lfp}(NT_P) $  are defined  as usual.
$$NT_P (I)=\{h \ \vert \vert \ H \leftarrow B \in P, 
\ \,  (h  \vee   \Delta) \in H_\Delta,  \ \,  
 I   \models    B,   \ \,  I   \fmodels \,\Delta  \}.$$ 
 
 \vspace{-.7cm}
 \qed
\end{definition}

\noindent {\bf Remark.} Just as $ T_P $ for classical NP programs,   
$ NT_P $   is monotonic, 
i.e. $I \subseteq J$ implies 
$NT_P(I ) \subseteq NT_P(J )$. Since 
$NT_P(I ) \subseteq B_P$ for any 
$I \subseteq B_P$,
there is a fixpoint $NT_P(I' ) = I'$ 
for some $ I'  \subseteq B_P$. 
Further,   in an  NNP program $P$,
 $NT^ k_P $ becomes
a fixpoint for some $  k$ and it coincides with   $LM(  P)$.  

\begin{example} Let us consider  program $P$ in Example \ref{ex:NNP-SUPPORTED}, where 
we denote by $B$ the body of the first rule.
It is trivial  that $NT^0_P=NT_P(\emptyset)=\{m,g_2\}$.
Besides, we easily could obtain $\Delta(c)=\Delta(a)
=\wedge \, [  \,   \overline{m}  \     \overline{d}  ]$. 
Then  we have:
$$  \{m,g_2\} \models B, \quad 
\{m,g_2\}  \fmodels \, \Delta(c) = \Delta(a).$$ 
$$NT^1_P=NT_P(\{m,g_2\})=\{m,g_2,c,a\}.$$

\noindent  So we have the following powers $NT^k_P$:
\begin{itemize}
\item $NT^0_P=NT_P(\emptyset)=\{m,g_2\}$.
\item $NT^1_P=NT_P(\{m,g_2\})=\{m,g_2,c,a\}$.
\item $NT^2_P=NT_P(\{m,g_2,c,a\})=\{m,g_2,c,a\}$.
\item $NT^k_P=NT^2_P$ \quad for $ k > 2 $. \qed
\end{itemize}
\end{example}

\begin{example} Consider  program  $ P $ formed by the single rule in Example \ref{ex:INTERPRE-DN-asp-3}. We give the values of the operator $ NT_P $ for two interpretations:
\begin{itemize}

\item $ NT_P(\{b,f,m\})=\{m,c\} $.

\item $ NT_P(\{b,f,m,d\})=\{m,c,a\} $ \qed
\end{itemize}
 
\end{example}

\begin{definition}  \label{def:ANSWER_SET_NNP}  Let    $ P $ be an 
NNP   program 
 and    
 $ I  $ be   an interpretation, and let $ H \leftarrow B \in P $.  
 The    $I$-reduct of $P$, denoted by $ P^I $, is
defined below (with a slight abuse of notation). An answer set and consistency of $P$  
  are
 defined as in Definition \ref{def:ANSWER-I}.
 $$H^I_\Delta=   \bigwedge \,[\,(h \vee \Delta^I) \parallel   (h \vee \Delta) \in H_\Delta, \ \Delta^I \not \equiv \top  \, ].$$
$$P^I=\{H_\Delta^I \leftarrow B^I
\ \ \vert \vert \ \ H \leftarrow B \in P, 
\  \ B^I \not \equiv \bot, \ \ H_\Delta^I \neq \wedge []  \}$$

\end{definition}

\begin{example} Let us consider the program $P=\{r\}$, where  $r=H'' \leftarrow \top$ and   $H''$ 
being from Example \ref{ex:MANY-NOTs}.
     We determine the sub-formulas $\Delta^I$ for $I=\emptyset$:
$$(\Delta(\neg y^2))^\emptyset=(\ \overline{\mbox{\em not} \, x} \ )^\emptyset=\overline{\, (\mbox{\em not} \, x)^\emptyset 
\, }=\bot$$ 
$$(\Delta(\neg y^1))^\emptyset=( \, \vee (\overline{\mbox{\em not} \, x} \ \ \overline{\mbox{\em not} \, \neg y}) \ )^\emptyset=
\vee (\ \ \overline{ (\mbox{\em not} \, x)^\emptyset  } \ \ \  \ \overline{ (\mbox{\em not} \, \neg y})^\emptyset \ ) =\bot$$ 
$$(\Delta(\neg x))^\emptyset=( \ \vee (\overline{\mbox{\em not} \, x} \ \ \overline{v}) \ )^\emptyset=
\vee   (\, \overline{(\mbox{\em not} \, x)^\emptyset \ }  \ \ \overline{v}^\emptyset \ )= \overline{v}.$$

Hence $H_\Delta''^{\emptyset}=\bigwedge_2 [ \neg y \ \ \vee (\neg x \ \ \overline{v})]$ 
which entails  $P^{\emptyset}=H_\Delta''^{\emptyset} \leftarrow \top$.
 \qed
\end{example}


We now  show that claims
from Theorem \ref{TH:Tarski55-Endem76}
to Proposition \ref{prop:BASIC-Propoerties-ANSWER-SET} 
also hold for NNPs.  Previously,
we state in Propositions \ref{Prop:EQUIV-NNP-ND-1} 
and \ref{Prop:EQUIV-NNP-ND-2}  the relationships  
 connecting semantically    NNP 
and  NP programs. The proofs of the claims below
are given in the Appendix.


\begin{proposition} \label{Prop:EQUIV-NNP-ND-1} 
Let  $P$ be an NNP  program and $I$ be an interpretation. We have:
  \begin{itemize}
 
 \item [{\em (1)}] 
  $P$ is head-consistent iff 
 so is  $NN(P) $.
 
 \vspace{-.2cm}
 \item [{\em (2)}]
 $I$ is closed under $P$ iff 
$I$ is closed  under  $NN(P) $.

 \vspace{-.2cm}
 \item [{\em (3)}] $I$ is supported by $P$ iff 
  $I$ is supported by  $NN(P) $.
 
 \end{itemize} 
\end{proposition}

\begin{proposition} \label{Prop:EQUIV-NNP-ND-2}  Let   $P$ be any   not-free,
 constraint-free and   head-consistent NNP program  and let us denote by $Cn(P)$
     the smallest interpretation $I$  closed under $P$. We have:
 \begin{itemize}
 
 \item [{\em (4)}] $ Cn(P)=Cn(NN(P)) $.
  
 \vspace{-.2cm}
 \item [{\em (5)}] $NT_P(I) = T_{NN(P)}(I)$.
 
 \vspace{-.2cm}
 \item [{\em (6)}] $P \Leftrightarrow  NN(P)$.
 
 \end{itemize} 
\end{proposition}


\begin{theorem} \label{TH:Tarski55-Endem76-NFNP}
 For any     not-free, constraint-free and  
  head-consistent  NNP  program $ P $: 
  $$LM(P)=\bigcup_{0 \leq i} \,NT^i_P=\mbox{lfp}(NT_P) .$$
\end{theorem}

\vspace{-.5cm}
\begin{proposition} \label{prop:BASIC-Properties-not-free-NFNP}  
If $  P$ is a {not-}free and constraint-free  NNP$_\mathcal{A} $  program, then $P$, $LM(P)$ and
  $ Cn(P) $ verify the properties (a) to (c)
   in Proposition \ref{prop:BASIC-Properties-not-free-NP}.
\end{proposition}

 \begin{proposition} \label{prop:BASIC-Propertiesprograms-not-free-FXNP}  
	If $  P$ is a {not-}free, constraint-free and head-consistent 
NNP$_\mathcal{L} $ program, then $P$, $LM(P)$ and $Cn(P) $ verify
 properties (b) and (c) specified in Proposition \ref{prop:BASIC-Properties-not-free-NP}.
\end{proposition} 


\begin{proposition} \label{prop:BASIC-Propoerties-ANSWER-SET-NFNP} 
Let   P be an  NNP program, S  be  an answer set of P
and  $\Sigma$ be  a set of constraints. Then P, S and $\Sigma$
verify the properties expressed in Proposition \ref{prop:BASIC-Propoerties-ANSWER-SET}.


\end{proposition}

 \section{ Nested Unit-Resolution} \label{sec:PUR-HPUR}

\subsection{Introduction}

The logical calculi   {\em Nested  Unit-Resolution,} 
   $\mathcal{UR}$ for short, and  {\em Nested  Hyper Unit-Resolution,} 
   $\mathcal{HUR}$ for short, are  early defined 
  for possibilistic and regular many-valued logics 
 in  \cite{Imaz2023a,Imaz2023b}.  
 The result  of first extracting from the latter logical calculi the  one for propositional logic  
  and then of suitably adapting it  to the NASP framework  is described below.

  \vspace{.1cm}
We will prove  that $\mathcal{UR}$ is   complete
  for {\em not}-free NNP programs, just as   unit-resolution is  for {\em not}-free NPs.
   Without loss of generality,  
         programs are considered non-empty.

   \begin{tcolorbox}
  Within the execution of the 
  methodology  known 
  as "Guess and Check" or Generate-and-Test \cite{Lifschitz02} 
  (see also \cite{LeonePFEGPS06,EiterIK09}), 
  we arrive at the moment  the ASP-solver has chosen interpretation $I$ as 
  the candidate for an answer set of  program $P$ at hand, and  then,
   resorts to $\mathcal{UR}$ to compute  $LM(P^I)$.
  This way, {\em in this section, 
  we   handle programs without  defaults, 
  i.e. with only   positive $\mathcal{L} \cup \{\bot\} $    
    and   negative 
    $\{\overline{x} \, \vert \vert \, x \in \mathcal{E}^1_\mathcal{L}   \}$} elements.
     \end{tcolorbox}
     


Altogether, we work only with {\em not}-free or purely {\bf propositional programs} 
$P$ and we do so  in order to determine $LM(P)$. Consequently:

\vspace{.15cm}
$\bullet$   $ H \leftarrow B $  is  better represented
by  $ H \, \vee \, \mbox{NNF}(\neg{B}) $, 
where $\mbox{NNF}(\neg{B})$ stands for 
  $\neg{B}$ 
  
  \ \ \ in NNF and is determined
by solely  using  De Morgan's laws on ${B} $.

\vspace{.1cm}
$\bullet$ We substitute underlined literals  
$\overline{\ell}$ by negated ones $\neg \ell$ since, by Definitions 
\ref{def:interpretation-Model-BASIC} and 

\ \ \ \ref{def:LITERAL-HEAD-SEMAN},  any interpretation $I$ verifies:
$I \models \ell$ iff $I \fmodels \,\overline{\ell}$, and hence $  \neg \ell \equiv \overline{\ell}$.

\begin{example} For $  \{e,f\} \subset I$, any $I$-reduct of the expression $ H \vee \mbox{NNF}(\neg{B}) $ corresponding to
   the nested rule 
in Example \ref{ex:INTRO-compact} in the introduction section is:
$$ \bigvee_2 ( \ \bigwedge_3 \ [ \ \vee (\neg {b}   \ \ \neg c )  \quad \
  b \ \quad  \bigvee_2 \, ( g   \ \   \neg {d} \, )  \  ]  
      \qquad   \wedge [ \neg{a} \ \,   \neg{m} ] \ ).$$ 
 
 \vspace{-.7cm}     
 \qed
\end{example}

As we are in propositional logic strong-equivalence $\Leftrightarrow$ is not required.  We  recall that   $\mathcal{UR}$ is applied  towards knowing whether  $I=LM(P^I)$ for a given  $I$. 
 $\mathcal{UR}$ encompasses
the main inference rule,  called NUR (nested UR), 
plus  a group of simplification inference rules. 
 The latter  are simple,  however,  NUR is quite elaborate and so is presented    progressively. 
 
 \vspace{.1cm}
 After this introduction, Subsection \ref{subsec:ALMOST-CLAUSAL} presents NUR  
 for simple   almost-clausal  expressions and
 Subsection \ref{subsec:RULE-NUR}   for general  expressions.
 Subsection \ref{subsec:Running-Example} illustrates the steps carried out by  $\mathcal{UR}$ 
  running on an example. Subsection \ref{subsec:HYPER-UNIT-RESOL}
 is devoted to $\mathcal{HUR}$.

\subsection{Almost-Clausal Expressions} \label{subsec:ALMOST-CLAUSAL}

The syntactical  pattern of what we consider almost-clausal  expressions is shown below,  wherein
 $ {\textcolor{black} {\ell}} \in \mathcal{L}$    
  and the $ \varphi $'s and $ \phi $'s are  expressions:  
$$\Phi={\textcolor{black} {\wedge}}  \,[\, \varphi_1  \, \ldots  \, \varphi_{l-1} 
 \ {\textcolor{black} \ell} \  \varphi_{l+1}  \, \ldots  \, \varphi_{i-1} \
{\textcolor{black} {\vee}} (\phi_1 \, \ldots \, \phi_{j-1} \ 
{\textcolor{black} {\neg {\, \ell \, }}} \ \phi_{j+1} \, \ldots \, \phi_k) \ \varphi_{i+1} \, \ldots \, \varphi_n \, ]$$

 In fact,  we say that $ \Phi $ is almost-clausal because   
if the $\varphi$'s and   $\phi$'s were clauses and  literals, 
respectively, then  $ \Phi $ would  be  clausal.  
  Thus we  obtain that 
$ \Phi $ is equivalent to $ \Phi' $:  
$$\Phi \equiv \Phi'= {\textcolor{black} {\wedge}}  [\, \varphi_1  \, \ldots  \, \varphi_{l-1} 
 \ {\textcolor{black} \ell} \  \varphi_{l+1}  \, \ldots  \, \varphi_{i-1} \ 
{\textcolor{black} {\vee}} (\phi_1 \, \ldots \, \phi_j  
\, \phi_{j+1} \, \ldots \, \phi_k) \ \varphi_{i+1} \, \ldots \, \varphi_n \}$$
 and then,   the     simple   inference rule below, on the left,  can be derived.
This inference is
more concisely rewritten   on the right, below, wherein $\Sigma=
\vee (\phi_1 \, \ldots \, \phi_j \, \phi_{j+1}  \, \ldots \, \phi_n)$:   
$$\frac{{\textcolor{black} \ell}     \ {\textcolor{black} {\wedge}} \ \  
{\textcolor{black} {\vee}} (\phi_1 \ \ldots \phi_j \ 
{\textcolor{black} {\neg {\, \ell \ }}} \ \phi_{j+1} \ \ldots \ \phi_k) }
{{\textcolor{black} \ell}     \ {\textcolor{black} {\wedge}} \ \
{\textcolor{black} {\vee}} (\phi_1 \ \ldots \phi_j  
 \ \phi_{j+1} \ \ldots \ \phi_k) } {{\mbox{\,NUR}}} \qquad \quad 
 \frac{ {\textcolor{black} \ell} \ \, {\textcolor{black} \wedge} \ \, 
 ({\textcolor{black} {\neg {\, \ell \, }}} \ \, {\textcolor{black} {\vee}}  \ \, 
 \Sigma \,) }
{{\textcolor{black} \ell}     \ {\textcolor{black} {\wedge}} \  \Sigma  }{{\mbox{\,NUR}}}$$

 {\bf Remark.} {\em  
The  above rule, particularized to   CNFs,  recovers  
clausal unit-resolution. }

\vspace{.1cm}
   We now expand our analysis to more complex expressions by assuming that 
   ${\textcolor{black} {\neg{\, \ell \ }}}$ is not isolated anymore
   and that it is conjunctively linked to a sub-expression denoted by 
   $\Delta({\textcolor{black} {{\neg {\, \ell \ }}}})$.    
   More specifically,   the  pattern of the  considered almost-clausal expressions
 is:  
 $$\Phi={\textcolor{black} {\bigwedge}} [\, \Pi \ \ {\textcolor{black} \ell} \ \ 
 (\Delta({\textcolor{black} {{\neg {\, \ell \ }}}}) \ \ {\textcolor{black} {\vee}} \ \ 
 \Sigma  ) \ \ \ \Pi' ],$$ 
 
 where $ \Pi =\varphi_1 \ldots \varphi_{i-1}  $ 
 and $ \Pi'=\varphi_{i+1} \ldots \varphi_k $
  are concatenations of
expressions and $\Delta({\textcolor{black} {{\neg {\, \ell \ }}}})$ is defined as
the maximal sub-expression of the input, conjunctively   linked
to ${\neg {\, \ell \ }}$, i.e. 
$\Delta({\textcolor{black} {{\neg {\, \ell \ }}}})$ {\bf becomes false when
${\neg {\, \ell \ }}$  is false.} More formally, 
$\Delta({\textcolor{black} {{\neg {\, \ell \ }}}})$ verifies:
\begin{itemize}

\item [(a)]  $ \Delta({\textcolor{black} {{\neg {\, \ell \ }}}}) 
\equiv {\neg {\, \ell}} \wedge \Psi$,

\item  [(b)]    $\nexists \, \Delta'$: (i) $\Delta'$ includes $\Delta$; and (ii) $\Delta'$ verifies (a).
\end{itemize}

\begin{example}
    Assume  we have the following expression:
     $$\varphi  =  \bigvee_3 (\varphi_1 \quad  \bigwedge_2 [\, \phi_ 1 \ \bigwedge_3 
   [{\textcolor{black} {{\neg {\, \ell \ }}}} \ 
 \vee (\phi_2 \ \neg {a}) \ \phi_3] \, ] \quad \varphi_2\,).$$ 
 
 Then we have:
 $\Delta({\textcolor{black} {\neg {\, \ell}}})=
 \bigwedge_2 [\,\phi_ 1 \ \ \bigwedge_3 [{\textcolor{black} {\neg {\, \ell \ }}} \  \  \vee (\phi_2 \ \neg {a}) \ \phi_3] \, ]$ 
 because: 
 
 \vspace{.3cm}
 (a)   
 $\Delta({\textcolor{black} {\neg {\, \ell \ }}}) \equiv 
  {\textcolor{black} {\neg {\, \ell \ }}} 
  \wedge  [\bigwedge_2 [\phi_ 1 \ \  \bigwedge_2 [   \vee (\phi_2 \ \neg {a}) \ \ \phi_3] \, ] \, ] = 
  {\textcolor{black} {\neg {\, \ell \ }}} \wedge \Psi$; 
 and 
 
 \vspace{.2cm}
 (b)    $\nexists \, \Delta' $ including $\Delta $ and
 fulfilling $\Delta'({\textcolor{black} {\neg {\, \ell \ }}}) \equiv {\textcolor{black} {\neg {\, \ell \ }}} \wedge \Phi' $.
\end{example}

 Hence,  the augmented inference rule UR    is:    
  $$\frac{ {\textcolor{black} \ell}  \, {\textcolor{black} \wedge} \  
  (\Delta({\textcolor{black} {\neg {\, \ell \ }}}) \ \
  {\textcolor{black} {\vee}} \ \ 
   \Sigma  \,)} 
  {{\textcolor{black} \ell}  \, {\textcolor{black} \wedge} \ \Sigma}
  {{\mbox{\,NUR}}}$$ 
%


\subsection{ General Expressions}  \label{subsec:RULE-NUR}

It is not hard to check that, by  generalizing the previous almost-clausal expressions to general unrestricted expressions  to which
     NUR  should be indeed    applied, one reaches the next syntactical 
 pattern (see Definition \ref{def:NC-formulas} for notation
  $   \odot \langle\varphi_1 \ldots \varphi_k \rangle $):
 $$\bigwedge_4 [\Pi_0 \ \  {\textcolor{black} \ell} \ \ 
 \bigvee_3 (\Pi_1 \  \ldots  \   \bigwedge_3 [ \Pi_{k} \ \,  
  (\Sigma  \ \, {\textcolor{black} {\vee}} \ \, 
 \Delta({\textcolor{black} {\neg {\, \ell \ }}}) \,) \ \,  \Pi'_k \,] \ \ 
    \ldots  \ \ \Pi'_1 \,) \ \ \Pi_0' \, ]$$

 where  as before 
the $\Pi_j$'s and  $\Pi_j'$'s are concatenations of expressions. 
Thus    NUR is:
$$\frac{ {\textcolor{black} \ell}  
\ {\textcolor{black} {\wedge}} \  \bigvee_3 (\Pi_1 \  \ldots  \   
\bigwedge_3 [ \Pi_{k} \ \,    (\Sigma  \ \, {\textcolor{black} {\vee}} \ \, 
 \Delta({\textcolor{black} {\neg {\, \ell \ }}}) \,) \ \,  \Pi'_k \,] \ \    \ldots  \ \ \Pi'_1 \,) }
 {{\textcolor{black} \ell}  \ {\textcolor{black} {\wedge}} \  \bigvee_3 (\Pi_1 \  \ldots  \   
 \bigwedge_3 [ \Pi_{k} \ \,   \Sigma  \ \,  \Pi'_k \,] \ \ 
    \ldots  \ \ \Pi'_1 \,) } 
 {\ \mbox{NUR}}$$

  Therefore,  applying NUR  amounts to simply  removing 
expression  $\Delta({\textcolor{black} {\neg {\, \ell}}})$.

\vspace{.1cm}
   By   denoting  the right conjunct of the numerator by $\Pi$  and   
by denoting  by $\Pi \cdot \varphi$ that $\varphi$ 
is a sub-expression inside $\Pi$,  the inference rule   NUR 
can be more schematically written:
$$\tcboxmath{\frac{{\textcolor{black} \ell} \ \ {\textcolor{black} {\wedge}} 
\ \ \Pi \cdot  (\Sigma  \ \, {\textcolor{black} {\vee}} \ \, 
 \Delta({\textcolor{black} {\neg {\, \ell \ }}}))}
{{\textcolor{black} \ell} \  {\textcolor{black} {\wedge}} \ \Pi \cdot  \Sigma     }
{\mbox{\ \ (NUR)}}}$$


$\bullet$ \underline{\bf Simplification Rules.}  The simplification
 rules given below must  accompany  NUR in order to complete the logical calculus 
 $\mathcal{UR} $: 
\begin{itemize}
\item $\vee^\bot$  and $\wedge^\bot$ rules  
  simplify  expressions by (upwards) propagating 
  the truth constant $\bot$   from sub-expressions to expressions.  
  $ \varphi $ stands for any sub-expression  
    and we recall that $ \varphi \cdot \phi $ means that  $ \phi $ is a sub-expression of $ \varphi $.

 \item  $\odot^{k+1} $ and $\odot^{k+n} $ rules  remove  connectives whenever they become redundant.
 The first rule, called $\odot^{k+1} $, 
removes $ \odot \in \{\wedge,\vee\}$ if it is applied to a single expression,
i.e. $   \odot  \, \langle  \phi_1 \rangle $,
and the second rule, called $\odot^{k+n} $, removes a connective  $\odot_2$ that is inside
another one $\odot_1$ when both are equal $ \odot_2=\odot_1 $.
\end{itemize} 

\begin{tcolorbox}
$$\frac{  \varphi \cdot {\textcolor{black} {\vee}} (\phi_1 \ldots   \phi_{i-1} 
\ {\bot} \ \phi_{i+1} \ldots \phi_k  \,)  }
{ \varphi \cdot {\textcolor{black} {\vee}} (\phi_1 \ldots   \phi_{i-1}  \ \  
\,  \phi_{i+1} \ldots \phi_k  \,)  }{\vee^\bot}$$ \qquad 
$$\frac{
\varphi \cdot {\textcolor{black} {\wedge}} [\varphi_1 \ldots   \varphi_{i-1} 
\ {\bot} \ \varphi_{i+1} \ldots \varphi_k  \,]  }
{ \varphi \cdot    {\textcolor{black} {\bot}}   }{\wedge^\bot}$$
$$\frac{ \varphi \cdot  {\textcolor{black} {\odot_1}} 
\langle \varphi_1 \ldots  \varphi_{i-1} 
\   \odot_2 \langle \, \phi_1 \,\rangle \ \varphi_{i+1} \ldots \varphi_k  \,\rangle  }
{\varphi \cdot  {\textcolor{black} {\odot_1}} \langle \varphi_1 \ldots   \varphi_{i-1} 
 \,  \phi_1 \,  \varphi_{i+1} \ldots \varphi_k  \,\rangle }{\ \odot^{k+1}}$$
 
$$\frac{ \varphi \cdot \langle{\textcolor{black} {\odot_1}} 
\  \ \varphi_1 \ldots   \varphi_{i-1} \  
\odot_2 \langle \phi_1 \ldots \phi_n \,\rangle \ 
\varphi_{i+1} \ldots \varphi_k  \,\rangle, \odot_1=\odot_2  }
{\varphi \cdot \langle{\textcolor{black} {\odot_1}} \  \ \varphi_1 \ldots   \varphi_{i-1} 
 \,  \phi_1 \ldots \phi_n \,  \varphi_{i+1} \ldots \varphi_k  \,\rangle }{\ \odot^{k+n}}$$
\end{tcolorbox}

\vspace{.1cm}
 Next, we state the logical and 
computational properties of $\mathcal{UR}$ and
   relegate   their proof to the Appendix.
   
\begin{definition} We  define    $\mathcal{UR}$ as the  calculus
 formed by  the NUR rule  and the above described  simplification rules, i.e.  
     $\mathcal{UR}=\{\mathrm{NUR},\vee^\bot, \wedge^\bot, \odot^{k+1}, \odot^{k+n}\}$.
\end{definition}

The next statement confirms that the proposed logical calculi $\mathcal{UR}$ is indeed 
the legitimate generalization of classical unit-resolution for the nesting framework.

\begin{proposition} \label{pro:UR-CNF-COINCIDES} $\mathcal{UR}$  particularized to CNF coincides with unit-resolution.
\end{proposition}


Next, we denote by $LM(H)$ the least model of a not-free Horn expression $H$.

\begin{lemma} \label{Lem:Correctness-UR} When  $\mathcal{UR}$  is  applied to a not-free Horn expression $H$ 
until   no  inference is applicable or $\bot$ is derived,
    $\mathcal{UR}$ derives   $  LM(H)$   if $H$ is consistent  and  $\bot$ otherwise.
     $  LM(H)$ is the set of   literals  in the last derived expression $H'$ s.t.
  $\ell \in LM(H)$ iff $H'=\ell \wedge H''$.
\end{lemma}

   See the example in Subsection 
\ref{subsec:Running-Example}.

\begin{lemma}  \label{lem:POLYNOmial-UR}  Given a not-free Horn expression $H$, 
the number of  $\mathcal{UR}$  inferences  required 
to determine  either $LM(H)$ or the inconsistency of $H$
is bounded polynomially.
\end{lemma}

\begin{theorem} \label{the:NNP-Programming-COMPLEXITY} We have the following complexities 
related to NNP programs:

\vspace{.25cm}
(1)  Determining  $LM(P)$ of a not-free NNP  $P$
    is a  problem in $\mathcal{P}$. 
    
    \vspace{.1cm}
    (2) {\em not-}free NNP logic programming is $\mathcal{P}$-complete. 
    
    \vspace{.1cm}
   (3) Determining whether an NNP  $P$ has an answer set is a $\mathcal{NP}$-complete problem.
\end{theorem}

\subsection{Running Example} \label{subsec:Running-Example}

\begin{example} \label{ex:RUNNING-UR} 
Assume the {\em not}-free NNP program $P$ whose unique rule  is: 
$$ \bigwedge_3 [ \, \vee (c \ \ \overline{a}) \qquad
\bigvee_2 (\overline{a} \ \ \bigwedge_3 [ \, \vee   (n \  \ \overline{m} \,) 
\quad \bigvee_2 (e \ \   
\wedge [\overline{a}  \ \ {\textcolor{black} {\overline{e}}} \,] \, ) \ \ c \, ] \, ) 
\qquad {\textcolor{black} a} \, ] \longleftarrow \top.$$

 Since $B=\top$, then   $H \vee \mbox{NNF}(\neg {B})=H$ whose associated tree  is depicted  in {\bf Fig. 2}, on the left,
 following the classical modeling   by trees  stipulated
 in Subsection \ref{subsec:BACKground-Nested-Expressions}.

\begin{center}
\begin{adjustbox}{valign=t}
\begin{tikzpicture}[]
\Tree
[.$\wedge$   [.$\vee$ [.$c$\\ ][.$\neg {\, a \, }$\\ ] ] 
             [.$\vee$  [.{\color{black}$\neg {\, a \, }$}\\ ] 
                       [.$\wedge$  [.$\vee$ [.$n$\\ ] [. \  $\neg {\, m \, }$\\ ]  ] 
                                   [.$\vee$ [.$e$\\ ] 
                                              [.$\wedge$ [.$\neg {\, a \, }$\ \ \ \\ ] 
                                              [.{\ \ \ $\neg {\, e \, }$}\\ ] ]
                                              ]
                                   [.$c$\\ ] 
                       ]
              ]
             [.{\color{black} $a$}\\ ]
        ]
\end{tikzpicture}
\end{adjustbox} 
\begin{adjustbox}{valign=t} \hspace{.5cm}
\begin{tikzpicture}[]
\Tree
[.$\wedge$   [.$\vee$ [.$c$\\ ][.$\neg {\, a \, }$\\ ] ] 
             [.$\vee$  [.{\color{black}$\neg {\, a \, }$}\\ ] 
                       [.$\wedge$  [.$\vee$ [.$n$\\ ] [.$\neg {\, m \, }$\\ ]  ] 
                                   [.$\vee$ [.$e$\\ ] 
                                              ]
                                   [.$c$\\ ] 
                       ]
              ]
             [.{\color{black} $a$}\\ ]
        ]
\end{tikzpicture}
\end{adjustbox}

\vspace{-.3cm}
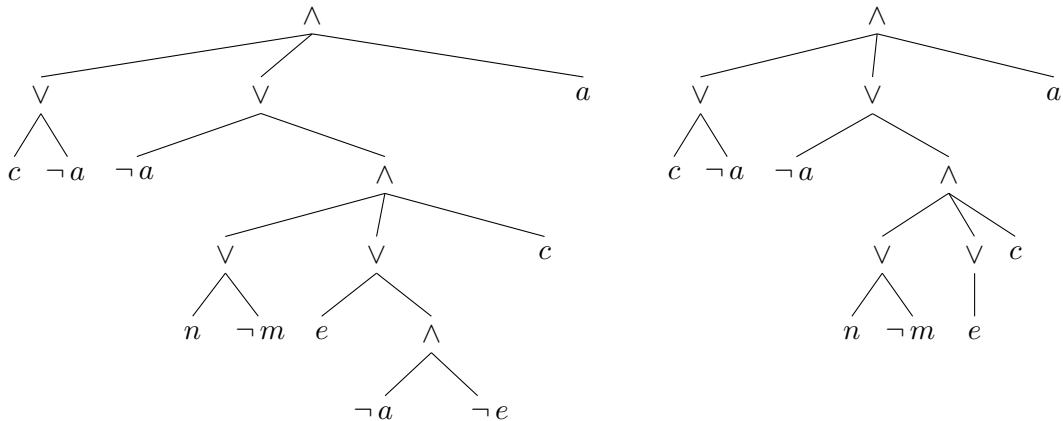
{\small {\bf Fig.  2.} { Expressions $ H $ (left) and  $ H' $ (right)}.}
\end{center}

 
The positive occurrences in $H$ are 
the set $\{c^1,n,e,c^2,a\} $, where $c^1$ and $c^2$ refer, respect., 
to the left- and right-most occurrences of  $c$.
Thus $P$  is consistent as  is head-consistent. 
By  repeatedly applying  $\mathcal{UR} $, we will compute  $LM(P)$ according 
to the    properties above.

\vspace{.1cm}
For the first NUR application, we pick up the occurrence of
literals  ${\textcolor{black} a}$ and the right-most 
${\textcolor{black} {\neg { \, a \, }}}$ within $H$.
This leads to   the following  expressions in the NUR numerator:

\vspace{.3cm}
 $\bullet  $ 
 $  \Delta({\color{black} {\neg {\, a  }}})=
 \wedge [{\neg {\, a \, }} \ \ {\textcolor{black} {\neg {\, e}}} ].$ 
 
  \vspace{.3cm}
 $\bullet  $ 
 $  \Sigma=e$.
 
  \vspace{.3cm}
$\bullet  $ $   \Delta({\color{black} {\neg {\, a  }}}) \, \vee  \, \Sigma
=\bigvee_2 (e \ \   \wedge [{\neg {\, a \, }} \ \ {\textcolor{black} {\neg {\, e }}} \,] \, ).$

\vspace{.3cm}
$\bullet  $ $\Pi= \bigvee_2 (\neg {a} \ \ \bigwedge_3 [ \, \vee   (n \  \ \neg {m} \,) 
\quad \bigvee_2 (e \ \   
\wedge [\neg {a}  \ \ {\textcolor{black} {\neg {e}}} \,] \, ) \ \ c \, ] \, ).$

\vspace{.3cm}
  Then, applying   NUR to $ H $
yields $ H' $ below, whose   tree is  on the right of  {\bf Fig. 2}: 
$$H'=\bigwedge_3 [ \, \vee (c \ \ \neg {\, a \, }) \ \ 
\bigvee_2 (\neg {\, a \, } \ \ \bigwedge_3 [ \, \vee   (n \  \ \neg {\, m \, } \,) 
\quad \vee (e ) \ \ c \, ] \, ) 
\quad {\textcolor{black} a} \, ].$$

We can now apply to  $ H' $ the simplification rule $\odot^{k+1} $ and  obtain:
$$H''=\bigwedge_3 [ \, \vee (c \ \ \neg {\, a \, }) \ \ 
\bigvee_2 (\neg {\, a \, } \ \ \bigwedge_3 [ \, \vee   (n \  \ \neg {\, m \, } \,) 
\quad  e  \ \ c \, ] \, ) 
\quad {\textcolor{black} a} \, ].$$

With the same atom $a$ as before and this time picking   the right-most  
${\color{black}\neg {\, a \, }}$,   we have the following expressions
in the numerator of NUR:

\vspace{.3cm}
$\bullet  $  $\Delta({\color{black}\neg {\, a \, }})={\color{black}\neg {\, a \, }}.$  

\vspace{.2cm}
$\bullet  $
 $\Sigma=\bigwedge_3 [ \, \vee   (n \  \ \neg {m} \,) \  \ e  \ \  c \, ].$

\vspace{.2cm}
$\bullet  $ 
$   \Sigma  \,  \vee \, \Delta({\color{black} \neg {\, a  }})= 
\bigvee_2 (\neg {\, a \, } \ \ \bigwedge_3 [ \, \vee   (n \  \ \neg {\, m \, } \,) 
\quad  e  \ \ c \, ] \, ).$ 

\vspace{.2cm}
$\bullet  $ $\Pi=    \Sigma \, \vee \, \Delta({\color{black}\neg {\, a  }}).$

 \vspace{.3cm}
 By applying    NUR to $ H'' $,   the obtained expression
 corresponds  to the tree  in    {\bf Fig. 3}, on the left.
 After two  applications of rule $\odot^{k=1}$ and  one of $\odot^{k+n}$, 
 one finds the expression associated with the  right tree depicted in  {\bf Fig. 3.}

\vspace{-.2cm} 
\begin{center}
\begin{adjustbox}{valign=t}
\begin{tikzpicture}[]
\Tree
[.$\wedge$   [.$\vee$ [.$c$\\ ][.$\neg {\, a \, }$\\ ] ] 
             [.$\vee$   
                       [.$\wedge$  [.$\vee$ [.$n$\\ ] [.$\neg {\, m \, }$\\ ]  ] 
                                   [.$\vee$ [.$e$\\ ] 
                                              ]
                                   [.c\\ ] 
                       ]
              ]
             [.{\color{black} {$a$}}\\ ]
        ]
\end{tikzpicture}
\end{adjustbox}
\begin{adjustbox}{valign=t} \hspace{2.cm}
\begin{tikzpicture}[]
\Tree
[.$\wedge$   [.$\vee$ [.$c$\\ ][.$\neg {\, a \, }$\\ ] ]   
                        [.$\vee$ [.{\textcolor{black}{$n$}}\\ ] 
                        [.{\textcolor{black}{$\neg {\, m \, }$}}\\ ]  ] 
                                    [.$e$\\ ] 
                                   [.{\textcolor{black} {$c$}}\\ ] 
             [.{\textcolor{black} {$a$}}\\ ]
        ]
\end{tikzpicture}
\end{adjustbox}

\vspace*{-.3cm}
{\small {\bf Fig. 3.} {\em Example \ref{ex:RUNNING-UR} (continued).}} 
\end{center}

It is immediately seen that the expression modeled by the last tree is:
$$\bigwedge_5 [\, \vee (c \ \ \neg {\, a \, } ) \ \ 
\vee (n \ \ \neg {\, m \, } ) \quad c \quad e \quad a ] $$

Another application of rule NUR and then one of rule $\odot^{k=1}$ leads to expression: 
$$\bigwedge_5 [ c  \ \ \vee(n \ \ \neg {\, m \, } ) \quad c \quad e \quad a ] .$$

At this point, no more $\mathcal{UR} $ rules are applicable. 
The least model of the last deduced expression is $\{c,e,a\} $. Since, $\mathcal{UR} $ 
is sound,   the least model   is preserved through the 
  search process and so 
$\{c,e,a\} $ is the sought $LM(P)$. \qed
\end{example}

The previous example contained  a consistent, even  a head-consistent program.
Now we examine the opposite case, an inconsistent program obtained 
from the above example by incorporating   a slight modification, i.e.  switching
  one literal in the head expression:

\begin{example} We consider the program obtained from Example \ref{ex:RUNNING-UR}  
wherein the right-most occurrence  
of   $c$ in the head is exchanged by $\neg c$. 
The head literals of the new program $P$ are $\{\neg c,n,e,c,a\} $  and so $P$ is not  head-consistent,
and thus is potentially inconsistent.

\vspace{.1cm}
Then, one can check that the same inference rules applied in Example \ref{ex:RUNNING-UR}
are  applicable   to the new program $P$ as well and so  the whole search process 
of the previous example   can be mimicked;  now  
  the last obtained expression is:
 $$\bigwedge_5 [ c  \ \ \vee(n \ \ \neg {\, m \, } ) \quad \neg c \quad e \quad a ] ,$$

The above expression is inconsistent  and so is the initial program given that
all the  inferences integrating the logical calculus $\mathcal{UR} $ are  sound.
\qed
\end{example}

\subsection{    Nested Hyper-Unit-Resolution $\mathcal{HUR} $   } \label{subsec:HYPER-UNIT-RESOL}


 
The previous rule NUR serves us as the base  to  
define the main nested hyper unit-resolution (NHUR) rule. 
 Below the generalization  from NUR to NHUR  is explained.
The logical calculus $ \mathcal{HUR} $
  also encompasses hyper simplification rules that are established by generalizing  the 
 given simplification rules  in a similar fashion 
as  NHUR is from NUR. However, 
for the sake of brevity, we omit  the details of the generalization of the simplification rules.

\vspace{.1cm}
 Assume that an expression  $ \varphi $ contains a conjunction 
 of $ \textcolor{black} \ell $ with  
 two sub-expressions  $ ({\textcolor{black} {\vee}} \ 
 \ \Sigma^1  \ \ 
 \Delta({\textcolor{black} {\neg {\, \ell^1 \, }}})  \,) $ and
 $ ({\textcolor{black} {\vee}} \ 
 \ \Sigma^2 \ \ 
 \Delta({\textcolor{black} {\neg {\, \ell^2 \, }}})  \,) $,
 where ${\textcolor{black} {\neg {\, \ell^i \, }}} $ denotes a specific occurrence
 of  $ {\textcolor{black} {\neg {\, \ell \, }}} $. 
 The simultaneous application of NUR to both of them 
 is  formalized as: 
 $$\frac{\textcolor{black} \ell 
\ \ {\textcolor{black} {\wedge}} \ \ \Pi^1 \cdot \, (\Sigma \ {\textcolor{black} {\vee}} 
 \  \Delta({\textcolor{black} {\neg {\, \ell^1 \, }}})  \,)    \ \
 {\textcolor{black} {\wedge}} \ \  
  \Pi^2 \cdot \, (  \Sigma  \ {\textcolor{black} {\vee}} \ 
 \Delta({\textcolor{black} {\neg {\, \ell^2 \, }}})  \,)}
{ \textcolor{black} \ell 
\ \ {\textcolor{black} {\wedge}} \ \  \Pi^1 \cdot  \Sigma^1 
\ \  \wedge \ \ 
  \Pi^2 \cdot \Sigma^2   }{\mbox{ NHUR}}$$

\noindent We introduce compact notations  by denoting  
 $  \Pi^i \cdot ( \Sigma  \ {\textcolor{black} {\vee}} \ 
 \Delta({\textcolor{black} {\neg {\, \ell^i \, }}})  \,) $ 
  by $  \Pi^i \cdot 
 \Sigma \Delta({\textcolor{black} {\neg {\, \ell^i \, }}})  $,
 and thus rule  NHUR  for $ k $ sub-formulas is formally expressed  as follows:
$$\tcboxmath{\frac{ 
  {\textcolor{black} {\ell }} \ \  {\textcolor{black} {\wedge }} \ \
 \Pi^1 \cdot \Sigma \Delta({\neg {\, \ell^1 \, }})
\ {\textcolor{black} {\wedge}}  \,\ldots \, {\textcolor{black} {\wedge}}
\ \Pi^k \cdot \Sigma \Delta({\textcolor{black} {\neg {\, \ell^k \, }}})  }
{{\textcolor{black} {\ell }} \ {\textcolor{black} {\wedge }} \ \Pi^1 \cdot \Sigma^1 \ 
{\textcolor{black} {\wedge}} \ \ldots  \ {\textcolor{black} {\wedge}} \
 \Pi^k \cdot \Sigma^k  } 
{\mbox{ \ \underline{NHUR}}}}$$

\noindent Since
the  $ {\neg {\, \ell^i \, }}$'s   
    are pairwise different, so are the   sub-formulas 
 $ \Sigma \Delta^i $ and $ \Sigma ^i $. However, the formulas $ \Pi^i $
are not necessarily pairwise different as   shown in the next example:

  \begin{example} \label{ex:HUR}  Let us reconsider  the following not-free expression:
$$\varphi=\bigwedge_3 [\vee (c \ \ \phi_1) \ \ 
\bigvee_2 ({\textcolor{black} {\neg {\, {a}^1 \, }}} \ \ 
\bigwedge_3 [\vee ({\textcolor{black} {\neg {\, {a}^2 \, }}} \  \ \neg {c} \,) 
\quad \bigvee_2 (\phi_2 \ \   \wedge [\neg {b}  \ \ 
{\textcolor{black} {{\neg {\, {a}^3 \, }}}} \,] \, ) \ \ c \, ] \, ) \quad {\textcolor{black} a} \, ].$$

One can apply NHUR with $ {\textcolor{black} a}  $ and the three occurrences 
${\textcolor{black} {\neg {\, a^i}}}$. It is not hard to check 
that  $ \Pi $ in the numerator
of NHUR is   the same for the three  ${\textcolor{black} {\neg {\, a^i \, }}}$, 
so below $ \Pi $ is denoted by $ \Pi^{1,2,3}$. 
However,  expressions $ (\vee \ \ \Sigma^i \ \ \Delta(\neg {\, {a}^i \, }))$
 are different. Specifically, the expressions associated to   the NHUR numerator are as follows:

\vspace{.25cm}
  $\bullet \ \  \Pi^{1,2,3}= \bigvee_2 ({\textcolor{black} {\neg {\, {a}^1 \, }}} \ \ 
\bigwedge_3 [\vee ({\textcolor{black} {\neg {\, {a}^2 \, }}} \  \ \neg {c} \,) 
\quad \bigvee_2 (\phi_2 \ \   \wedge [\neg {b}  \ \ 
{\textcolor{black} {{\neg {\, {a}^3 \, }}}} \,] \, ) \ \ c \, ] \, )$

\vspace{.2cm}
  $\bullet \ \  \Sigma^1 \, \vee  \, \Delta({\neg {\, {a}^1 \, }})=\Pi^{1,2,3}.$

\vspace{.15cm}
  $\bullet \ \  \Sigma^2  \, \vee  \, \Delta({\neg {\, {a}^2 \, }})= 
\vee ({\textcolor{black} {\neg {\, {a}^2 \, }}} \  \ \neg {c} \,). $

\vspace{.2cm}
  $\bullet \ \  \Sigma^3 \, \vee  \, \Delta({\neg {\, {a}^3 \, }})= 
\bigvee_2 (\phi_2 \ \   \wedge [\neg {b}  \ \ 
{\textcolor{black} {{\neg {\, {a}^3 \, }}}} \,] \, ). $

\vspace{.25cm}
 Applying   NHUR to the  above expression  yields:
$$\varphi=\bigwedge_3 [\vee (c \ \ \phi_1) \ \ 
\bigvee_1 ( \bigwedge_3 [\vee ( \neg {c} \,) 
\quad \vee (\phi_2  \, ) \ \ c \, ] \, ) \quad {\textcolor{black} a} \, ].$$

 That after simplifying leads to:
$$    \bigwedge_5 [\vee (c \ \ \phi_1) \ \ 
       \neg {c} \ \  \phi_2    \ \ c   \,\ {\textcolor{black} a} \, ] .$$ 
       
Clearly,   a further   NHUR   deduces $   \bot  $.
Altogether,  at least in this particular case,  rule   NHUR accelerates considerably
the proof of inconsistency. \qed
\end{example}

 An even more general rule than the anterior
 ${\mbox{NHUR}}$    can be devised which could be capable of handling simultaneously 
$  n \geq 1   $      unit-clauses, instead of just one as  done above.
Thus, assume that we have $n$ unit-clauses with literals
 $   \ell_j, \ 1 \geq j \geq n  $,  and    so that for each  $   \ell_j  $, 
   many  $  n_j    $ sub-formulas  with the previous pattern:  
   $$  \Pi^{j,i} \cdot (   \Sigma^{j,i} 
 \ \vee  \Delta({\textcolor{black} {\neg {\, \ell^{j,i} \, }}}) ), \ 1 \geq i \geq n_j, $$  
   can be covered by   $   \ell_j$. 
 That is, we can compute simultaneously $ n \geq 1 $ 
 applications of the previous   NHUR rule, one  for each unit-clause $   \ell_j  $ 
 and, for each one of these unit-clauses $   \ell_j  $, we can apply $n_j$ NUR rules  as the one 
 devised in Subsection  \ref{subsec:RULE-NUR}. 
 
 \vspace{.1cm}
 Thus  the expression in the previous NHUR numerator:
$$  
 \Pi^1 \cdot \Sigma \Delta({\neg {\, \ell^1 \, }})
{\textcolor{black} {\wedge}} \ldots {\textcolor{black} {\wedge}} \ 
 \Pi^i \cdot \Sigma \Delta({\textcolor{black} {\neg {\, \ell^i \, }}}) \
 {\textcolor{black} {\wedge}} \,\ldots  {\textcolor{black} {\wedge}}
\Pi^k \cdot \Sigma \Delta({\textcolor{black} {\neg {\, \ell^k \, }}}) $$
 will be compacted into
 $ \Pi^{ K} \cdot \Sigma \Delta({\textcolor{black} {\neg {\, \ell \, }}})^{ K} $ and
similarly  the sub-expression 
in the denominator into
$ \Pi^{K} \cdot \Sigma^{K} $. Then
for $ n  \geq 1 $ unit-clauses, 
inference rule NHUR is formally defined as:
$$\tcboxmath{\frac{{\textcolor{black} {\ell_1}} , \ldots ,
 {\textcolor{black} {\ell_n}} ,   
 \Pi^{ K_1} \cdot \Sigma \Delta({\textcolor{black} {\neg {\, \ell_1 \, }}})^{ K_1}, \ 
\ldots , \
\Pi^{ K_n} \cdot \Sigma \Delta({\textcolor{black} {\neg {\, \ell_n \, }}})^{ K_n}}
{{\textcolor{black} {\ell_1}} , \ldots ,
 {\textcolor{black} {\ell_n}} ,   
 \Pi^{ K_1} \cdot \Sigma^{K_1}, \
\ldots , \
\Pi^{ K_n} \cdot \Sigma^{ K_n}} {\, \mbox{NHUR}}} $$

  As said at the beginning of this subsection, the simplification
  rules described in Subsection \ref{subsec:RULE-NUR} can  also be generalized
  and new hyper simplification rules can be devised   simultaneously executing several simplifications as the ones given in   Subsection \ref{subsec:RULE-NUR}.
  
  \vspace{.1cm}
  If   applying NHUR to  expression $ \varphi $ results in $ \varphi' $, then $ \varphi \equiv \varphi' $. 
  The proof  follows straightforwardly from the soundness of NUR, since 
  the latter is the base of NHUR as  discussed above.  Analogously, completeness and
  polynomiality are also kept.

\vspace{.1cm}
 Clearly    $\mathcal{HUR} $   makes it possible to find    shorter proofs than those 
 found via  $\mathcal{UR} $. 
Therefore,   the suitable management of  $\mathcal{HUR} $  inferences  can increase the overall deductive efficiency
  just as happens in the  unnesting framework with hyper unit-resolution.


\section{Related Work: Nested ASP} \label{sec:related-work}

A comprehensive   related work on nested (or non-clausal)
reasoning is provided in \cite{Imaz2021horn} which embraces the following topics: 
  satisfiability, incomplete solvers, QBFs, Max-SAT, theorem proving,
resolution, knowledge compilation,   polynomial   classes, Horn functions, 
implicational   systems, 
directed hyper-graphs,
Horn approximations, abduction and Petri nets. 
On the other hand, in references \cite{Imaz2023a,Imaz2023b}, 
one can find moreover related work on non-clausal reasoning based on both  regular-many valued logic and on possibilitic logic.
Below,  chronologically organized related work on  NASP    is discussed. 

\vspace{.25cm}
{\bf Nested  Completion Semantics.} The first declarative semantics
for nested logic programs (Prolog programs) was proposed by Lloyd and Topor \cite{LloydT84},
who extended the completion semantics \cite{Clark77} to   programs with nested
first-order formulas  in the bodies of rules and whose heads were a single literal.
In their work, the  main result  was the presentation
of the soundness of SLDNF-resolution for their studied nested programs.

\vspace{.25cm}
{\bf Nested ASP.}  The answer set semantics is extended
for the first time to programs with nested expressions in \cite{LifschitzTT99}. 
A set of transformation rules is provided and shown that any nested
program is equivalent to a set of disjunctive rules, possibly
with negation as failure in the heads. This transformation is exponential in the worst-case. 
A major difference  between our approach and that of  \cite{LifschitzTT99}   is that,
in  the latter, the {\em not} connective
is allowed to be unlimitedly nested. 
Besides, in that reference,  a transformation  for  propositional logic  between the Lloyd-Topor 
programs \cite{LloydT84} and the   NASP  programs of \cite{LifschitzTT99}
  is given. It is also proven
in \cite{LifschitzTT99} that the semantics of \cite{LloydT84}
is equivalent to the answer set semantics, modulo replacing $\neg$ 
employed in the language 
of \cite{LloydT84} by {\em not} from the language of \cite{LifschitzTT99}.

\vspace{.2cm}
{\bf Strong-Equivalence.} The notion of strong-equivalence 
(Definition  \ref{def:BACKGROUND-STRONG-EQUIV}) is applied 
to nested programs    in \cite{LifschitzPV01}.
Its associated proof  is determined in terms of the Here-and-There logic (HT) or G\"odel logic ($G_3$),  
which was adapted to ASP by D. Pearce in \cite{Pearce96, Pearce06} and who 
also set up Equilibrium Logic on the basis of HT. The latter results
signified an important breakthrough that meant a purely logical
treatment of ASP.  It was also
proven in \cite{LifschitzPV01} that two nested programs  are strongly-equivalent
if they have the same equilibrium models.   The latter discovery has been    highly
influential  within the ASP field. 

\vspace{.2cm} 
{\bf Polynomial Translation.} The first polynomial translation 
of nested programs into unnested ones is due to the authors in \cite{PearceSSTW02},
as previous translations incurred in an exponential blow-up in the worst-case.
The proposed translation is based on the introduction of new atoms   substituting
sub-formulas, until getting an unnested program. Its original
idea was set up in \cite{Tseitin83} and later was theoretically studied
in depth in \cite{plaisted1986structure,egly1996different,egly1997definitional}. 
This transformation has been studied in automated
reasoning, wherein  
only satisfiability needed to be preserved. 
However, transformations 
within ASP have higher complexity due to the need of respecting the minimality
criterion of answer sets. This translation has the unquestionable merit of making possible solving 
nested programs through running traditional, available ASP-solvers. 
Nevertheless, as discussed in Subsection \ref{subsec:RESEARCH-FRAME}, 
using translators has severe drawbacks   compromising the global efficiency of the system. 
We have proposed here an alternative
means consisting in directly computing nested programs and omitting translators, 
just as  done to computing   ASP programs.

\vspace{.2cm}
{\bf  Three-Valued Logic.} While nested programs are usually computed 
by translating them to an unnested program, the central idea  of   \cite{Cabalar02} is 
the proposal of a new characterization of strong-equivalence of programs, which is 
different from the original one in \cite{LifschitzPV01} and  is based 
on a tree-valued logic. Strong-equivalence of nested programs is also
analyzed under this logic and  verified that this perspective yields different
results from that of the Here-and-There logic of the original one \cite{LifschitzPV01}.

\vspace{.2cm} 
{\bf Intuitionistic Logic.} In the line of the above work and given two programs,
the main result of the authors in \cite{OsorioNA01} is showing that 
it is possible to construct in linear time 
a propositional formula $F$ and in linear size with respect to the  programs, 
such that these two programs are strongly-equivalent if and only if $F$
is a theorem in intuitionistic logic.  In \cite{OsorioNA04}, further results are proven 
 that enable
one  to simplify programs and to transform nested into unnested programs. 
More specifically: 
(ii)  strong equivalence between nested programs is   expressed in terms
of the G\"odel three-valued  logic; 
(iii) a new notion 
of a program that is a conservative extension
of another program  is introduced which is a less restrictive
concept than that of strong equivalence and serves also to simplify  programs;
(iv) the concept min-answer set, which is simultaneously  
 a minimal model and an answer set, is defined and its helpful role in the checking
 of strong equivalence is justified.

\vspace{.2cm} 
{\bf Fages' Theorem.}  Fages' theorem \cite{Fages94}  connecting 
 the completion semantics and the answer set semantics
of logic programs is extended in  \cite{ErdemL01,ErdemL03} to programs with nested expressions permitted  
in the bodies of rules and whose heads are single literals. 
First, the syntactic condition, so-called {\em tightness}, expressed by Fages  
and then his proper theorem are both  generalized
to such a class of nested programs. 
This result gave rise to syntactically identify a class of nested programs
such that the computation of their answer sets   
could be assured by SAT-solvers, which are experimentally proven to be faster than ASP-solvers.
Actually, these results led to the design of answer set solvers ASSAT \cite{LinZ04} 
and CMODELS \cite{GiunchigliaLM04}.
Furthermore, the authors provide some evidence of how nested programs can be used
to represent and solve real life problems.

\vspace{.2cm} 
{\bf Weak Tightness.} The work in \cite{YouYM03} presents
a  weaker condition than  that of previous approaches,  called weak tightness, 
in order to determine the class 
of nested programs whose answer sets can be computed through their completion
models and SAT solvers, as discussed above in the "Fages' Theorem " item. 
Weak tightness is proven to be a necessary and sufficient
condition for a program to have the same answer sets and completion models.
The considered nested programs 
 integrate a nested expression in their bodies only.
Taking advantage of weak tightness, a transformation claimed linear
(although
it is demonstrated that the  transformation is indeed linear in size or space,
no algorithm with linear complexity is devised) converts a nested program into an unnested one. 

 \vspace{.2cm} 
{\bf Loop Formulas.} The concept of loop formulas 
of normal programs due to Lin and Zhao \cite{LinZ04}
is extended in  \cite{LeeL03,FerrarisLL06} to disjunctive and nested programs. 
This extension is then 
employed to generalize  the influential Clark's definition \cite{Clark77} of
completed program to  nested programs. Two dual (conjunctive and disjunctive) loop formulas
are set up, which added to the completion formula of the program, form
an augmented formula whose models are exactly the answer sets of the initial nested 
program. This principle is hence an alternative manner to that based on  the tightness concept.

\vspace{.2cm}
{\bf   Strong-Equivalence: Complexity.} A new 
concept named SE-model of a nested program is defined in \cite{Turner03} and it is proven 
that two nested programs are strongly equivalent if and only if they have the same SE-models.
A relevant result of \cite{Turner03} is proving that, contrary to what expected,
the problem of proving  equivalence of nested programs is   harder  
than that of proving their strong equivalence. Indeed, the former is 
$\Pi^P_2$-hard \cite{EiterG93}, whereas the latter is proven 
by the author in \cite{Turner03} to be co-NP. The new theorem
of strong equivalence based on SE-models is furthermore applied to the transformation of programs
and to the test of strong equivalence of weight constrains \cite{NiemelaSimons00}.

\vspace{.2cm}
{\bf Polynomial Transformation.} 
The authors in \cite{LinkeTW04} define the class of head-cycle free nested logic  programs (HCF)
and its proper sub-class of acyclic nested programs.  The main result 
of their article  is proving that any HCF program can be rewritten
into a normal ASP program in  polynomial time.
This outcome is notable as until then, 
all nested programs were translated to disjunctive ASP programs.
Such a translation interestingly entails that determining answer sets of HCF programs
has NP complexity, taking advantage of the lower NP complexity of normal ASP
programs \cite{MarekT91}  with respect to that of disjunctive ASP programs, which is
 in the second level of the polynomial hierarchy \cite{EiterG95}. 
This made  class HCF  be the first class of nested
programs to have NP complexity, viz. lower than that of arbitrary nested programs. 
The definition of the HCF class and the demonstration of its relative low complexity are
inspired by the existence of similar classes in the ASP context
which were described by Ben-Eliyahu and Dechter \cite{Ben-EliyahuD94}.

\vspace{.2cm}
{\bf Weight Constraints.} 
That logic programs  with weight constraints
can be viewed as shorthands for programs with  
 nested expressions (negation connectives are freely nested) of a special form  was
 proven in \cite{FerrarisL05}.
The authors also showed that  nested programs can be translated
into non-disjunctive nested programs, i.e.
with a single literal as head 
(see Subsection \ref{subsec:EXPONEN-SUCCINCT}). Afterwards, using fresh atoms,
non-disjunctive nested programs are converted into unnested ones.
This idea was initially explored in \cite{MarekR02} where, using new atoms,
programs with cardinality constraints were transformed into unnested programs.  
Transformations in \cite{FerrarisL05} are more  general than those in \cite{MarekR02} and
 avoid in some occasions the exponential
grow of the size of transformed programs.

\vspace{.2cm}
{\bf  Nested Epistemic Programs.} 
Classical NASP and epistemic logic programs, wherein belief operators of the kind  
$F$ is known or may believed to be true 
can be explicitly presented under the semantics 
of world views, are both seen under a formal unifying framework in \cite{WangZ05}.
This perspective is then generalized by defining the nested epistemic programs. 
A characterization of strong-equivalence of   nested epistemic logic programs is 
also put forward. For that purpose, the epistemic 
Here-and-There logic is formally defined and  forward steps  are taken, 
which are 
inspired by  those followed in \cite{LifschitzPV01}.

\vspace{.2cm}
{\bf Alternative Semantics.} An alternative  definition of reduct 
w.r.t. the traditional one is proposed in  \cite{Ferraris05a,Ferraris11}. 
This new reduct definition allows the author
to extend the definition of an answer set of programs with nested expressions to general
unrestricted propositional theories and so to be comparable 
to that of equilibrium logic  \cite{Pearce96}, which defined for the first time
an answer set for arbitrary propositional theories. 
The author studies the relationships between
the application of the classical reduct and the new one, checking that sometimes
they lead to different outcomes. Importantly, the new definition
of an answer set  is equivalent to the original one 
  \cite{Lifschitz99} if applied to nested programs.  
The new definition of an answer set is equivalent to the definition 
of a model in equilibrium logic, so that it shares 
important properties of equilibrium logic such as the characterization 
of strong equivalence in terms of the logic
of Here-and-There. Then programs with aggregates are studied, in which aggregates
are seen under two perspectives, as primitive constructs and as abbreviations
 for propositional formulas (see weighted constraints). It is proven that the complexity 
 of the existence of  an answer set in programs with arbitrary 
 propositional theories is $\Sigma_2^P $-complete, while it is NP-complete
 for non-disjunctive nested programs.

\vspace{.2cm}
{\bf Arbitrary Theories.} The main result
in \cite{CabalarF07}  is that, under the answer set semantics, 
any theory can always be re-expressed
as a strongly equivalent disjunctive logic program, possibly with negation in the head.
The authors  provide two different proofs for this result: one involving a syntactic transformation
and one that constructs a program starting from the counter-models of the theory in the
intermediate logic of Here-and-There.
From an LP reading, this result is pointing out that any generic propositional theory
can be seen as ``shorthand" for a logic program. Another possible reading of the main result is that the form of logic programs is
a kind of {\em normal form} for arbitrary formulas from the point of view of answer set semantics
(Equilibrium Logic). 
The main focus in  \cite{CabalarF07} was just to prove the existence of a strongly
equivalent program for any propositional theory,  the methods  to
obtain the resulting program are mostly thought to achieve simple proofs, 
rather than to obtain an efficient computation of the final program itself, which  is tackled in  \cite{CabalarPV05} 
and discussed next.

\vspace{.2cm}
{\bf Transformations.} In reference \cite{CabalarPV05},
 the authors add to the approach in \cite{CabalarF07}
a pair of alternative syntactic transformations to reduce a propositional 
theory into a strongly equivalent logic program.
One of the transformations preserves the vocabulary of the original theory, but
is exponential in the obtained program size, whereas the other is polynomial, but
with the cost of adding auxiliary atoms. The authors in  \cite{CabalarPV05} further confirm this
complexity  for nonnested programs, viz. proving that no polynomial time
transformation to nonnested programs exists if we preserve the original
vocabulary. The existence of polynomial space/time strongly equivalent translations
to programs with nested expressions is an open question.
 Another proof of the strong equivalence
between propositional theories and logic programs is provided in \cite{LeeP07}. Unlike the
other approaches based on the logic of Here-and-There, the proof in \cite{LeeP07} uses
properties of classical logic.

\vspace{.2cm}
{\bf   NASP with Variables.} 
All nested programs studied so far were propositional and did not allow for variables,
which limits the suitability of NASP in many application domains, especially
when reasoning is to be done on large numbers of input facts. When processing
 programs with variables,  a  requirement of ASP systems is that rules must be safe
 (domain independent),
 that is, each variable in a rule must occur in a positive body literal.
 Applying published translations e.g. \cite{PearceSSTW02} to nested programs 
 with variables, one easily obtains unsafe rules. 
 In \cite{BriaFL08,Bria09c,BriaFL09b,BriaFL09}
 the authors analyze safety and semantic properties of nested programs with variables, 
 wherein  rule heads are expressions in DNF and rule bodies are
 expressions in CNF. Also a polynomial translation to convert these nested
 programs into disjunctive ASP programs is furnished which is implemented
 in a compiler tool   available on-line \cite{BriaFL09b}. Furthermore, it is shown that: 
 (1) the proposed semantics  coincides with that of \cite{Lifschitz99} 
 on the common language fragment; (2)  the answer sets
 of the proposed nested programs coincides with the standard one 
 \cite{GelfondL91}; and (3) the introduced notion 
 of safety is more general than others previously defined.
 The concept of safety is generalized by the authors in reference \cite{CabalarPV09} to arbitrary
 first-oder formulas.

 \vspace{.2cm}
{\bf Equilibrium Logic and NASP.} In the paper \cite{PearceTW09}, the authors present
polynomial reductions of the main reasoning tasks associated with equilibrium
logic and nested logic programs into quantified propositional logic (QBFs).
In particular, the presented encodings map a given decision problem into some QBF 
such that the latter is valid precisely in case the former holds. The reasoning
tasks dealt with are the consistency problem,  brave reasoning and skeptical reasoning.
Additionally, encodings are provided  for testing the equivalence
of theories or programs under different notions of equivalence, viz. ordinary, 
strong and uniform equivalence. For all considered reasoning tasks, the authors
analyze their computational complexity and give strict complexity bounds.
These complexity bounds show that, w.r.t. the considered reasoning tasks,
equilibrium logic and nested logic   programs
behave complexitywise precisely as disjunctive logic programs.

 \vspace{.2cm}
{\bf Reducts and Semantics.}
  The  author  in   references  \cite{Truszczynski09,Truszczynski10} develops generalizations
  of the Faber, Leone and Pfeifer semantics \cite{FaberLP04} to the full language
  of propositional logic using appropriate variants of methods proposed earlier by 
  Pearce \cite{Pearce96} and Ferraris \cite{Ferraris05a,Ferraris11} for the stable model semantics.
  This allows the author to cast the three semantics in the same uniform framework,
  which eases comparisons and offers insights into their properties.

  \vspace{.2cm}
{\bf Preferences.}
  Preferences in  ASP \cite{BrewkaNS04} are
 envisaged   by allowing a new connective $ \times $ in the rule heads: 
   $ a \times b \leftarrow B $ means ``if  $ B $ holds,    option $ a $ is preferred
   to $ b $". This approach was lifted to NASP in references \cite{confalonieri2010nested,ConfalonieriN11}  whose 
   disjunctive options
   can be nested expressions $ F $. NNP programs cannot represent any subclass
   of nested programs with preferences as the latter are intrinsically  disjunctive, whereas 
   the former are normal or deterministic. Yet, NNP programs can be helpful as  
   nested programs with preferences can be
     decomposed  into a set of NNP programs. 
  For instance, 
    rule $ F \times F' \leftarrow B $ can be split into 
   $ F  \leftarrow B $  and $ F'  \leftarrow B $, and   both can be recursively decomposed until  only NNP programs remain.  
   The NNP programs   inheriting $ F'  \leftarrow B $ must inherit 
   a penalty higher than that  inherited by those programs inheriting $F  \leftarrow B $.
   After the decomposition,   mechanisms based
   on nested unit-resolution  to find the answer sets can be employed. 
   An appropriate management of penalties by the   decomposition
   process needs   to be studied.

\vspace{.2cm}
   
{\bf Possibilistic NASP.} 
Basic concepts such as closed interpretation,
reduct, answer set,   strong-equivalence, etc.,  introduced 
 for propositional NASP in \cite{LifschitzTT99,LifschitzPV01}       are extended
to  possibilistic NASP in  \cite{NievesL12, NievesL15}.  
 Towards computing  answer sets, possibilistic NASP programs
are first  translated  into 
   possibilistic unnested-disjunctive programs.  Then,  
 mechanisms  anteriorly developed  
 in \cite{NicolasGSL06,NievesOC13,BautersSCV15} are applied to the latter programs.
The authors showed  that their proposal has practical 
utility by successfully solving real-world medical problems. 
Concretely,   programs in such references look   in fact flat nested (FNASP). Their underlying possibilistic logic is based 
on non-numerical, partially-ordered  values since, and as the authors justify,
such characteristics are convenient within real-world applications.

\vspace{.2cm}
{\bf   Functional NASP}. Contrarily to the recurrent
  studies of programs equipped with functions that are carried out in ASP, 
e.g.  \cite{AlvianoFL10, Cabalar11,BartholomewL19},
such programs have been only occasionally envisaged,   as far as we know, within NASP.
In reference \cite{PearceV04} their authors integrate function symbols in Equilibrium Logic. 
NASP embedded with function symbols is viewed under the general theory of stable models in \cite{FerrarisLL11}.
The authors of \cite{WangYZ13} perform a somehow complete revision 
of traditional,  widely studied  ASP concepts by generalizing them to 
a NASP  with functions  framework  and they establish that answer sets of  such an NASP program
can be characterized by the models of its completion and loop formulas.

\vspace{.2cm}
{\bf Non-NNF expressions.} In the seminal work on NASP \cite{LifschitzTT99},
negation by default was allowed to be freely nested,
whereas the scope of the classical negation 
connective was  atomic. This   limitation has been recently overcome
  in \cite{AguadoCF0PV19} wherein  
  the scope of classical negation includes general expressions. 
The authors provide a set of transformation rules  
that   preserve strong-equivalence and
lead   to disjunctive ASP programs.   
Besides,     the principles
of the  Here-and-There (HT)
and Equilibrium Logic \cite{Pearce96} are extended to  provide a complete new characterization of 
classical negation within NASP.

\vspace{.2cm}
{\bf Austere ASP.} The work in \cite{Fandinno21} interestingly 
shows that all  unnested and nested programs 
can be transformed into an unnested program   having an extremely simple structure,
called austere form of ASP.
Indeed, such simple programs are constituted  by only a set of 
choice, implication and constraint rules. It 
is advocated that such a form represents  a kind of normal form for ASP
and is of especial interest for ASP methodology learners.


\vspace{.2cm}
{\bf Nested Regular Many-Valued.} 
Horn expressions  were proposed in 
\cite{Imaz2023b,  Imaz2023a} under the name of Horn non-clausal formulas 
in the context of  regular many-valued logic \cite{Hahnle01a,  Hahnle03} 
and possibilistic logic \cite{ DuboisP94, DuboisP14}.
  The satisfiability of regular many-valued Horn expressions  
 was proven to be decidable in polynomial time  \cite{Imaz2023b}.
 Tractable  classes had existed in many-valued logic e.g.  \cite{Hahnle01a,  Hahnle03}, 
  however, all of them were clausal,
  and so, the class of regular Horn expressions    turned out
to be the solely  non-clausal many-valued  polynomial class.

\vspace{.2cm}
{\bf Nested Horn  Possibilistc   Bases.} 
 The work of \cite{Imaz2023a}  aims at
 determining  the inconsistency  degree 
  of Horn non-clausal possibilistic bases, which were newly introduced in such a reference.  
A correct   logical calculus is proposed and  proven that 
it allows for  the polynomial computation  of the inconsistency degree of such 
 Horn non-clausal possibilistic bases. This latter feature
entitles the   Horn non-clausal possibilistic bases as being
the unique class to be polynomial
in non-clausal possibilistic reasoning,  
which  generalizes other outcomes
in clausal possibilistic reasoning, see e.g. \cite{Lang00}.

\subsection{Discussion}

 From the above discussed related work, one can derive that
  transforming a nested program into a disjunctive ASP program,   generally 
containing also negation by default  in the head, has been contemplated up to now
as the unique possible fashion 
   to compute the answer sets of a  nested program.
 Such a  transformation
was initially proposed in the seminal work     \cite{LifschitzTT99},
    and  indeed, since then it
    has been a repeatedly resorted means  as  the 
great number of published variants of translations witnesses. 

\vspace{.1cm}
Translating has been  employed not only in classical NASP
but also in the solution of possibilistic nested programs 
\cite{NievesL12, NievesL15}.
All state-of-the-art translations are
 based on the introduction of fresh atoms, because, otherwise, the existence
 of polynomial-size   translations was proven not to be guaranteed \cite{CabalarF07}.
  Unfortunately, 
this kind of translation was discussed in  
Subsection \ref{subsec:RESEARCH-FRAME}  to be extremely inconvenient
towards achieving overall system efficiency on the grounds on
 important disadvantages  pointed out there.

\vspace{.1cm}
  Therefore, the  proposal  presented in this research arises as a new 
  promising and innovative manner to compute answer sets of nested programs,
   as it permits to treat them
  in their original nested form and so allows for getting rid of translators.
  Since, one can proceed with NASP by using the same methodology as with classical ASP,
    the presented work is a starting point to   develop
    effective ways to computing nested programs by mimicking 
  existing techniques developed so far within the traditional ASP.

\section{Future Work} \label{sec:future-work}

Possible future research lines on NASP are many and heterogeneous and they encompass 
the generalization to the NASP environment of any concept, technique, analysis and
study of any type and nature, approached  within ASP. Additionally,
 new proper NASP lines,   lacking any meaning within ASP, can be introduced. 
 Among the possible future research lines, we just point out some  of them representing 
 the existing variety:


\vspace{.2cm}
{\bf  Disjunctive NASP.} In Subsection \ref{subsec:SPLITING},  
we have tackled the decomposition of 
DNP programs into a set of NNPs, 
but we did it  informally, just through  specific examples, our focus being 
to show that this operation
is now an available and realistic
option to efficiently compute    disjunctive NASP (DNP) programs. 
In future work, we will construct 
a formal methodology for performing the aforesaid decomposition  and will analyze  
the involved logical and computational issues, 
comparing them with those   arisen in ASP. 

\vspace{.2cm}
{\bf Beyond NNF.} We have  
suggested that nested programs that are not in NNF form can benefit from the 
translation to NNF given in \cite{AguadoCF0PV19}.
As the latter has been devised
for general nested programs, we think that some extra simplifications
may exist   for the subclass of  programs whose corresponding NNF form is
an NNP program. In other words, dealing with sub-classes 
of non-NNF programs, and not with the whole class of them,
 makes it possible to simplify   transformations.
Finding them  is envisaged as future work.

\vspace{.2cm}
{\bf Programs with Variables.} We have not considered   variables
in programs, and yet, their expressive power is  really needed
in practice. 
In such programs, the grounding process
is of course 
a  pivotal issue. To the best of our knowledge, no proposal
for grounding nested programs exists in the literature. Lifting  some of the state-of-the-art
 grounding ASP tecniques  \cite{kaminski_schaub_2022} 
to the NASP level arises as a future worthwhile challenge. 

\vspace{.2cm}
{\bf NASP with Functions.} Computing ASP with function symbols
is a periodical concern in the literature that  yields regular  findings of new  insights  e.g. 
 \cite{ Cabalar11,BartholomewL19}. Nonetheless,
integrating functions within the   more general NASP context  
 belongs to a distinct landscape,
taking into account that, as far as  we know, this logical framework 
has been seldom studied, and concretely, is done only by the authors in reference \cite{PearceV04,FerrarisLL11,WangYZ13}
(see related work). Our research effort  will aim at
finding classes  of NNP programs with one function symbol where decidability
is warranted and high standards of efficiency are achievable.


\vspace{.2cm}
{\bf Tight NNP Programs.} Within unnested ASP, the completion semantics 
due to K. Clark \cite{Clark77} and the answer set semantics \cite{GelfondL88} 
coincide for the so-called   tight programs \cite{Lifschitz96}  as was proven
by a theorem by F. Fages \cite{Fages94}. This result
has been generalized  \cite{ErdemL01,ErdemL03}  
to the so-called non-disjunctive nested programs, i.e.
to those admitting  nested expression only in the rule bodies.
One of our future objectives will be firstly the definition of the notion of 
"tightness" applied to our NNP programs, which by generalizing previous studies  also admit 
nested expressions  in rule heads, and secondly, the upgrading from normal ASP
 to normal NASP  of the aforementioned theorem by F. Fages  \cite{Fages94}.

\vspace{.2cm}
{\bf Splitting  NASP.} Towards computing an elaborated program, a possible
solving tactic consists in decomposing the program into smaller and simpler ones and
separately computing the solutions of the latter programs, which turns out to be
an easier task \cite{FerrarisL05b, Lifschitz96}. The global 
solution is often quite straightforwardly recovered  from the partial solutions.
 So far as we know, the unique solution proposed
for solving a nested program  consists in translating it into an unnested
one, whereas the principle followed in this paper is compatible
with the aforementioned tactic consisting in fragmenting programs into smaller programs. Thus,  
we plan studying how to split an  NNP program into several smaller sub-programs 
probably by using techniques  
inspired by those existing within ASP.

\vspace{.2cm}
{\bf Constraints.} Unnested programs with constraints e.g. \cite{CabalarKOS16} is a field of ASP 
that has   been  abundantly studied theoretically and practically since 
the  seminal works \cite{NiemelaSimons00,SoininenNTS01}. Moreover, 
 as we can deduce from the related work section,
 theoretical aspects underlying the semantics    
 of nested programs  incorporating  constraints have been considerably developed
 e.g. \cite{CabalarFSS19,CabalarFSW20},
whereas the practical side of solving  such programs has not received much  attention.
We plan to integrate  constraints in the NNP programs presented here. We will
start with simple ones, more specifically, with constraints  $x \odot a$
and $x \odot y$, where $\odot \in \{<,>,=,\geq, \leq\} $,  $x,y$ are variables
and $a$ is a constant and all of them range over the same domain. 
An open question is to check whether the   complexity of $\mathcal{NP}$-completeness
of   the propositional case is respected 
if some  simple constraints (as the above ones) are embedded into the language, i.e. whether
there exist NP-complete classes of nested programs incorporating constraints.

\vspace{.2cm}
{\bf Nested SLDNF. } Bottom-up deducing as expressed by the immediate consequence 
operator makes use of  inference such as  nested unit-resolution 
or nested hyper-unit-resolution. However, they are inappropriate 
under certain contexts. On the other hand, the analogous to  
SLDNF for classical ASP  \cite{MarpleBMG12,AriasCCG22}  
 is missing thus far for NASP. It is well known that, within certain contexts,
reasoning in a top-down way  is suitable. For instance,
whenever it is asked  whether a specific   
goal is an answer set  and  this 
only depends on a  small portion of the rules in the knowledge 
base. We intend for future work to formalize
nested SLDNF in a technical manner similar to that followed here for nested 
unit-resolution and nested hyper-unit-resolution. Showing that nested SLDNF calculus,  
 if  particularized to ASP, coincides with the classical SLDNF 
would be  a symptom that the proposed nested SLDNF is indeed the legitimate 
extension of the classical one.

\vspace{.2cm}
{\bf Possibilistic Logic.} This  is the most popularly employed logic
to reason with incomplete knowledge and partial inconsistency \cite{ DuboisP94, DuboisP14}
and its suitable embedding  into ASP has been  argumented by a number of works \cite{NicolasGSL06,NievesOC13,BautersSCV15}.
As mentioned in related work, the possibilisic Horn expressions are 
already characterized and analyzed in \cite{ Imaz2023a}. As future work,
we will study their adaptation  to possibilistic NASP,  or in other words, we will lift
the nested concepts developed here to possibilistic logic.
The obtained possibilistic NASP setting  will  be contrasted
with the existing approaches  \cite{NievesL12,NievesL15}.


\vspace{.2cm}
{\bf Disjunction with Preferences.} In related work, we argued that 
the applications  of ASP notably would increase  if 
the language  was equipped with  connective $\times $ for ordered disjunction \cite{BrewkaNS04}.
Nested programs including it have already been 
 proposed and developed \cite{confalonieri2010nested,ConfalonieriN11}. 
Nevertheless, we think that we can contribute to this area with new 
outcomes based on the fact that NNP programs are now characterized. 
NNP programs  can indeed be of great help in several directions: (i) to separate classical  from preferred disjuncts
in nested expressions in heads;
 (ii) to determine the complexity of the overall computation;  and
  especially, (iii) to open a new research line consisting of searching for answer sets by  splitting  the nested program 
into
 several NNP programs wherein each of the latter  inherits a penalty-value.
 The  value of each NNP program should be  a function
of the  order within the ordered global disjunction occupied by the  disjuncts inside the NNP program.


\section{Discussion and Conclusions}

 Answer set programming (ASP), rooted in logic programming  and non-monotonic reasoning, 
     has become  a  prominent  artificial intelligence discipline thanks to
  its powerful problem-solving methodology and its versatility for knowledge representation.  
  Indeed, on the one hand, ASP  is supported by 
  rich (including first-order 
  and non-classical \cite{CabalarPV18}) logics and  permits the incorporation of
 non-Boolean elements \cite{Lierler22,Lierler23}
  which altogether
  confer on ASP   great    expressiveness 
 and flexibility.  On the other hand, 
 ASP is equipped  with    state-of-the-art reasoning techniques 
whose high  effectiveness has already been shown by
   solving recognized hard commonsense and combinatorial problems.
   Actually,
a non-negligible pragmatical advantage of ASP is that
its theoretical advances   have been gathered   and implemented in
   diverse    ASP solvers, e.g.  \cite{GebserHKLS15,LeonePFEGPS06,AdrianACCDFFLMP18}, that are available online.
 Another relevant virtue    of ASP   is that
 its  main principle  relies on implication rules,
which  aligns ASP  with users' intuition  and, thereby,    connects it 
 to a large amount of  applications  \cite{BryEEFGLLPW07}. 
     To put it another way,  ASP   is connected with  
  multiple   computing problems  dealt by  artificial intelligence, 
and so,  improvements   in ASP are primary for AI research. 
  
  \vspace{.1cm}
  Our presented research  helps  render pragmatically useful  
 the NASP language originally proposed in \cite{LifschitzTT99}.
In pursuit of this goal, we have
characterized a highly interesting class of nested programs,   
 demonstrated  its  logical and computational properties 
  and equipped it with complete inference  mechanisms relying 
  on newly designed nested  logical calculi. 
      Our presented research  
 is    synthesized as follows.

 \vspace{.1cm}
After the proposal of NASP   in 1999 \cite{LifschitzTT99}, 
heavy research  on it has been conducted, allowing for the
discovery of significant insights.
Nevertheless,  this research has been  fundamentally focused
on theoretical aspects, without
     efficient reasoning being a primary goal. 
This was perhaps a consequence  of  
the unawareness of the existence  of  
  normal  NASP programs, which probably
      signified a serious barrier and
  blocked any progress in NASP reasoning.
 Such a claim is partially supported by the undeniable fact that 
  achievements in (unnesting) 
  ASP would hardly have  been  achieved without the
 assistance  of the normal  programs (NPs) and their suitable properties.
  Here, we have filled 
the aforementioned gap by defining the normal nested programs 
(NNPs) and by analyzing them syntactically,  semantically and computationally.

\vspace{.1cm}
  By taking as the starting point 
  the recently presented Horn expressions \cite{Imaz2023b, Imaz2023a}, 
we first originally  defined  its  
 subclass of positive-Horn expressions, and then, postulated 
 that  NNP   rules should be those  having 
   a positive-Horn  expression  in their head and  
  a  (unrestricted) general expression in their body.
  This   proposal for  rule syntax has been inspired by 
  the    ASP policy, since as a matter of fact, 
   positive-Horn clauses  are the cornerstone of NP rules.  
This conceptualization,
mimicking the NP working environment  
has allowed us to lift to and establish within NNP many recognized interesting  properties of NP  
and so it was demonstrated that such properties 
were   exportable to the  general nesting level.
%


\vspace{.1cm}
Subsequently, we  faced the definition of 
 semantical NNP concepts and obtained them
by generalizing existing  NP ones. Thus we introduced the following  concepts 
for NNP programs: 
  answer set, minimal and least  models, model closed  under and  supported by a program,
 immediate consequence operator  and program consistency. 
 Conversely, and with the help 
 of an intermediate language  that we called   normal  flat NASP (NFNP), 
   we   demonstrated that  our NNP concepts are natural expansions 
 of the NP ones. 
 
 \vspace{.1cm}
 More concretely, concerning the last issue,
 we verified that the mathematical functions representing  
   semantical concepts are  the same ones for  
  NFNP and NNP and that     NNP variables integrating
   such functions are just  the general nested version  of    NFNP variables.
  Also,
  we showed that NNP notions  
  coincide with NP ones when particularized to ASP. We think 
  that this along with
    other relationships shown throughout the article 
    suggest that  {\em our proposed 
 NNP  language is indeed the correct generalization  of the  NP one.}

\vspace{.1cm} 
A list of the considerable number of properties 
 enjoyed by   NNP programs proven here is given
  in the Introduction (page 7), which reflect that:  (1) NNP  programs compact 
NP ones   up to an exponential factor;  (2) NNP programs 
retain the nice semantical  properties of NP programs, e.g.
 {\em not}-free NNP programs possess a least model;
and (3) NNP and NP  exhibit the same type of computational complexities. 
On the other hand,  the  difference regarding   
NP and NNP rule heads is noteworthy because 
only single literals are allocatable to the former 
 while    to  the latter so are   expressions with an unlimited number of literals.

\vspace{.1cm}   
Hopefully, the characterization  of the  NNP programs  
and the proofs of its nice properties  
will pave the way  for future research  on NASP  oriented towards reaching 
high  efficiency while exploiting its great expressiveness.  Since
no technical reasons exist to doubt that NASP capabilities for problem-solving
  are comparable and even superior  to those already exhibited by ASP, 
    there exists huge room   for improving NASP through  research, 
    and   we think that the outcomes presented 
    in this research arise as  a fair step forward in such a direction.

    \vspace{.1cm}  
    Another  additional contribution of this research  is the presentation 
    of logical calculi, specifically nested unit-resolution 
    and nested hyper-unit-resolution, permitting the computation of NASP programs
    in their original nested version. Actually, the NASP  literature 
    has contemplated  thus far, 
    as the unique possible implementable way of computing nested programs,
     the translation
    of nested programs into unnesting ones while preserving some suitable equivalence.
    Nonetheless, the consequences of this procedure are far from being negligible and have
    been made explicit at the end of  Subsection \ref{subsec:RESEARCH-FRAME}. Contrarily,
    the introduced logical calculi allow for skipping translators 
    and  the translation task while  taking advantage of the structural
    information embedded in the original nested form of the input program,
     which altogether facilitates the achievement of high    efficiency.
 
 \vspace{.1cm} 
 As claimed at the beginning of future work section, there is indeed a great number
 of possible directions for future research, some of them arisen  by just  mimicking
 what is or has been done in the unnested ASP framework, while others
 being exclusively or properly NASP lines as they lack of any meaning within the ASP setting.
 %
 As just examples, next we mention 
several   research  objectives that are already discussed in Section \ref{sec:future-work}:
  
    \vspace{.1cm} 
  (i) decomposing a disjunctive NASP program into  NNP programs; 
(ii) specifying the stratified and tight NNP programs; 
(iii) facing the grounding process of NNP programs with variables;
(iv) setting up Nested SLDNF-Resolution, undefined thus far;  
(v) syntactically and semantically describing 
  NNP programs in other logics, e.g. possiblistic logic, many valued logic, rough sets, etc;
(vi) coping  with NNP programs enriched with preferences; 
(vii) extending NNP programs with constraints and non-Boolean elements; 
(viii) generalizing Clark's completion to NNP programs; 
and  more.

\vspace{.5cm}
\noindent \hspace{4.5cm} {\Large \bf Appendix: Formal Proofs.}

\vspace{.3cm}
\noindent \hspace{5.cm} \underline{\large \bf  
Proofs from Section \ref{sec:Positive-Horn-Nested-Form}.}

\vspace{.3cm}
\noindent {\bf Theorem  \ref{lem:HyNegD=a}.} Let  
$H \in  \mathcal{H}_{\mathcal{L}}^+$, 
 ${\textcolor{black} {h}} \in \mathcal{L} \cup \{\bot\}$ be  any  positive occurrence     
  in $H$ and  $\Delta(H,h) $
  be its associated sub-expression   specified  by Definition \ref{def:DELTA}.
  We have:  
  $$\forall I, \ H^I \models (h \ \vee \Delta(H,h))^I.$$
 
 \begin{niceproof} Obviously, interpretations $I$ 
 such that  $I \models H^I $ and $I \models \Delta(H,h)^I  $ trivially verify 
   $(h \ \vee \Delta(H,h))^I$. Hence  $  \mbox{let us assume that} \ I \models H^I \ \ \mbox{and} \ \
 I \fmodels  \, {\Delta(H,h)^I}$ and  
  prove that $h \in I$
 by structural induction on the depth $dp(\Delta(H,{\textcolor{black} {h}}))$ 
 of expression ${\Delta(H,h)}$. 
 
 \vspace{.3cm}
$\bullet $ Base Case $dp(\Delta(H,{\textcolor{black} {h}}))=0$:

\vspace{.3cm}
 $\mbox{\qquad If \ } dp(\Delta(H,{\textcolor{black} {h}}))=0 \mbox{\ then \ } 
\Delta(H,{\textcolor{black} {h}})=\bot$

\vspace{.3cm}
  $\qquad \Delta(H,{\textcolor{black} {h}})=\bot \ \mbox{entails} \ H =(h \vee \bot) \wedge H'.$

\vspace{.3cm}
 $\mbox{\qquad  Hence}, \ I \models  H^I  \quad  \mbox{implies} \quad 
 h \in I. $

 \vspace{.3cm}
$\mbox{\qquad  Thus}, \ H^I \models (h \ \vee \Delta(H,h))^I. $

\vspace{.4cm}
$\bullet $ Inductive Step $dp(\Delta(H,{\textcolor{black} {h}})) > 1$: 

\vspace{.35cm}
$- $  {\bf Case} $H=\bigwedge [ H_1 \ldots H_i(h) \ldots H_k ]$.

\vspace{.3cm}
\mbox{\qquad By Definition \ref{def:DELTA},  
\quad ${\Delta(H,h)}=   \Delta(H_i,h).$} 

\vspace{.3cm}
$\mbox{\qquad Thus, \ } I \fmodels  \, \Delta(H,h)^I  \quad \mbox{entails} 
 \quad I \fmodels  \,  \Delta(H_i,h)^I.$

\vspace{.3cm}  
$ \mbox{\qquad By induction,} \quad 
 \ \ I \models H^I \ \ \mbox{and} \ \
 I \fmodels  \, {\Delta(H_i,h)^I} \quad \mbox{entails} \quad h \in I.$
 
 \vspace{.3cm}  
$ \mbox{\qquad Therefore we have:} \quad 
 \ \ I \models H^I,   \ \
 I \fmodels  \, {\Delta(H,h)^I} \quad \mbox{implies} \quad h \in I.$  
 
 \vspace{.3cm} 
  $\mbox{\qquad Therefore}, \ \forall I, \, H^I \models (h \ \vee \Delta(H,h))^I. $
 
\vspace{.5cm}
$- $  {\bf Case}   $H=\bigvee ( H_1 \ldots H_i(h) \ldots H_k )$.

\vspace{.3cm}
 \mbox{\qquad By Definition \ref{def:DELTA}, \quad  ${\Delta(H,h)}= \bigvee ( H_1 \ldots \Delta(H_i,h) \ldots H_k ).$} 
 
 \vspace{.3cm}
 $\mbox{\qquad So,} \quad I \fmodels  \, \Delta(H,h)^I  \quad \mbox{entails} 
 \quad I \fmodels  \,  \Delta(H_i,h)^I$ 
 
  \vspace{.3cm}
 $ \mbox{\qquad By induction,} \quad 
 \ \ I \models H^I \ \ \mbox{and} \ \
 I \fmodels  \, {\Delta(H_i,h)^I} \quad \mbox{entails} \quad h \in I.$
 
 \vspace{.3cm}
 $ \mbox{\qquad Then we have,} \quad 
 \ \ I \models H^I,  \ \
 I \fmodels  \, {\Delta(H,h)^I} \quad \mbox{entails} \quad h \in I.$  
 
  \vspace{.3cm} 
  $\mbox{\qquad Therefore}, \ \forall I,  H^I \models (h \ \vee \Delta(H,h))^I. $
 \end{niceproof}

\vspace{.25cm}
\noindent {\bf Theorem \ref{THE:CLAUSAL-PATTERN}.}  
If $H  \in  \mathcal{H}_{\mathcal{L}}^+$, $ h_1,   
\ldots, h_k \in \mathcal{L} \cup \{\bot\}$ 
are all the  positive occurrences   in $H$ and 
 $ \Delta_1,   \ldots, \Delta_k $ are by Definition \ref{def:DELTA} their associated sub-expressions, then:
$$  {H}  \Leftrightarrow   H_\Delta=
\bigwedge \, [ \, {  (h_1 \vee \,  \Delta_1(H,h_1))}  
\ \ldots \
   { (h_k \vee \,  \Delta_k(H,h_k))} \, ].$$

   \begin{niceproof} Below, we prove both senses of relationship $ \Leftrightarrow$:
   
   \vspace{.25cm}
   {$\bullet \ \forall I, \ {H}^I \models  H^I_\Delta$}. Since 
   $H  \in  \mathcal{H}_{\mathcal{L}}^+$,
   by    Theorem  \ref{lem:HyNegD=a}, 
   we have   $\forall I, \, H^I \models (h_i \vee \Delta_i(h_i,H))^I $ for all positive occurrences $h_i$
   in $H$.   Therefore, we have that:
   $$\forall I, \, {H}^I \models  \bigwedge \, [ \, {  
   (h_1 \vee \,  \Delta_1(h_1,H))^I}  \ \ldots \  \   { (h_k \vee \,  \Delta_k(h_k,H))^I} \, ]=H^I_\Delta .$$
   
   {$\bullet \ \forall I, \, H^I_\Delta \models  H^I$}. Without lost of generality,    
   we can assume that   the input  $ H  $ is a conjunction
   $H=\bigwedge [H_1 \ldots H_k] $ because if $H$ is a disjunction, 
  then  we can consider $H'=\bigwedge [H]$ as input program.   
   From Definition \ref{def:Positive-Horn-Nested},  
 each conjunct $H^i_n$ of a Horn expression $H_n=\bigwedge [H^1_{n} \ldots H^k_{n}] $ with depth $n$  has the form: 
    $$H^i_n=\vee (\varphi_n^- \ \  
     \bigwedge [H^1_{n-1} \ldots H^k_{n-1}] \, ) \ \mbox{with} \ \varphi_n^-\in \mathcal{N}_\mathcal{L}.$$ 
    
    and Horn expressions of depth 1  have the form:  
    $$H_{1}=   \bigwedge    [\, (\vee \ \varphi_{1}^- \ C_1^+)   \ldots (\vee \ \varphi_{k}^- \ C_k^+)\, ] 
    \quad \mbox{so} \quad H^i_{1}=(\vee \ \varphi_{i}^- \ C_i^+)$$
    
   where:   
$$\varphi_{i}^-\in \mathcal{N}_\mathcal{L} , \ {C}_i^+=\wedge [h_{i,1} \ldots h_{i,n_i}] \
  \mbox{and} \   h_{i,j} \in \mathcal{L} \cup \{\bot\} , \ 1 \leq i \leq k, \ 1 \leq j \leq n_i .$$
  
  The proof is by induction on the depth $n$. 
  
  \vspace{.25cm}
  $\bullet $ Base Case ($n=1$):  $H_1 $ has the form given above.
  
  \vspace{.25cm}
   \  By applying Definition \ref{def:DELTA} and Theorem \ref{THE:CLAUSAL-PATTERN} 
  to $H_1$, we obtain that:
  $$H_{\Delta,1}=\bigwedge [\,(h_{1,1} \vee \varphi_{1}^- ) \ldots 
  (h_{1,n_1} \vee \varphi_{1}^- )
  \ldots (h_{k,1} \vee \varphi_{k}^- ) \ldots (h_{k,n_k} \vee \varphi_{k}^- )\,] $$
 

 \vspace{.0cm}
 Therefore, the base case is clearly proven: 
 $$\forall I, \ (H_{\Delta,1})^I \models H_1^I .$$

  \vspace{.15cm}
  $\bullet $ Induction Case ($n$): 
    
    \vspace{.25cm}
    Assume that  a conjunct $H^i_{\Delta,n-1}$ associated with  a conjunct $H_{n-1}^i$  is:
    $$H^i_{\Delta,n-1}=\bigwedge [\,(h_{i,1} \vee \Delta_{i,1} ) \ldots 
    (h_{i,n_i} \vee \Delta_{i,n_i} )\,] $$
    
     and that the induction hypothesis holds for $n-1$, namely: 
  $$\forall  I, \, (H^i_{\Delta,n-1})^I=\bigwedge [\,(h_{i,1} \vee \Delta_{i,1} ) \ldots 
    (h_{i,n_i} \vee \Delta_{i,n_i} )\,]^I   \  \models (H^i_{n-1})^I.$$
    
    Therefore, for the conjunction  $H_{n-1}^{1,k}=\bigwedge [H^1_{n-1} \ldots H^k_{n-1}]$ inside $H^i_n$ we have:   
 $$
  H_{\Delta,n-1}^{1,k}=\bigwedge [\,(h_{1,1} \vee \Delta_{1,1} ) \ldots 
    (h_{1,n_1} \vee \Delta_{1,n_1} ) \ldots \ldots (h_{k,1} \vee \Delta_{k,1} ) \ldots 
    (h_{k,n_k} \vee \Delta_{k,n_k} )\,]
   $$
    
    Since by induction $\forall  I, \, (H^i_{\Delta,n-1})^I  \  \models (H^i_{n-1})^I,$ obviously:
    \begin{equation} \label{eq:Prova}
     \forall  I, \, (H_{\Delta,n-1}^{1,k})^I  \  \models (H_{n-1}^{1,k})^I.
     \end{equation}

   Thus,   by applying Definition \ref{def:DELTA} and 
  Theorem \ref{THE:CLAUSAL-PATTERN}  to: $$H^i_n=\vee (\varphi_n^- \ \  
     \bigwedge [H^1_{n-1} \ldots H^k_{n-1}] \, )= \vee (\varphi_n^- \ \ H_{n-1}^{1,k}),$$
  
  we obtain without difficulty:
  $$H^i_{\Delta,n}=\bigwedge [\,\vee (h_{1,1} \ \varphi^-_n \ \Delta_{1,1})  \ldots 
    \vee (h_{1,n_1} \ \varphi^-_n \ \Delta_{1,n_1}) \ldots \ldots
    \vee (h_{k,n_k} \ \varphi^-_n \  \Delta_{k,n_k} )\,]$$ 
  
  Since, from  (\ref{eq:Prova}), 
  it is immediate that:
  $$\forall  I, \, (H^i_{\Delta,n})^I \models (H^i_{n})^I=\bigvee (\varphi_n^- \ \ H_{n-1}^{1,k} \, )^I . $$  
  
  From this relation and as the input   is $H=\bigwedge [H^1_{n} \ldots H^i_{n} \ldots H^k_{n} ] $, one obtains:  
  $$ \forall I, H_\Delta^I=\bigwedge [H^1_{\Delta,n} \ldots H^i_{\Delta,n} \ldots H^k_{\Delta,n} ]^I \models H^I.$$
   \end{niceproof}

\noindent \hspace{5.5cm} \underline{\large \bf  Proofs from Section \ref{sec:DefiniteNLP-Syntax}.}

\vspace{.3cm}
\noindent {\bf Lemma  \ref{pro:HORN-Exp-HORN-CNF}  } 
 Applying distributivity to any   $ \varphi \in  \mathcal{H}^+_{\mathcal{L}}$
 leads to a positive-Horn  CNF.

\begin{niceproof}  We consider Definition \ref{def:Positive-Horn-Nested}  
 of $ \mathcal{{H}^+_{L}} $. We denote by $cl(\varphi)$ the CNF obtained by repeatedly applying
distributivity to  $\varphi$ and we denote by 
$\mathcal{H}_\mathcal{L}^+$   the  class of  positive-Horn CNFs. 
Thus, the lemma amounts to state, formally:
$$ \varphi \in  \mathcal{H}^+_{\mathcal{L}}, \ \ cl(\varphi) \in \mathcal{H}_\mathcal{L}^+.$$ 
 
 The proof is done by  structural induction  on   
    the depth  $r(\varphi)$    of any $\varphi \in \mathcal{{H}^+_{L}} $ which
    formally it is  defined   as follows:
\[r(\varphi)= \left[
\begin{array}{l l l}

\vspace{.2cm}
  0   &   \varphi \in \mathcal{L} \cup \{\bot\}.\\
  
1+max\,[r(\varphi_1), \ldots, \,r(\varphi_k)]    &  
\varphi=  \odot \langle  \varphi_1  \ldots    \varphi_k \rangle. \\

\end{array} \right. \]

\vspace{.15cm}
 $\bullet$     {\em Case Base:} 
By definition of  $  \mathcal{H}^+_{\mathcal{L}} $,  if $ r(\varphi)= 0 $  then
  $\varphi \in \mathcal{L} \cup \{\bot\}$
  and so    $ cl(\varphi) \in \mathcal{H}_\mathcal{L}^+$.
  
 \vspace{.3cm} 
 $\bullet$     {\em Induction Step:}  
 Assume  that  $\forall \varphi \in  \mathcal{{H}^+_{L}} $ and     $ r(\varphi) \leq n $, 
   we have $ cl(\varphi) \in \mathcal{H}_\mathcal{L}^+$ 
and let us prove  the lemma for $ n+1 $. 
From Definition \ref{def:Positive-Horn-Nested}, lines (2) and (3) below are distinguishable:

\vspace{.35cm} 
$\bullet$   (2)   $\varphi=\wedge \, [  \, \varphi_1   \ldots \varphi_{i}  \ldots  \varphi_k]
 \in  \mathcal{{H}^+_{L}}$.

  \vspace{.3cm} 
  $-$  By definition of $r(\varphi)$,   $ \ r(\varphi_i) \leq n$ 
for  $1 \leq i \leq k$.

\vspace{.3cm}
  $-$  By induction, \  $   cl(\varphi_i)=\mathrm{H}_i \in \mathcal{H}_\mathcal{L}^+$ for  $1 \leq i \leq k$.

\vspace{.3cm}
   $-$ It is obvious that \ 
%
  $  cl(\varphi) = \wedge  [  cl(\varphi_1) \  
 \ldots  \ cl(\varphi_{i})   \ldots  cl(\varphi_k) \,].$
    
 \vspace{.3cm}
  $-$  Therefore, \ 
%
$cl(\varphi) =   \wedge  [ \mathrm{H}_1 
\ldots \mathrm{H}_{i}  \ldots \mathrm{H}_k]= \mathrm{H} \in \mathcal{H}_\mathcal{L}^+$.

\vspace{.4cm}
 $\bullet$   (3)  
 $\varphi= \vee(  \varphi_1  \ldots \varphi_{i} \ldots \varphi_j \ldots \varphi_k) 
 \in \mathcal{{H}^+_{L}}$. 
 
 \vspace{.3cm} 
 $-$ By definition of  $  \mathcal{H}^+_{\mathcal{L}} $, 
 $\varphi_i \in \mathcal{H}^+_{\mathcal{L}}$ and
$\vee (  \varphi_1 \ldots \varphi_{i-1} \, \varphi_{i+1} \ldots \varphi_{k})=\varphi^- \in  
\mathcal{N_L}$.

 \vspace{.3cm} 
 $-$ By  definition of $r(\varphi)$, \
%
   \,$r(\varphi_i), r(\varphi^-) \leq n$.

 \vspace{.3cm}
\vspace{.0cm} $-$ To obtain $ cl(\varphi) $, one  needs firstly $cl(\varphi^-)$
and $cl(\varphi_i)$, and so

 \vspace{.2cm}
\hspace{1.cm}
 $(i) \ \ cl(\varphi) = cl(\,\vee(\varphi^-  \  \varphi_i) \,)= 
 cl(\,\vee (cl(\varphi^-)  \ \ cl(\varphi_i)\,)\,).$

\vspace{.3cm}
 $-$  By definition of $\varphi^- \in \mathcal{N_L}$, 
 
 \vspace{.2cm}
\hspace{0.8cm} $(ii) \ \  cl(\varphi^-)=[\wedge \ D^-_1   \ldots  \,D^-_m]$, where          
 the $D^-_i$'s are negative clauses.

\vspace{.3cm}
  $-$  By induction hypothesis, 

\vspace{.2cm}
\hspace{.8cm} $(iii) \ \ cl(\varphi_i)=\mathrm{H}=\wedge [ Ch_1 \  \ldots  Ch_n ]$,  
  where the $h_i$'s are positive-Horn clauses.

\vspace{.3cm}
 $-$ By   ($i$) to ($iii$), 
 
 \vspace{.2cm}
\hspace{.8cm}  $ cl(\varphi) = cl(\,\vee ( \ \wedge[  D^-_1 \   \, \ldots  \,D^-_m] 
 \ \ \wedge [Ch_1 \ \ldots  \,Ch_n \, ]\,)\,).$

\vspace{.3cm}
 $-$ Applying $\vee\//\wedge$ distributivity to $cl(\varphi)$ and denoting 
 $C_{i}=\vee (  D^-_1 \ Ch_{i} \,)$, 

\vspace{.2cm}
\hspace{1.cm} $cl(\varphi) = 
 cl(\,\wedge [ \ \wedge [  C_1   \ldots C_i \ldots C_n] \ \  
    \vee (  \, \wedge [  \ D^-_2 \ldots   \,D^-_m \, ] 
    \  \ \wedge  [ \ Ch_1 \ \ldots  \,Ch_n \, ] \,)  \,] \ ).$

\vspace{.3cm}
 $-$ Since the  $C_i=(\vee \ D^-_1 \ h_{i} \,)$'s are  positive-Horn clauses,  

\vspace{.2cm} 
\hspace{1.cm} $cl(\varphi) = cl (\,[\wedge \ \ \mathrm{H}_1 \ \ 
(\vee \ \ [\wedge \ D^-_2 \ldots D^-_{m-1} \,D^-_m \, ]   
\ \ [\wedge \ Ch_1 \ \ldots  \,Ch_n \, ]\, ) \,] \ ).$

\vspace{.3cm}
 $-$ For any $  j < m$ we have,
 
\vspace{.2cm} 
\hspace{1.cm} $cl(\varphi) = cl( \ [\wedge \ \mathrm{H}_1 \ \ldots   \mathrm{H}_j   \  \ (\vee \ \ [\wedge \ D^-_{j+1} \ldots \,D^-_m]  \ \  
[\wedge \ Ch_1 \ \ldots  \,Ch_n \, ] \,) \, ] \ ).$

\vspace{.3cm}
 $-$ For $j = m$ and recalling that $ [\wedge \ Ch_1 \  \ldots  Ch_n ]=cl(\varphi_i)=\mathrm{H}$,

\vspace{.2cm} 
\hspace{1.cm} 
   $cl(\varphi) = [\wedge \ \mathrm{H}_1 \  \ldots \mathrm{H}_{m-1} \,\mathrm{H}_m \     
\mathrm{H} \,]=\mathrm{H}' \in \mathcal{H}^+_{\mathcal{L}}.$

\vspace{.3cm}
 $-$  Hence   $cl(\varphi)   \in  \mathcal{H}^+_{\mathcal{L}}$.
\end{niceproof}

\vspace{.15cm}
\noindent{\bf Proposition  \ref{prop:NN-1=NN}. }  Let  $P$ be an NNP program. 
We have $NN_1(P) \Leftrightarrow NN(P) $.

\begin{niceproof} If $H \leftarrow B$ is an NNP rule, then by Theorem \ref{THE:CLAUSAL-PATTERN},
 $H_\Delta \Leftrightarrow H $, and so: 
 \begin{equation}  \label{EQ:NN=NN-1}
(H \leftarrow B) \Leftrightarrow (H_\Delta \leftarrow B).  
\end{equation}

On the one hand, by Proposition \ref{pro:STRONG-PRESERVED}, line 3,
  $$(H_\Delta \leftarrow B) \Leftrightarrow \{h \vee \Delta   \leftarrow B \parallel h \vee \Delta   \in H_\Delta\}.$$
  
  From this right set of rules and by applying   Definition \ref{def:DEFINITION-of-OVERLINE-F},
one deduces: 
$$H_\Delta \leftarrow B \Leftrightarrow
\{h \leftarrow C \ \vert \vert \ \  h \vee \Delta  \in H_\Delta, \ \  
C \in dnf(B \wedge {\Delta}_B)  \   \}$$

Then by Definition \ref{pro:ALTERNTIVE-FP-PROGRAM},
\begin{equation} \label{eq:NN0NN-2}
H_\Delta \leftarrow B \Leftrightarrow \mbox{N}(H_\Delta \leftarrow B)=NN(H_\Delta \leftarrow B)
\end{equation}

On the other hand, obviously we have:
$$H \leftarrow B \Leftrightarrow cnf(H) \leftarrow dnf(B) $$

and by claim (6), Proposition    \ref{Prop:EQUIV-NFNP-ND-2}, 
$$H \leftarrow B \Leftrightarrow FN(cnf(H) \leftarrow dnf(B))=NN_1(H \leftarrow B). $$

Therefore, by  (\ref{EQ:NN=NN-1}), (\ref{eq:NN0NN-2}) and the last statement, we have 
$\forall r \in P, \  NN_1(r) \Leftrightarrow NN(r) $, which entails Proposition  \ref{prop:NN-1=NN}. 
\end{niceproof}


\vspace{.4cm}
\noindent   \hspace{5.5cm} \underline{\large \bf Proofs from Section \ref{sec:DefiniteNLP-Semantics}.}

\vspace{.3cm}
 \noindent {\bf Proposition \ref{Prop:MODEl_NNP-CLOSED}.} Any model of an NNP program $P$ is closed under $P$. 
 
 \begin{niceproof} We take $H \leftarrow B \in P$.
  By Definition  \ref{cor:NNP-CLOSED}, $I$  is  closed by $P$ when, 
 $$     \forall  \, {(h \vee \Delta)} \in H_\Delta: \quad  h \in I \quad 
\mbox{whenever} \quad   
I \models B 
\quad \mbox{{and}} 
\quad I  \fmodels \, \Delta.  $$

\qquad By Definition  \ref{cor:NNP-CLOSED-SUPPORTED}, $I$ is a model of $P$ when:
$$ I  \not \models    B  \quad 
\mbox{\underline {or}} \quad \forall \,  {(h \vee  \Delta)} \in H_\Delta:
\quad  I \not \fmodels  \,  \Delta   \quad \mbox{\bf \underline {or}} \quad 
 h \in I.$$ 
 
 For each $(h \vee  \Delta) \in H_\Delta$, either $h \in I$  or  $h \notin I$. 
 
 \vspace{.25cm}
 (1) $h \in I$:  
 $I$ can be a model of $ H \leftarrow B$ and $I$ can be closed under $ H \leftarrow B$.
 
 \vspace{.25cm}
 (2) $h \notin I$:   $I$ can be a model of $ H \leftarrow B$
 if $ I  \not \models    B $ or $ I \not \fmodels  \,  \Delta  $. 
 
 \vspace{.2cm}
 \qquad \qquad \ \ If $ I  \not \models    B $ or $ I \not \fmodels  \,  \Delta  $ then $I$ 
 can be  closed under
 $ H \leftarrow B$.
 
 \vspace{.2cm}
 By (1) and (2) $I$ is a model of  $H \leftarrow B$  iff 
 $I$ is closed under $H \leftarrow B$.
 
 \vspace{.2cm}
 Since $H \Leftarrow B$ is any rule in $P$, the proposition holds.
 \end{niceproof}
 
 \noindent 
 {\bf Proposition \ref{Prop:MINIMAL_MODEL_SUPPORTED_NNP}.} A  minimal model of an NNP program $P$ is supported by $P$.

\begin{niceproof} If $I$ is a minimal model of $P$ then 
if $\forall \ell \in I$: 
\begin{equation}   \label{EQ:SEPTIMA}
\quad \exists (H \leftarrow B) \in P, \quad \exists 
\,{ (h \vee \Delta)} \in H_\Delta: \qquad \mbox{} h =\ell, \quad 
 I \models  B 
  \quad \mbox{{and}} \quad
I  \fmodels \,\Delta.
\end{equation}  

One can check that by Definition \ref{cor:NNP-SUPPORTED},   $I$ is supported by $P$.

\vspace{.2cm}
Assume $ \ell \in I$   does not verify (\ref{EQ:SEPTIMA})
and prove that $I-\ell$ is a model.

\vspace{.2cm} Then, if $ \ell \in I$ not verifying (7) then: 
$$\quad \forall (H \leftarrow B) \in P, \quad \forall 
\,{ (h \vee \Delta)} \in H_\Delta: \qquad \mbox{} \ell \neq h \in I, \quad 
 I \not \models  B 
  \quad \mbox{{or}} \quad
I \not  \fmodels \,\Delta.$$

Then, we have the following entailment:

\vspace{.2cm}
\qquad If $I \not \models  B $ trivially $I-\ell \not \models  B .$ 

\vspace{.2cm}
\qquad If $I \not  \fmodels \,\Delta$  then  $I-\ell \not  \fmodels \,\Delta.$

\vspace{.2cm}
\qquad If $\ell \neq h \in I$ then $\ell \neq h \in I - \ell$.

\vspace{.2cm}
Hence if $ \ell \neq h \in I - \ell, \quad 
 I - \ell \not \models  B 
  \quad \mbox{{or}} \quad
I- \ell \not  \fmodels \,\Delta$, \ then 

\vspace{.2cm}
by Definition \ref{cor:NNP-CLOSED-SUPPORTED}, $I - \ell$ is a model of  $P$.

\vspace{.2cm}
Hence $I$ is  not minimal. 
\end{niceproof}

\vspace{.3cm}
\noindent {\bf Proposition \ref{Prop:EQUIV-NNP-ND-1}.} 
Let  $P$ be an NNP  program and $I$ be an interpretation. We have:
  \begin{itemize}
 
 \item [{\em (1)}] 
  $P$ is head-consistent iff 
 so is  $NN(P) $.
 
 \vspace{-.2cm}
 \item [{\em (2)}]
 $I$ is closed under $P$ iff 
  so is under  $NN(P) $.

 \vspace{-.2cm}
 \item [{\em (3)}] $I$ is supported by $P$ iff 
  so is by  $NN(P) $.
 
 \end{itemize} 
 
 \begin{niceproof} We prove the next previous statements:
 
 \vspace*{.2cm}
 (1) Definitions
  \ref{def:FNASP-EXTENDED-rule},
 \ref{pro:ALTERNTIVE-FP-PROGRAM},  \ref{cor:SECT-APPROACH:pos}  
 and  \ref{def:TR1} establish the relationship between $P$ and $NN(P) $, 
  and according to them, programs $P$ and  $NN(P) $ have the same head elements 
 $h \in \mathcal{L} \cup \{\bot\}$. By the latter claim
 and by Definitions   \ref{def:FNASP-EXTENDED-rule} and 
 \ref{def:syntac-sug-NP} of
 head-consistency of programs $P$ and $NN(P)$, 
 we conclude that  statement (1) holds.
 
 \vspace{.2cm}
 (2) Let us consider an NNP rule 
 $H \leftarrow B \in P$ and its associated set of NP rules
 $ N( H \leftarrow B)=
\{h \leftarrow C \ \vert \vert \ \  h \vee \Delta  \in H_\Delta, \ \  
C \in dnf(B \wedge {\Delta}_B)  \,   \}$.
 Assume that interpretation $I$ is closed under program $P$. 
 If $I$ is closed under $H \leftarrow B$ then by Definition  \ref{cor:NNP-CLOSED}:
    $$     \forall  \, {(h \vee \Delta)} \in H_\Delta: \quad  h \in I \quad 
\mbox{whenever} \quad   
I \models B 
\quad \mbox{{and}} 
\quad I  \fmodels \, \Delta.  $$

By Definition \ref{def:DEFINITION-of-OVERLINE-F}: \quad 
$ \forall  \, {(h \vee \Delta)} \in H_\Delta: \quad  h \in I \quad 
\mbox{whenever} \quad   
I \models B \wedge {\Delta}_B.$
$$ \forall  \, {(h \vee \Delta)} \in H_\Delta: 
\quad  h \in I \quad 
\mbox{whenever} \quad   
I \models dnf(B \wedge {\Delta}_B).$$
$$ \forall  \, {(h \vee \Delta)} \in H_\Delta, \, \exists  C \in dnf(B \wedge {\Delta}_B):  
\quad h \in I \quad 
\mbox{whenever} \quad   \ I \models C.$$

The latter claim entails that $I$ is closed under  $ N( H \leftarrow B)$.

\vspace{.1cm}
Conversely, it is analogously proven that: 

\vspace{.08cm}
\qquad $I$ is closed under  $ N( H \leftarrow B)$ if $I$ is 
closed under  $H \leftarrow B$.

\vspace{.1cm}
Therefore $I$ is closed under $P$ 
iff so $I$ is closed is under $NN(P)$.

\vspace{.2cm}
(3) An interpretation $I$ is supported by  an NNP rule 
 $H \leftarrow B$ and if and only if it is  supported by any rule in the set of NP rules
 $ N( H \leftarrow B)=
\{h \leftarrow C \ \vert \vert \ \  h \vee \Delta  \in H_\Delta, \ \  
C \in dnf(B \wedge {\Delta}_B)  \,   \}$.  
If  $\ell \in I $ is supported by $H \leftarrow B$ then, by Definition \ref{cor:NNP-SUPPORTED}, we have:
$$    \exists 
\,{ (h \vee \Delta)} \in H_\Delta: \qquad \mbox{} h =\ell, \quad 
 I \models  B 
  \quad \mbox{{and}} \quad
I  \fmodels \,\Delta.$$ 

By Definition \ref{def:DEFINITION-of-OVERLINE-F}: \quad $    \exists 
\,{ (h \vee \Delta)} \in H_\Delta: \qquad \mbox{} h =\ell, \quad 
 I \models  B \wedge {\Delta}_B.$
$$    \exists 
\,{ (h \vee \Delta)} \in H_\Delta: \qquad \mbox{} h =\ell, \quad 
 I \models  dnf(B \wedge {\Delta}_B).$$ 
 $$    \exists 
\,{ (h \vee \Delta)} \in H_\Delta, \exists  C \in dnf(B \wedge {\Delta}_B):
 \qquad \mbox{} h =\ell, \quad 
 I \models  C.$$ 
 
 which entails that $I$ is supported by any rule in $ N( H \leftarrow B)$.
 
 \vspace{.1cm}
 Conversely  it is analogously proven that:   
 
 \vspace{.08cm}
 \qquad $I$ is supported by  $ N( H \leftarrow B)$ if $I$ is 
supported by   $H \leftarrow B$.

\vspace{.1cm}
Therefore $I$ is supported by  $P$ 
iff so is supported by $NN(P)$.
\end{niceproof}

\noindent {\bf Proposition \ref{Prop:EQUIV-NNP-ND-2}.}  Let   $P$ be any   not-free, 
 constraint-free and   head-consistent NNP program  and let us denote by $Cn(P)$
     the smallest interpretation $I$  closed under $P$. Then:
 \begin{itemize}
 
 \item [{\em (4)}] $ Cn(P)=Cn(NN(P)) $.
  
 \vspace{-.2cm}
 \item [{\em (5)}] $NT_P(I) = T_{NN(P)}(I)$.
 
 \vspace{-.2cm}
 \item [{\em (6)}] $P \Leftrightarrow  NN(P)$.
 
 \end{itemize}

 \begin{niceproof} We prove below the previous statements:

 \vspace{.2cm}
 (4) The proof of this claim follows immediately from previous claim (2) since the same 
 
 set  of interpretations  is closed under both programs $P$ and $NN(P)$.
 
  \vspace{.2cm}
 (5) We have that by Definition \ref{def:OPERATOR:NT}:
 $$NT_P (X)=\{h \ \vert \vert \ H \leftarrow B \in P, 
\ \,  (h  \vee   \Delta) \in H_\Delta,  \ \,  
 I   \models    B,   \ \,  I   \fmodels \,\Delta  \}.$$
 
 \qquad By Definition \ref{def:DEFINITION-of-OVERLINE-F}:
 $$ NT_P (I)=\{h \ \vert \vert \ H \leftarrow B \in P, 
\ \,  (h  \vee   \Delta) \in H_\Delta,  \ \,  
 I   \models    B \wedge \overline{\Delta } \}.$$
 
\qquad Then  we have:
 $$ NT_P (I)=\{h \ \vert \vert \ H \leftarrow B \in P, 
\ \,  (h  \vee   \Delta) \in H_\Delta,  \ \,  C \in  dnf( B \wedge \overline{\Delta }), \
 I   \models  C \}.$$
 
 \qquad By Definition   \ref{pro:ALTERNTIVE-FP-PROGRAM},     \quad 
 $$ NT_P (I)=\{h \ \vert \vert \ H \leftarrow B \in P, \ (h \leftarrow C) \in  NN(H \leftarrow B), \
 I   \models  C \}.$$
 
 \qquad By Definition \ref{def:CONSEQ-IMM-OPE},  \quad 
 $$NT_P (I)= T_{NN(P)} (I).$$
 
  \vspace{.2cm}
 (6) Let us take 
   $(H \leftarrow B) \in P$ and prove: 
   $$H \leftarrow B \Leftrightarrow N(H \leftarrow B)$$
   
   \qquad where $ N(H \leftarrow B)$ is specified in Definition \ref{pro:ALTERNTIVE-FP-PROGRAM}.
   
   \vspace{.25cm}
 \qquad   By Theorem \ref{THE:CLAUSAL-PATTERN}, \quad $H \leftarrow B \Leftrightarrow H_\Delta \leftarrow B.$

  \vspace{.25	cm}
   \qquad By expanding $H_\Delta$, \ \,$H \leftarrow B \Leftrightarrow \{(h  \vee   \Delta)  \leftarrow B \parallel (h  \vee   \Delta) \in H_\Delta \}$

  \vspace{.2cm}
  \qquad By  Definition \ref{def:DEFINITION-of-OVERLINE-F}, \quad $H \leftarrow B \Leftrightarrow \{h    \leftarrow B \wedge   \overline{\Delta} \parallel (h  \vee   \Delta) \in H_\Delta \}$
  
  \vspace{.2cm}
  \qquad \qquad \qquad \qquad \qquad \quad \,$H \leftarrow B \Leftrightarrow \{h    \leftarrow cnf( B \wedge   \overline{\Delta}) \parallel (h  \vee   \Delta) \in H_\Delta \}$
   
   \vspace{.2cm}
  \qquad \qquad \qquad \qquad \qquad \quad \,$H \leftarrow B \Leftrightarrow \{h    \leftarrow  C  \parallel (h  \vee   \Delta) \in H_\Delta, \ 
  C \in dnf( B \wedge   \overline{\Delta}) \}.$
  
  \vspace{.2cm}
  \qquad By Definition  \ref{pro:ALTERNTIVE-FP-PROGRAM},  \ \ $H \leftarrow B \Leftrightarrow N(H \leftarrow B ).$
  
   \vspace{.2cm}
  \qquad Therefore, \qquad \qquad \ $P \Leftrightarrow NN(P)$.
 \end{niceproof}

\vspace{.3cm}
\noindent {\bf Theorem \ref{TH:Tarski55-Endem76-NFNP}.}
 For any     not-free, constraint-free and  
  head-consistent  NNP  program $ P $: 
  $$LM(P)=\bigcup_{0 \leq i} \,NT^i_P=\mbox{lfp}(NT_P) .$$

  \begin{niceproof} 
  By Definition  \ref{pro:ALTERNTIVE-FP-PROGRAM}, $NN(P)$ is clearly NP, so by
    applying Theorem \ref{TH:Tarski55-Endem76} to $NN(P)$, 
  $$LM(NN(P))=\bigcup_{0 \leq i} \,T^i_{NN(P)}=\mbox{lfp}(T_{NN(P)}) .$$
  
   By statement (6), 
   
   \hspace{3.cm}$P \Leftrightarrow NN(P)$ and
  hence $LM(P)=LM(NN(P))$.
  
  \vspace{.2cm} Hence, we have:
  $$LM(P)=\bigcup_{0 \leq i} \,T^i_{NN(P)}=\mbox{lfp}(T_{NN(P)}) .$$
  
  \vspace{.2cm}
  By statement (5), \qquad \quad $NT_P(I) = T_{NN(P)}(I)$.
  
   \vspace{.2cm}
  Thus, we obtain:
   $$LM(P)=\bigcup_{0 \leq i} \,NT^i_P=\mbox{lfp}(NT_P) .$$ 
  \end{niceproof}

\noindent{\bf Proposition \ref{prop:BASIC-Properties-not-free-NFNP}.}  
If $  P$ be a {not-}free and constraint-free  NNP$_\mathcal{A} $  program, then $P$, $LM(P)$ and
  $ Cn(P) $ verify the properties (a) to (c)
   in Proposition \ref{prop:BASIC-Properties-not-free-NP}.

\begin{niceproof} (Sketch) Below we sketch the proofs of the claims:

\vspace{.3cm}
(a) Trivial, since $P$  because all  head elements of $P$ are atoms.

\vspace{.3cm}
(b) $Cn(P)$ is consistent by the same reason  in (a). 

\vspace{.2cm}
\qquad Below we show that $Cn(P)$ is supported by $P$. 

\vspace{.2cm}
\qquad  By Proposition \ref{Prop:MODEl_NNP-CLOSED}:

\vspace{.2cm}
\qquad \qquad \qquad $I$ is a model of $P$ iff $I$ is closed by $P$.

\vspace{.25cm}
\qquad Therefore, $LM(P)=Cn(P)$.

\vspace{.2cm}
\qquad By  {Theorem \ref{TH:Tarski55-Endem76-NFNP}}, \quad $Cn(P)=\bigcup_{0 \leq i} \,NT^i_P$ and $NT^0_P=NT_P(\emptyset)$. 

\vspace{.25cm}
\qquad Case Base.  \quad Trivially, $\forall \ell \in NT_P(\emptyset), \ \ell  \mbox{\ is supported by \ }  P.$

\vspace{.25cm}
\qquad Induction Step.  By  definition of $NT_P(I)$:
  $$\mbox{Clearly, \ \ }\forall \ell \in NT_P(NT^{i}_P) \mbox{ is  supported by $P$ if so are }
  \forall \ell' \in NT^{i}_P.$$

\vspace{.1cm}
(c) By Proposition \ref{Prop:EQUIV-NNP-ND-2}, $P \Leftrightarrow NN(P)$.

\vspace{.25cm}
\qquad  By Definition  \ref{pro:ALTERNTIVE-FP-PROGRAM}, $NN(P)$ is clearly NP.

\vspace{.25cm}
\qquad Hence $LM(NN(P))=LM(P)$ is the answer set of $P$.
\end{niceproof}

 \noindent {\bf  Proposition \ref{prop:BASIC-Propertiesprograms-not-free-FXNP}.}  
	If $  P$ is a {not-}free, constraint-free and head-consistent 
NNP$_\mathcal{L} $ program, then $P$, $LM(P)$ and $Cn(P) $ verify
the properties (b) and (c) specified in Proposition \ref{prop:BASIC-Properties-not-free-NP}.

\begin{niceproof} The proofs are  as those in the proof 
of Proposition \ref{prop:BASIC-Properties-not-free-NFNP}, cases (b) and (c). 
\end{niceproof}

\vspace{.3cm}
\noindent {\bf Proposition \ref{prop:BASIC-Propoerties-ANSWER-SET-NFNP}.} 
Let   P be an  NNP program, S  be  an answer set of P
and  $\Sigma$ be  a set of constraints. Then P, S and $\Sigma$
verify the properties expressed in Proposition \ref{prop:BASIC-Propoerties-ANSWER-SET}.

\begin{niceproof} Below, we sketch the proofs of statements (1) to (4).

\vspace{.2cm}
  (1) If $S$ is an answer set of $P$ then $S$ is also a model of $P^S$.
  
  \vspace{.2cm}
  \quad \ Hence by Proposition \ref{Prop:MODEl_NNP-CLOSED}, $S$ is closed under $P^S$.
  
  \vspace{.2cm}
  \quad \ By definition of $P^S$, if $S$ is closed under $P^S$ then $I$  is closed under $P$.
    
\vspace{.3cm}
  (2) If $S$ is an answer set of $P$, then $S=LM(P^S)$.
  
  \vspace{.2cm}
  \quad \  So any subset of $S$ is not a model of $P^S$.
  
  \vspace{.2cm}
  \quad \ Hence no subset of $S$ is a model $P$, so $S$ is  minimal.
  
  \vspace{.3cm}
  (3) By Definition  \ref{def:ANSWER_SET_NNP}, $S=LM(P^S)$.

  \vspace{.2cm}
  \quad \    By Proposition \ref{Prop:MINIMAL_MODEL_SUPPORTED_NNP}, $S$ is supported by $P^S$.
  
  \vspace{.2cm}
  \quad \ By Definition of $P^S$, if $S$ is supported by $P^S$, then $S$ is supported by $P$.

  \vspace{.3cm}
  (4) By Definition \ref{def:NNP-rule} and By Theorem \ref{THE:CLAUSAL-PATTERN}, we have:
   $$H_\Delta =\wedge [ \, (\bot \ \vee \ \Delta_1) \ldots (\bot \ \vee \ \Delta_k) \, ]$$
   
   \vspace{.2cm}
  \quad \ By Definition \ref{cor:NNP-CLOSED-SUPPORTED}, $S$ is a model of $\Sigma$
  only if \ $\forall i, S \not \fmodels \Delta_i$. 
  
   \vspace{.2cm}
  \quad \ If $S$ is an answer set of $P$ and a model of $\Sigma$, then $S$ is an answer set 
  of $P \cup \Sigma$.   
\end{niceproof}

\vspace{.4cm}
\noindent \hspace{5.5cm} \underline{\large \bf  Proofs from Section \ref{sec:PUR-HPUR}} 

 \vspace{.2cm}
\noindent {\bf Proposition   \ref{pro:UR-CNF-COINCIDES}.} $\mathcal{UR}$  particularized to CNF coincides with unit-resolution.

\begin{niceproof} We consider first the rule NUR:
$$\frac{{\textcolor{black} \ell} \ \ {\textcolor{black} {\wedge}} 
\ \ \Pi \cdot  (\Sigma  \ \, {\textcolor{black} {\vee}} \ \, 
 \Delta({\textcolor{black} {\neg {\, \ell \ }}}))}
{{\textcolor{black} \ell} \  {\textcolor{black} {\wedge}} \ \Pi \cdot  \Sigma     }
{\mbox{\ \ (NUR)}}$$

It is trivial that, for CNF expressions, we have $\Pi \cdot  (\Sigma  \ \, {\textcolor{black} {\vee}} \ \, 
 \Delta({\textcolor{black} {\neg {\, \ell \ }}}))= (\Sigma  \ \, {\textcolor{black} {\vee}} \ \, 
 \Delta({\textcolor{black} {\neg {\, \ell \ }}}))$, $\Delta({\textcolor{black} {\neg {\, \ell \ }}})= {\neg {\, \ell \ }}$ 
 and $\Sigma=\vee (\ell_1 \ldots \ell_n)$. By substituting these values in NUR, we obtain:
 $$\frac{{\textcolor{black} \ell} \ \ {\textcolor{black} {\wedge}} 
\  {\textcolor{black} {\vee}} (\ell_1 \ldots \ell_n  \ \,   {\neg {\, \ell \ }})} 
{{\textcolor{black} \ell} \  {\textcolor{black} {\wedge}} \ (\ell_1 \ldots \ell_n)     }
{\mbox{\ \ (UR)}}$$

And the latter is obviously classical unit-resolution. By carrying out  similar  operations
with the simplification rules of $\mathcal{UR}$, one verifies that the proposition holds.
\end{niceproof}

\vspace{.3cm}
\noindent{\bf Lemma \ref{Lem:Correctness-UR}.} When  $\mathcal{UR}$  is  applied to a not-free Horn expression $H$ 
until   no  inference is inapplicable or $\bot$ is derived,
      $  LM(H)$  is derived if $H$ is consistent  and  $\bot$ otherwise.

\begin{niceproof} (Sketch) We analyze below both directions of the lemma.

\vspace{.15cm}
 $ \bullet \Rightarrow$ Assume that $H$ is consistent.     
 {\em NUR}  is iteratively applied until
 an expression $H' $ different from $ \bot$ is obtained. Clearly, 
 if the {\em NUR} numerator is not applicable then 
there is not a conjunction of a literal $\ell$ with  a disjunction 
 including   $\neg \ell$.
 Then we have: (i) since {\em NUR} is sound,  $ H \equiv H' $; and
(ii) if $H'$ has   complementary literals $\ell$ and  $\neg \ell$, then they are integrated in
 disjunctions. Thus   $H'$ is satisfied by assigning the value F to all its unassigned propositions, since,  by Definition \ref{def:Positive-Horn-Nested} of Horn expression, 
 all  disjunctions have at least one negative  disjunct 
 $\varphi^- \in \mathcal{N}_\mathcal{L}$. 
 Therefore, since $H'$ is consistent so is  $H$.
  $ LM(H)$ is formed by the  literals in the last derived expression $H'$, 
i.e. $\ell \in LM(H)$ iff $H'=\ell \wedge H''$, which are exactly those employed 
by the NUR rule and expressed by Theorem   \ref{TH:Tarski55-Endem76-NFNP}.

\vspace{.15cm}
$\bullet \ \Leftarrow $ Assume that $H$ is inconsistent. 
Then $H$ must have
a sub-formula verifying the {\em NUR}  numerator, since
otherwise, it implies that    all 
complementary pairs 
$\ell$ and  $\neg \ell$  in $H$ are 
included in disjunctions. In the latter case, 
by Definition \ref{def:Positive-Horn-Nested},
 all disjunctions of $ H $ have
at least one negative disjunct $ \varphi^- \in \mathcal{N}_\mathcal{L} $, and then, $ H $ is satisfied by assigning 
the value $F$ to all propositions, 
 contradicting the initial hypothesis. 
Therefore, {\em NUR} is applied to $ H $ with 
two complementary  literals $\ell$ and  $\neg \ell$ and the resulting Horn expression 
$H'$    is simplified with the simplification rules 
$\{\vee^\bot, \wedge^\bot, \odot^{k+1}, \odot^{k+n}\}$. 
The new  formula $H'$ both, it is logically equivalent to $ H $ 
and has at least one literal less than $H $. 
Hence, by induction on the number of literals
of $ H $, we  obtain that  $ \mathcal{UR}$   ends only
 when $  \bot$ is derived.
\end{niceproof}

\noindent {\bf  Lemma \ref{lem:POLYNOmial-UR}.}  Given a not-free Horn expression $H$, 
the time to determine either  $LM(H)$ or the inconsistency of $H$
is bounded polynomially.

\begin{niceproof} (Sketch) Each application of rule NUR deletes at least
one literal, and so the maxim  number required of such inference is bounded
by the   size of the input $H$. Each application
of the simplification rules $\{\vee^\bot, \wedge^\bot, \odot^{k+1}, \odot^{k+n}\}$ removes at least a connective, and 
hence, the   maxim number  required of such inferences is also limited
 by the formula size.
On the other hand, it is not difficult to find
data structures to execute polynomially each rule, and so, the lemma holds.
\end{niceproof}

\noindent {\bf Theorem \ref{the:NNP-Programming-COMPLEXITY}.} We have the following complexities 
related to NNP programs:

\vspace{.2cm}
(1) Determining  $LM(P)$ of a not-free NNP program $P$
    is in $\mathcal{P}$. 
    
    \vspace{.1cm}
    (2) {\em not-}free NNP programming is $\mathcal{P}$-complete. 
    
     \vspace{.1cm}
   (3) Determining whether an NNP program $P$ has an answer set is $\mathcal{NP}$-complete.

\begin{niceproof} (Sketch)   Below, we sketch the proofs of the statements:

\vspace{.15cm}
(1) As exposed at the beginning of Section \ref{sec:PUR-HPUR},
 each rule is immediately transformed in a Horn expression and this is done without incrementing
the size. Therefore, an input program is transformable into a Horn expression
preserving the size. Hence, (1)
 follows from the latter fact and from Lemma \ref{lem:POLYNOmial-UR}.
 
 \vspace{.1cm}
 (2)  is consequence of statement (1) and 
the facts that NNP programming subsumes classical NP programming and the latter
is well-known to be $\mathcal{P}$-complete \cite{DantsinEGV01}. 

\vspace{.1cm}
(3) One can select a potential answer set $I$ and  
obtain the $I$-reduct of program $P$ in polynomial time.
Then according to claim (1),   one can determine $LM(P^I)$ in polynomial time.
Finally,   whether $I$ verifies $I=LM(P^I)$ is also checkable  polynomially.
\end{niceproof}

\vspace{.3cm}
\noindent {\bf Acknowledgments.} This work was supported by the Spanish project ISINC (PID2022-139835NB-C21) funded by
MCIN/AEI/10.13039/501100011033.


\end{document}